\documentclass[11pt,a4paper]{article}
\usepackage[paperheight=14in,paperwidth=14in]{geometry}
\usepackage{keyval,amsfonts,slashed,bm}
\usepackage{graphicx,jheppub,textcomp}
\usepackage[colorlinks=true,linktocpage=true,linkcolor=blue,citecolor=blue]{hyperref}
\usepackage{epsfig,amsmath,amssymb}
\usepackage{subfig}
\usepackage{graphicx}

\usepackage{color}
\newcommand{\ba}{\begin{eqnarray}}
\newcommand{\ea}{\end{eqnarray}}

\newcommand{\nn}{\nonumber\\}

\definecolor{sinopia}{rgb}{0.8,0.25,0.04}

\definecolor{greenopia}{rgb}{0.3,0.65,0.14}

\newcommand{\mdot}{\hspace{-1.2mm}\cdot\hspace{-1.2mm}}

\title{NLO quark self-energy and dispersion relation using the hard thermal loop resummation}
\author[a]{Sumit,}
\author[b]{Najmul Haque}
\author[a]{and Binoy Krishna Patra}
\affiliation[a]{Department of Physics,
	Indian Institute of Technology Roorkee, Roorkee 247667, India}
\affiliation[b]{School of Physical Sciences, National Institute of Science Education and Research, An OCC of Homi Bhabha National Institute, Jatni-752050, India}
\emailAdd{sumit@ph.iitr.ac.in}
\emailAdd{nhaque@niser.ac.in}
\emailAdd{binoy@ph.iitr.ac.in}

\abstract{Using the hard-thermal-loop (HTL) resummation in real-time formalism, we study the next-to-leading order (NLO) quark self-energy and corresponding NLO dispersion laws. In NLO, we have replaced all the propagators and vertices with the HTL-effective ones in the usual quark self-energy diagram. Additionally, a four-point vertex diagram also contributes to the quark NLO self-energy. We calculate the usual quark self-energy diagram and the four-point vertex diagram separately. Using those, we express the NLO quark self-energy in terms of the three- and four-point HTL-effective vertex functions. Using the Feynman parametrization, we express the integrals containing the three- and four-point HTL effective vertex functions in terms of the solid angles. After completing the solid angle integrals, we numerically calculate the momentum integrals in the NLO quark self-energy and plot them as a function of the ratio of momentum and energy. Using the NLO quark self-energy, we plot the NLO correction to dispersion laws.}

\keywords{Hard thermal loop, dispersion relation, Next-to-leading order}

\allowdisplaybreaks
\begin{document}
	{\Large
		\maketitle 
		\flushbottom
\section{Introduction}\label{Int}
The standard perturbative loop expansion in quantum chromodynamics (QCD) encounters several issues at finite temperatures. One of those issues is that the physical quantities, for example, dispersion laws, become gauge-dependent. In Ref.~\cite{Kalashnikov:1979cy}, the authors have calculated the QCD polarisation tensor at finite temperature and chemical potential in one-loop order, and then the gluon dispersion laws to leading order in the QCD coupling $g$. Also, the authors of ~\cite{Kalashnikov:1979cy} have shown that the dispersion laws are gauge-invariant, but the damping rates for gluons at one-loop-order are not in the long-wavelength limit. One of the important outcomes of the above reference tells us that the Debye screening in the chromoelectric mode does occur in the leading order in the one-loop calculation. However, chromomagnetic screening is absent in the lowest order. In leading order, the non-screening of chromomagnetic fields is also studied in refs.~\cite{Linde:1978px,Linde:1980ts,Gross:1980br}. In ref.~in~\cite{Klimov:1981ka}, the massless spectrum of the elementary quark excitations at the lowest order in $g$ is studied, and the full quasiparticle spectra to leading order for the whole momentum range are given in~\cite{Klimov:1982bv} at the high-temperature limit. The quasiparticle spectra were also studied in ref.~\cite{Weldon:1982aq} for gluons and~\cite{Weldon:1982bn} for quarks.
		
The issue related to the gauge dependence of the gluon damping rates, which have been calculated up to one-loop order, particularly at zero momentum, in different gauges and schemes, has been studied, and different outcomes have been obtained in~\cite{Kobes:1987bi}. Later, it was concluded that the lowest-order result is incomplete, and higher-order diagrams can contribute to lower orders in powers of the QCD coupling~\cite{Pisarski:1988vd}. In other words, the standard loop expansion is no longer valid in powers of the QCD coupling $g$. In a series of works done by Braaten and Pisarski, the issue is resolved. They developed a systematic theory for an effective perturbative expansion that sums the higher-order terms into effective propagators and effective vertices~\cite{Braaten:1989mz,Braaten:1989kk,Frenkel:1989br,Bellac:2011kqa} and known as hard-thermal loop (HTL) resummation. Using the effective HTL propagators and vertices, the transverse part of the gluon damping rate $\gamma_{t}(0)$ at vanishing momentum was calculated in ref.~\cite{Braaten:1990it}, and it was found to be finite, positive, and gauge independent.
		
Once developed, the important thing to check out is the reliability of the HTL-summed perturbative calculations in gauge theories at high temperatures. If so, it can be considered an important candidate for describing the quark-gluon plasma (QGP) properties. One of the exciting questions is the infrared sensitivity of the massless gauge theories, which worsens at finite temperatures due to the Bose-Einstein distribution (B.E.) function. As the B.E. distribution function behaves like $1/k$ for very vanishing gluon energies $k$, physical quantities are more infrared sensitive at finite temperature~\cite{Burgess:1991wc,Rebhan:1992ca,Rebhan:1993az}.
		
Using the HTL resummed propagators and vertices, many studies have been performed in perturbative QCD to acknowledge the thermodynamic attributes of plasma. For example, the pressure and quark number susceptibilities up to two- and three-loop order have been studied using the thermodynamic potential\cite{Haque:2014rua,Haque:2013sja,Haque:2013qta,Haque:2012my,Andersen:2011sf,Andersen:2010wu,Jiang:2010jz}. In addition, the electric and magnetic properties of the plasma were investigated in~\cite{Liu:2011if}. Using the HTL summation, it has been found that massless quarks and gluons acquire the thermal masses of order $gT$, $m_q$ and $m_g$ respectively~\cite{Kalashnikov:1979cy,Klimov:1981ka,Klimov:1982bv,Weldon:1982aq,Weldon:1982bn}, which shows that for the lowest order $gT$ in effective perturbation, the infrared region is ‘okay.’ However, as we mentioned earlier, the static chromomagnetic field does not screen at the lowest order; instead, it gets screened at the next order, so-called magnetic screening~\cite{Linde:1978px,Linde:1980ts,Gross:1980br}. Thus, if we want to demonstrate the infrared sector of the HTL perturbative expansion,  we need to go beyond the leading-order calculations. Recently many other calculations have been performed at NLO in thermal field theory to probe the hot and dense QCD matter. For example, the transport coefficient at NLO, namely the ratio of shear viscosity to entropy density $\eta/s$ and the ratio of the quenching parameter to temperature $\hat{q}/T^{3} $ have been obtained in {\cite{Muller:2021wri}}, HTL lagrangian has been derived at NLO for the photon in \cite{Carignano:2019ofj}, two loops HTL’s have been derived for any general model recently in ref.~\cite{Ekstedt:2023anj}, explicit results for the gluon self-energy in semi-QGP at NLO have been obtained in \cite{Wang:2022dcw}. Also, the complete calculation of soft photon self-energy at NLO in QED is presented in \cite{Gorda:2022fci}. Using that cold and dense electron gas pressure at N$^{3}$LO has been obtained in \cite{Gorda:2022zyc}.
		
To probe the infrared sector of HTL perturbative expansion, the first calculation performed using NLO HTL-dressed perturbative expansion is the non-moving gluon damping rate~\cite{Braaten:1990it}. In a similar line, the non-moving quark damping rate has been calculated~\cite{Kobes:1992ys,Braaten:1992gd} in imaginary time formalism~\cite{Bellac:2011kqa,Mustafa:2022got,Kapusta:2006pm}. The quark damping rate of a non-moving quark was recently calculated in~\cite{Carrington:2006gb} using real-time formalism. After the study of non-moving gluon and quark damping rates in 1992, later, by using this formalism,  the damping rates of slow-moving longitudinal~\cite{Abada:1998ue,Abada:1997vm} and transverse gluons~\cite{Abada:2004dr} in the one-loop order HTL, quarks damping rate~\cite{Abada:2007opj,Abada:2005na,Abada:2000hh}, electrons~\cite{Abada:2007zz} and photons \cite{Abada:2011cc} damping rate in QED, and also quasiparticles energy up to NLO in scalar QED~\cite{Abada:2005jq} have been studied.
		
After the rigorous studies of boson and fermion damping rates, people also studied the quasiparticle energies up to NLO. That would come from studying the real part of the HTL effective one-loop self-energies. The NLO quasiparticle energy calculation is much non-trivial than the computation of the damping rate. It was started by determining the plasma frequency $\omega_g(0)$ for pure gluon case up to NLO in the long-wavelength limit in ref.~\cite{Schulz:1993gf}. A gauge-invariant plasma frequency in next-to-leading order for pure-gluon plasma is established with $N_{f}=0$ in imaginary time formalism as
\begin{equation}
\omega_{g}(0)=\frac{\sqrt{c_{A}}}{3} g T\left(1-0.09 \sqrt{c_{A}} g+\ldots\right), 
\label{omega_nlo}
\end{equation}
with Casimir number $c_A=N_c$ and $N_c$ represents the number of colors in QCD, $g$ is the QCD coupling constant, and $T$ is the temperature. The first term of eq.~\eqref{omega_nlo} represents the leading order part, whereas the second represents the NLO contribution. Later, fermion mass up to NLO in QCD at high temperature (and QED) was calculated in refs.~\cite{Carrington:2006xj,Carrington:2006gb,Carrington:2008dw}. The NLO quark mass for two-flavour QCD is obtained in ref.~\cite{Carrington:2008dw} in the real-time formalism~\cite{Bellac:2011kqa,Landsman:1986uw,Chou:1984es} as
\begin{equation}
m_q^{nlo}=\frac{g T}{\sqrt{6}}\left(1+\frac{1.87}{4 \pi} g+\ldots\right).
\label{m_q_nlo}
\end{equation}
The NLO contributions to the plasma frequency and quark mass are from the one-loop diagrams involving soft momenta. Indeed, using the general power-counting arguments done in~\cite{Mirza:2013ula}, it has been shown that, except for the photon self-energy, NLO corrections come only from the soft one-loop resummed graphs with effective vertices and propagators. While for the former case, two-loop graphs with hard internal momentum also contribute. In work \cite{Carrington:2006gb} using the power-counting arguments, it has been shown that in imaginary-time formalism, one would not get the correct information about the number of terms that gives subleading contributions. In ref.~\cite{Abada:2014bma}, the authors have set up a framework to calculate the NLO dispersion relation for a slow-moving quark. The authors used the setup to show how one can proceed to evaluate the different terms in the NLO part of quark self-energy.  
In the present work, we calculate all sixteen terms of NLO quark self-energy to complete the NLO dispersion relations, i.e., energy and damping rates, for quarks moving slowly within HTL approximation in real-time formalism. This has been done by utilizing the closed-time-path (CTP) approach of thermal field theory~\cite{Martin:1959jp,Keldysh:1964ud}. The benefit of avoiding the analytic continuation from Matsubara frequencies to real energies. For the latter case,  it is non-trivial to obtain the analytical responses of physical observables. Nevertheless, one issue is that, because of the doubling of degrees of freedom, each $N$-point function gets a tensor structure, in which we have to work with $2N$ components. Thus, there will be enough rise in graphs containing three and four-point vertex integrals. Also, this computation will not be easy even if one sets the quark momentum to be zero from the starting point, as done in~\cite{Chou:1984es}, where they replace the momentum contractions of effective vertices by effective self-energy differences using Ward identities.
		
This work is sketched in this way. In section~\ref{NLO}, we discuss the lowest-order dispersion relations for quarks and the expressions of the effective quark and gluon propagators. Then in section~\ref{sec:nlo_formalism}, we present the NLO formalism for slow-moving quarks. In this section, we present the expression for NLO dispersion laws from which one can extract the NLO quark energies and damping rates of order $g^{2}T$. These quantities directly correspond with the HTL effective NLO quark self-energy $\Sigma^{(1)}$. This contribution $\Sigma^{(1)}$ to the NLO quark self-energy is calculated further in this section, and we present a detailed analytic result of $\Sigma^{(1)}$ in terms of the three- and four-point HTL effective vertex integrals. In section~\ref{sec4}, we evaluate the different terms of NLO quark self-energy and describe how these terms have been evaluated numerically. In section~\ref{sec:result}, we have plotted the NLO quark self-energy w.r.t. the ratio of momentum and energy. Then we plotted the NLO correction to dispersion relations using the above quantity. We encapsulate the paper in the last section~\ref{sec:summary}.
		
In appendix~\ref{sec:appendixA},  we present the derivation of the three and four-point HTL vertices. After that, in appendix~\ref{appendixB}, by using the Feynman technique, the solid-angle integrals have been calculated, which are present in the vertex HTL. In appendix~\ref{sec:appendixC}, we give the derivation of effective gluon and quark propagators utilized in section four.

\section{Lowest order dispersion relations}\label{NLO} 
\begin{figure}[t]
	\centering
	\includegraphics[width=7cm,height=9cm,keepaspectratio]{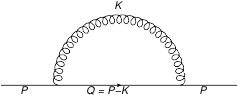}
	\caption{Feynman graph for the quark self-energy in leading order}
	\label{fig1}
\end{figure}
One-loop effective quark propagator can be written as
\begin{equation}
S(P)=\frac{1}{\slashed{P} - \Sigma(P)},\label{q_prop}
\end{equation}
Where $P = (p_0, \vec{p})\equiv (p_0, \hat{p}\,p)$ is the four-momentum of external quark and $\Sigma{(P)}$ is the one-loop HTL quark self-energy shown in figure~\ref{fig1}.
The zeros of the denominator of the propagator in eq.~\eqref{q_prop} give the dispersion  relations as
\begin{equation}\label{q_disp}
\text{det} [\slashed{P} - \Sigma(P)]=0,
\end{equation}
The self-energy $\Sigma(P)$ can be decomposed into helicity eigenstates as
\ba
\Sigma(P) &=& \gamma_{+p}\  \Sigma_{-}(P)+\gamma_{-p}\  \Sigma_{+}(P) .\label{q_self_energy}
\ea 
Here, $\gamma_{\pm p} \equiv\left(\gamma^{0} \mp \vec{\gamma} \cdot \hat{p}\right) / 2 $ , with $\hat{p} = \vec{p}/p $ and $\gamma^{\mu}$ are the Dirac matrices. In the lowest order, the quark self-energy $\Sigma_{{\pm}} (P)$ in eq.~\eqref{q_self_energy} is calculated as
\begin{equation}\label{q_self_lo}
\Sigma_{\pm}(\omega, p)= \frac{m_{q}^{2}}{p}\left[\pm 1+\frac{1}{2}\left(1 \mp \frac{\omega}{p}\right) \ln \frac{\omega+p}{\omega-p}\right] ,
\end{equation}
\noindent where $m_{q}^2= \frac{C_F}{8} g^2T^2 $ represents the square of the thermal mass of a quark in leading order at zero chemical potential with $ C_{F} = (N^{2}_{c}-1)/2N_{c}$. Equation~\eqref{q_disp} can be summarized in the following two dispersion relations: 
\begin{equation}\label{disp_two}
p_0 \mp p - \Sigma_{\pm}(P) = 0 .
\end{equation}
The solution of the dispersion relations in the lowest order is $p_0=\Omega^{(0)}_\pm$, and they are displayed in figure~\ref{fig2}. 
\begin{figure}[tbh]
	\centering
	\includegraphics[scale=.8,keepaspectratio]{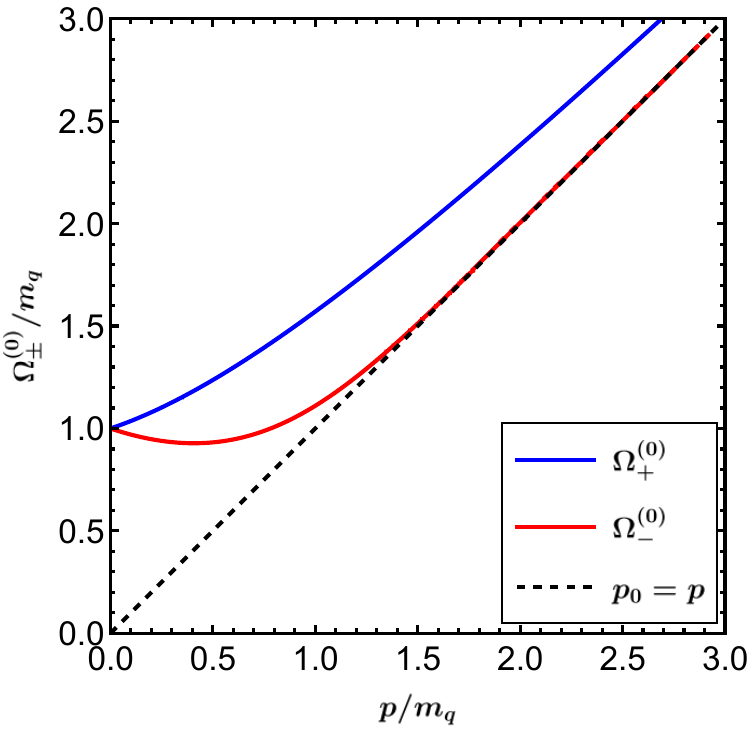}  
	\caption{Lowest-order dispersion laws for quark excitations}
	\label{fig2}
\end{figure}
The $\Omega^{(0)}_+(p)$ mode represents the propagation of an ordinary quark with a momentum-dependent thermal mass, and the ratio of chirality to helicity for this mode is $+1$. On the other hand, the $\Omega_-^{(0)}(p)$ mode represents the propagation of a quark mode for which the chirality to helicity ratio is $-1$. This mode is called the plasmino mode, and it is absent at zero temperature but appears as a consequence of the thermal medium due to the broken Lorentz invariance~\cite{Mustafa:2002pb}. At large momenta, two modes go to the light cone very quickly. Whereas the soft portion is effectively constricted. For soft external momenta $(p/m_{q} < 1)$, the solution of the dispersion relations can be expanded as

\begin{equation}\label{disp_sol}
\Omega^{(0)}_{\pm}(p)=m_{q}\bigg[1 \pm \frac{1}{3}\frac{p}{m_q}+\frac{1}{3} \frac{p^2}{m_q^2} \mp \frac{16}{135} \frac{p^3}{m_q^3}+\frac{1}{54} \frac{p^{4}}{m_q^4} \pm \frac{32}{2835} \frac{p^{5}}{m_q^5}-\frac{139}{12150} \frac{p^{6}}{m_q^6} \pm \ldots\bigg].
\end{equation}

Using the HTL self-energies $\Sigma_\text{HTL}$ defined in eq.~\eqref{q_self_lo}, one can write the one-loop effective quark propagator, which can also be decomposed into the helicity eigenstates as
\begin{equation}\label{q_prop_htl}
\begin{aligned}
\Delta(P) &=\gamma_{+p} \Delta_{-}(P)+\gamma_{-p} \Delta_{+}(P) ; \quad \quad
\Delta_{\pm}^{-1}(P) &=p_{0} \pm p-\frac{m_{q}^{2}}{p}\left[\mp 1+\frac{1}{2 p} m_{q}^{2}\left(p \pm p_{0}\right) \ln \frac{p_{0}+p}{p_{0}-p}\right] .
\end{aligned}
\end{equation}
Since the quark damping rate comes from the negative of the imaginary part of self-energy, there is no quark damping at the lowest order, and it starts to contribute from NLO.\\
In addition to the leading order quark self-energy, we also require an HTL-dressed gluon propagator to calculate the NLO contribution of the quark self-energy. In covariant gauge, a one-loop HTL resummed gluon propagator is 
\begin{equation}
D_{\mu \nu}(K)=\frac{\xi K_\mu K_\nu}{K^4}+D_{T}(K) A_{\mu \nu}+D_{L}(K) B_{\mu \nu},
\end{equation}
Here, $A_{\mu \nu}$ and $B_{\mu \nu}$ are the transverse and longitudinal projection operators, respectively, and can be expressed as

\begin{equation}
A_{\mu \nu}=g_{\mu \nu}-\frac{K_{\mu} K_{\nu}}{K^{2}}-B_{\mu\nu} ; \quad B_{\mu \nu}=-\frac{K^2}{k^{2}}\left(u_{\mu}-\frac{k_{0} K_{\mu}}{K^{2}}\right)\left(u_{\nu}-\frac{k_{0} K_{\nu}}{K^{2}}\right),
\end{equation}

Where $u^\mu$ is the four-velocity of the heat bath and in the plasma rest-frame, $u^\mu = (1, \vec{0}) $. The quantities $D_{L, T}$ are the longitudinal and transverse HTL effective gluon propagators, respectively, and given by
\ba
D_{L}^{-1}(K)&=&K^{2}+2 m_{g}^{2} \frac{K^{2}}{k^{2}}\left(1-\frac{k_{0}}{2 k} \ln \frac{k_{0}+k}{k_{0}-k}\right); \quad \quad
D_{T}^{-1}(K) = K^{2}-m_{g}^{2}\left[1+\frac{K^{2}}{k^{2}}\left(1-\frac{k_{0}}{2 k} \ln \frac{k_{0}+k}{k_{0}-k}\right)\right].\label{glu_prop}
\ea
In the above eq.~\eqref{glu_prop}, $ m_g = \frac{1}{6}\sqrt{(N_c + s_{F})}\ gT $ is the gluon thermal mass to leading order. Here, $s_{F} = N_{f}/2$, and  $N_{f}$ represents the number of quark flavors. 
\section{NLO formalism}\label{sec:nlo_formalism}
%
For on shell quarks, we write the (complex) quark energy $p_0 \equiv \Omega(p)$ 
as
\begin{equation}\label{disp_decom}
\begin{aligned}
\Omega(p)= \Omega^{(0)}(p) + \Omega^{(1)}(p) + \cdots \hspace{2mm} .
\end{aligned}
\end{equation}
A similar kind of approach is also relevant for self-energy $\Sigma$, as well
\begin{equation}\label{sigma_nlo}
\begin{aligned}
\Sigma(P)= \Sigma_\text{HTL}(P)+ \Sigma^{(1)}(P)+\cdots ,
\end{aligned}
\end{equation}
where $\Sigma_\text{HTL}$ is the lowest-order quark self-energy having $gT$ order, whereas $\Sigma^{(1)}$ the NLO contribution of quark self-energy, with order $g^{2}T$. Similarly, the first term $\Omega^{(0)}(p)$ and second term $\Omega^{(1)}(p)$ in eq.~\eqref{disp_decom} are of the same order as terms on r.h.s. of eq.~\eqref{sigma_nlo}. Thus, eq.~\eqref{disp_two} will take the form as
\begin{equation}
\Omega^{(0)}_{\pm}(p)+\Omega_{\pm}^{(1)}(p) +\cdots= \pm p + \left.\Sigma_{\mathrm{HTL} \pm}\left(\omega, p\right)\right|_{\omega\rightarrow\Omega_{\pm}(p)}+\left.\Sigma_{\pm}^{(1)}\left(\omega, p\right)\right|_{\omega\rightarrow\Omega_{\pm}(p)}+\ldots \hspace{2mm} . 
\end{equation}
\begin{figure}[t]
	\centering
	\includegraphics[width=8cm,height=8cm,keepaspectratio]{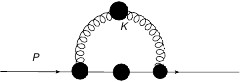}
	\caption{Feynman graph for the NLO HTL resummed quark self-energy $ \Sigma_{1}^{(1)} $. The black blobs indicate HTL effective quantities. All momenta are soft.}
	\label{fig3}
\end{figure}
Since we are interested in slow-moving quarks, we can take $p \sim gT$, and we get
\begin{equation}\label{disp_nlo0}
\Omega_{\pm}^{(1)}(p)=\frac{\Sigma_{\pm}^{(1)}\left(\Omega^{(0)}_{\pm}(p), p\right)}{1-\left.\partial_{\omega} \Sigma_{\mathrm{HTL} \pm}(\omega, p)\right|_{\omega=\Omega^{(0)}_{\pm}(p)}} .
\end{equation}
Here $\partial_{\omega}$ represents variation w.r.t $\omega$. Real values of above eq.~\eqref{disp_nlo0} give us the NLO corrections to the momentum-dependent quark energies, whereas the NLO contribution to the quark damping rate comes from the negative of the imaginary part of eq.~\eqref{disp_nlo0}. Also, by using the expression of $\Sigma_\text{HTL}$ mentioned in eq.~\eqref{q_self_lo}, the NLO dispersion relations will take the final form as
\begin{equation}\label{disp_nlo}
\Omega_{\pm}^{(1)}(p)=\frac{{\Omega^{(0)}_{\pm}}^{2}(p)-p^{2}}{2 m_{q}^{2}} \Sigma_{\pm}^{(1)}\left(\Omega^{(0)}_{\pm}(p), p\right) . 
\end{equation}
We need to determine the NLO quark self-energy to evaluate the above equation. For that, we have to consider two one-loop graphs with effective vertices, shown \footnote{The Feynman graphs are drawn using jaxodraw software~\cite{Binosi:2003yf}.} in figure~\ref{fig3} and figure \ref{fig4}. The graph in figure~\ref{fig3} can be written as
\begin{equation}
\begin{aligned}
\Sigma_{1}^{(1)}(P)=-i g^{2} C_{F} \int \frac{d^{4} K}{(2 \pi)^{4}} \Gamma^{\mu}(P, Q) \Delta(Q) \Gamma^{\nu}(Q, P) D_{\mu \nu}(K),
\end{aligned}
\end{equation}
where $Q = P - K$. Similarly, from the vertex graph in figure~\ref{fig4}, we can write
\begin{equation}
\begin{aligned}
\Sigma_{2}^{(1)}(P)=\frac{-i g^{2} C_{F}}{2} \int \frac{d^{4} K}{(2 \pi)^{4}} \Gamma^{\mu \nu}(P, K) D_{\mu \nu}(K).
\end{aligned}
\end{equation}
\begin{figure}[t]
	\centering
	\includegraphics[scale=1.9,keepaspectratio]{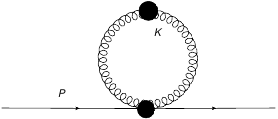}
	\caption{Feynman diagram for the NLO quark self-energy $ \Sigma_{2}^{(1)}$. The black blobs represent HTL effective quantities. All momenta are soft.}
	\label{fig4}
\end{figure}
We have three possible summation indices for the above equations: Lorentz (explicit), real-time-field (RTF), and Dirac. Firstly, we start with the Keldysh indices on the ``r/a" basis, utilizing the thermal field theory's closed-time-path (CTP) formulation. For the fermion,  the retarded (R), advanced (A), and symmetric (S) propagators can be defined as 
\ba
\Delta^{\mathrm{R}}(K) &\equiv& \Delta^{\mathrm{ra}}(K)=\Delta\left(k_{0}+i \varepsilon, \vec{k}\right); \quad \quad
\Delta^{\mathrm{A}}(K)  \equiv   \Delta^{\mathrm{ar}}(K)=\Delta\left(k_{0}-i \varepsilon, \vec{k}\right);\nn
\Delta^{\mathrm{S}}(K) &\equiv& \Delta^{\mathrm{rr}}(K)=N_{F}\left(k_{0}\right) \operatorname{sign}\left(k_{0}\right)\left[\Delta^{\mathrm{R}}(K)-\Delta^{\mathrm{A}}(K)\right].\label{RAS_F}
\ea
Similarly, for boson
\ba
D_{\mu\nu}^{\mathrm{R}}(K) &\equiv& D_{\mu\nu}^{\mathrm{ra}}(K)=D_{\mu\nu}\left(k_{0}+i \varepsilon, \vec{k}\right); \quad \quad
D_{\mu\nu}^{\mathrm{A}}(K)  \equiv  D_{\mu\nu}^{\mathrm{ar}}(K)=D_{\mu\nu}\left(k_{0}-i \varepsilon, \vec{k}\right);\nn
D_{\mu\nu}^{\mathrm{S}}(K) &\equiv& D_{\mu\nu}^{\mathrm{rr}}(K)=N_{B}\left(k_{0}\right) \operatorname{sign}\left(k_{0}\right)\left[D_{\mu\nu}^{\mathrm{R}}(K)-D_{\mu\nu}^{\mathrm{A}}(K)\right],\label{RAS_B}
\ea
where $N_{F}$ in eq.~\eqref{RAS_F} and $N_{B}$ in eq.~\eqref{RAS_B} are defined as follows
\begin{equation}\label{dist_func}
N_{B,F}(k_{0}) = 1\pm 2n_{B,F}(k_{0});   \quad  n_{B,F}(k_{0})= \frac{1}{e^{|k_{0}|/T} \mp 1}.
\end{equation}
The modulus value in the argument of Bose-Einstein (B.E) and Fermi-Dirac distribution function is required to avoid the blow-up of the function. Thus, for the two components of $\Sigma^{(1)}_{1}$, we have the following explicit expressions.
\ba
\Sigma_{(1) \pm}^{(1)}(P)&=& \frac{-i g^{2} C_{F}}{2} \int \frac{d^{4} K}{(2 \pi)^{4}} \operatorname{tr} \gamma_{\pm p}\bigg[\Gamma_{\operatorname{arr}}^{\mu}(P, Q) \Delta^{\mathrm{R}}(Q) \Gamma_{\operatorname{arr}}^{\nu}(Q, P) D_{\mu \nu}^{\mathrm{S}}(K)\nn
&+&\Gamma_{\mathrm{arr}}^{\mu}(P, Q) \Delta^{\mathrm{S}}(Q) \Gamma_{\operatorname{rar}}^{\nu}(Q, P) D_{\mu \nu}^{\mathrm{A}}(K) +\Gamma_{\mathrm{arr}}^{\mu}(P, Q) \Delta^{\mathrm{R}}(Q) \Gamma_{\text {aar }}^{\nu}(Q, P) D_{\mu \nu}^{\mathrm{A}}(K) \nn
&+&\Gamma_{\operatorname{arr}}^{\mu}(P, Q) \Delta^{\mathrm{R}}(Q) \Gamma_{\operatorname{arr}}^{\nu}(Q, P) D_{\mu \nu}^{\mathrm{R}}(K) +\Gamma_{\operatorname{ara}}^{\mu}(P, Q) \Delta^{\mathrm{A}}(Q) \Gamma_{\mathrm{rar}}^{\nu}(Q, P) D_{\mu \nu}^{\mathrm{A}}(K)\bigg] ,\label{sigma_1}
\ea
and for the two components of $\Sigma^{(1)}_{2}$, we get
\ba
\hspace{-5mm}\Sigma_{(2) \pm}^{(1)}(P)&=& \frac{-i g^{2} C_{F}}{4} \int \frac{d^{4} K}{(2 \pi)^{4}} \operatorname{tr} \gamma_{\pm p}\left[\Gamma_{\operatorname{arrr}}^{\mu \nu}(P, K) D_{\mu \nu}^{\mathrm{S}}(K)
+\Gamma_{\operatorname{aarr}}^{\mu \nu}(P, K) D_{\mu \nu}^{\mathrm{R}}(K)+\Gamma_{\operatorname{arar}}^{\mu \nu}(P, K) D_{\mu \nu}^{\mathrm{A}}(K)\right]. \label{sigma_2}
\ea
Equations~\eqref{sigma_1} and \eqref{sigma_2} are the results of the Mathematica program developed in ref.~\cite{Carrington:2006gb} 
which can take care of all the real-time-field indices of a given Feynman diagram.
We rederive all the three and four-point HTL-effective vertex integrals using their corresponding Feynman graphs. Thus, for the two-quarks-one-gluon effective vertices, we will get
\begin{equation}\label{vert_func_3}
\begin{aligned}
\Gamma_{\mathrm{arr}}^{\mu}(P, K) =\gamma^{\mu}+I_{--}^{\mu}(P, K); \quad \quad
\Gamma_{\mathrm{rar}}^{\mu}(P, K) =\gamma^{\mu}+I_{+-}^{\mu}(P, K); \quad \quad
\Gamma_{\mathrm{aar}}^{\mu}(P, K) =\Gamma_{\mathrm{ara}}^{\mu}(P, K)=0,
\end{aligned}
\end{equation}
and two-quarks-two-gluons effective vertices gives 
\begin{equation}\label{vert_func_4}
\Gamma_{\mathrm{arrr}}^{\mu \nu}(P, K)=I_{--}^{\mu \nu}(P, K) ; \quad \Gamma_{\mathrm{aarr}}^{\mu \nu}(P, K)=\Gamma_{\mathrm{arar}}^{\mu \nu}(P, K)=0.
\end{equation}
In eqs.~\eqref{vert_func_3} and \eqref{vert_func_4}, the four vectors $(I$'s) are HTL contribution which are given by
\ba 
I_{\eta_{1} \eta_{2}}^{\mu}(P, Q)&=& m_{q}^{2} \int \frac{d \Omega_{s}}{4 \pi} \frac{S^{\mu} \slashed{{S}}}{\left(P S+i \eta_{1} \varepsilon\right)\left(Q S+i \eta_{2} \varepsilon\right)} ;\label{solid_ang_int1} \\ 
I_{\eta_{1} \eta_{2}}^{\mu \nu}(P, K)&=& m_{q}^{2} \int \frac{d \Omega_{s}}{4 \pi} \frac{-S^{\mu} S^{\nu} \slashed{{S}}}{\left(P S+i \eta_{1} \varepsilon\right)\left(P S+i \eta_{2} \varepsilon\right)} 
\left[\frac{1}{(P+K) S+i \eta_{1} \varepsilon}+\frac{1}{(P-K) S+i \eta_{2} \varepsilon}\right] .
\label{solid_ang_int2}
\ea
Here $S\equiv (1,\hat{s}$) and the indices $\eta_{1} ,\eta_{2} = \pm 1$. Also, the quantity $PS $ and $QS $ are the dot product of four vectors defined as 
$PS= p_{0}-(\vec{p}\!\cdot\!\hat{s})$.
Using the above results, one can rewrite $\Sigma^{(1)}_{(1)\pm}$ from eq.~\eqref{sigma_1} as
\ba
\Sigma_{(1) \pm}^{(1)}(P)&=& \frac{-i g^{2} C_{F}}{2} \int \frac{d^{4} K}{(2 \pi)^{4}} \operatorname{tr} \gamma_{\pm p}\left[\left\{\gamma^{\mu}+I_{--}^{\mu}(P, Q)\right\} \Delta^{\mathrm{R}}(Q)\left\{\gamma^{\nu}+I_{--}^{\nu}(Q, P)\right\} D_{\mu \nu}^{\mathrm{S}}(K)\right.\nn
&+&\left.\left\{\gamma^{\mu}+I_{--}^{\mu}(P, Q)\right\} \Delta^{\mathrm{S}}(Q)\left\{\gamma^{\nu}+I_{+-}^{\nu}(Q, P)\right\} D_{\mu \nu}^{\mathrm{A}}(K)\right].\label{Sigma1pm}
\ea
Similarly, $\Sigma^{(1)}_{(2)\pm}$ can be rewritten from eq.~\eqref{sigma_2} as
\ba
\hspace{-10mm}\Sigma_{(2) \pm}^{(1)}(P)&=&\frac{-i g^{2} C_{F}}{4} \int \frac{d^{4} K}{(2 \pi)^{4}} \operatorname{tr} \gamma_{\pm p} I_{--}^{\mu \nu}(P, K) D_{\mu \nu}^{\mathrm{S}}(K).\label{Sigma2pm}
\ea
Equation~\eqref{Sigma1pm} can be divided into three kinds of terms as
\ba
\Sigma_{(1)\pm}^{(1)}(P)&=& \frac{-i g^{2} C_{F}}{2} \int \frac{d^{4} K}{(2 \pi)^{4}}\left[F_{\pm ; 0}^{\mathrm{SR}}(P, K)+F_{\pm ; 0}^{\mathrm{AS}}(P, K)+2 F_{\pm ;--}^{\mathrm{SR}}(P, K)+F_{\pm ;--}^{\mathrm{AS}}(P, K)\right.\nn
&+&\left.\ F_{\pm ;-+}^{\mathrm{AS}}(P, K)+F_{\pm ;--;--}^{\mathrm{SR}}(P, K)+F_{\pm ;--;+-}^{\mathrm{AS}}(P, K)
\right].\label{Sigma1pm2}
\ea
The first two terms within the square bracket in eq.~\eqref{Sigma1pm2} are due to the bare part of the vertex and can be written in a general form as
\ba 
F_{\epsilon_{p} ; 0}^{\mathrm{XY}}(P, K) &\equiv& \operatorname{tr}\left(\gamma_{\epsilon_{p}} \gamma^{\mu} \gamma_{\epsilon_{q}} \gamma^{\nu}\right) D_{\mu \nu}^{\mathrm{X}}(K) \Delta_{-\epsilon_{q}}^{\mathrm{Y}}(Q)\nn
&=&-2\left(1-\hat{p}_{\epsilon}\mdot\hat{k} \,\hat{q}_{\epsilon}\mdot\hat{k}\right) D_{T}^{\mathrm{X}}(K) \Delta_{-\epsilon_{q}}^{\mathrm{Y}}(Q) \nn
&-&\left[k_{0}^{2}\left(1-\hat{p}_{\epsilon}\mdot\hat{q}_{\epsilon}+2 \hat{p}_{\epsilon}\mdot\hat{k} \,\hat{q}_{\epsilon}\mdot\hat{k}\right)- 2k_{0} k\left(\hat{p}_{\epsilon}\mdot\hat{k}+\hat{q}_{\epsilon}\mdot\hat{k}\right)+k^{2}\left(1+\hat{p}_{\epsilon}\mdot\hat{q}_{\epsilon}\right)\right] \tilde{D}_{L}^{\mathrm{X}}(K) \Delta_{-\epsilon_{q}}^{\mathrm{Y}}(Q) \label{bare_terms}
\ea
Here, $\tilde{D}_{L}(K)={D}_{L}(K)/K^{2}$. Also, $\hat{p}_{\epsilon}= \epsilon_{p}\hat{p} $ with $\epsilon_{p} = \pm $ and similarly for $\hat{q}_{\epsilon}$, with summation over $\epsilon_{q}$. The superscripts $X$ and $Y$ can take the RTF indices values $(R, A,$ and $S)$. The third, fourth, and fifth terms within the square bracket in eq.~\eqref{Sigma1pm2} are the contribution that involves one HTL vertex function and can be written in a general form as
\ba
F_{\epsilon_{p};\eta_{1}\eta_{2}}^{\mathrm{XY}}(P,K) &\equiv& \operatorname{tr}\left(\gamma_{\epsilon_{p}} I_{\eta_{1} \eta_{2}}^{\mu} \gamma_{\epsilon_{q}} \gamma^{\nu}\right) D_{\mu \nu}^{\mathrm{X}}(K) \Delta_{-\epsilon_{q}}^{\mathrm{Y}}(Q) \nn
&&\hspace{-2cm}= m_{q}^{2} \int \frac{d \Omega_{s}}{4 \pi}\frac{1}{\left(PS+i \eta_{1} \varepsilon\right)\left(QS+i \eta_{2} \varepsilon\right)}
\times\bigg[D_{T}^{\mathrm{X}}(K) \Delta_{-\epsilon_{q}}^{\mathrm{Y}}(Q)\nn
&&
  \hspace{-2cm}  \times\Big\{1-\hat{p}_{\epsilon}\mdot\hat{q}_{\epsilon}-\hat{p}_{\epsilon}\mdot\hat{s}-\hat{q}_{\epsilon}\mdot\hat{s}+\hat{p}_{\epsilon}\mdot\hat{k} \,\hat{k}\mdot\hat{s}+\hat{q}_{\epsilon}\mdot\hat{k} \,\hat{k}\mdot\hat{s}-(\hat{k}\mdot\hat{s})^{2}+\hat{p}_{\epsilon}.\hat{q}_{\epsilon}\,(\hat{k}\mdot\hat{s})^{2}
+  2 \hat{p}_{\epsilon}.\hat{s} \,\hat{q}_{\epsilon}.\hat{s}-\hat{p}_{\epsilon}.\hat{k} \,\hat{q}_{\epsilon}\mdot\hat{s}\,\hat{k}\mdot\hat{s}-\hat{q}_{\epsilon}\mdot\hat{k}\,\hat{p}_{\epsilon}\mdot \hat{s}\,\hat{k}\mdot\hat{s}\Big\} \nn
&&\hspace{-1.8cm}-\Big\{k_{0}^{2}\left(\hat{p}_{\epsilon}\mdot\hat{k} \, \hat{k}\mdot\hat{s}+\hat{q}_{\epsilon}\mdot\hat{k} \,\hat{k}\mdot\hat{s}-(\hat{k}\mdot\hat{s})^{2} +\hat{p}_{\epsilon}\mdot\hat{q}_{\epsilon} \,(\hat{k}\mdot\hat{s})^{2} -\hat{p}_{\epsilon}\mdot\hat{k} \,\hat{q}_{\epsilon}\mdot\hat{s} \,\hat{k}.\hat{s}-\hat{q}_{\epsilon}.\hat{k} \,\hat{p}_{\epsilon}\mdot\hat{s}\, \hat{k}\mdot\hat{s}\right) +  k^{2}\left(1+\hat{p}_{\epsilon}\mdot\hat{q}_{\epsilon}-\hat{p}_{\epsilon}\mdot\hat{s}-\hat{q}_{\epsilon}\mdot\hat{s}\right) \nn
&-&  k_{0} k \left(\hat{p}_{\epsilon}\mdot\hat{k}+\hat{q}_{\epsilon}\mdot\hat{k}+2 \hat{p}_{\epsilon}\mdot\hat{q}_{\epsilon} \,\hat{k}\mdot\hat{s} -\hat{q}_{\epsilon}\mdot\hat{k} \,\hat{p}_{\epsilon}\mdot\hat{s}-\hat{p}_{\epsilon}\mdot\hat{k} \, \hat{q}_{\epsilon}\mdot\hat{s}-\hat{q}_{\epsilon}\mdot\hat{s} \, \hat{k}\mdot\hat{s}-\hat{p}_{\epsilon}\mdot\hat{s} \,\hat{k}\mdot\hat{s}\right)
\Big\} \tilde{D}_{L}^{\mathrm{X}}(K) \Delta_{-\epsilon_{q}}^{\mathrm{Y}}(Q)\bigg].\label{htl_terms_1}
\ea
Since $\eta_{1},\eta_{2} = \pm 1$ and because of symmetry in $D_{\mu\nu}$, the other contributions with one HTL vertex integral are the same as above when changing $\eta_{1}$ into $\eta_{2}$, namely,
\begin{equation}
\begin{aligned}
\operatorname{tr}\left(\gamma_{\epsilon_{p}} \gamma^{\mu} \gamma_{\epsilon_{q}} I_{\eta_{1} \eta_{2}}^{\nu}\right) D_{\mu \nu}^{\mathrm{X}}(K) \Delta_{-\varepsilon_{q}}^{\mathrm{Y}}(Q)=\operatorname{tr}\left(\gamma_{\epsilon_{p}} I_{\eta_{2} \eta_{1}}^{\mu} \gamma_{\epsilon_{q}} \gamma^{\nu}\right) D_{\mu \nu}^{\mathrm{X}}(K) \Delta_{-\epsilon_{q}}^{\mathrm{Y}}(Q).
\end{aligned}
\end{equation}
The sixth and seventh terms inside the square bracket of eq.~\eqref{Sigma1pm2} contribute to two HTL vertex functions. It can be written in a general form as
\ba
F_{\epsilon_{p} ; \eta_{1} \eta_{2} ; \eta_{1}^{\prime} \eta_{2}^{\prime}}^{\mathrm{XY}}(P, K) &\equiv& \operatorname{tr}\left(\gamma_{\epsilon_{p}} I_{\eta_{1} \eta_{2}}^{\mu} \gamma_{\epsilon_{q}} I_{\eta_{1}^{\prime} \eta_{2}^{\prime}}^{\nu}\right) D_{\mu \nu}^{\mathrm{X}}(K) \Delta_{-\epsilon_{q}}^{\mathrm{Y}}(Q)\nn
&&\hspace{-2.5cm}=m_{q}^{4} \int \frac{d \Omega_{s}}{4 \pi} \frac{1}{\left(P S+i \eta_{1} \varepsilon\right)\left(Q S+i \eta_{2} \varepsilon\right)}\times \int \frac{d \Omega_{s^{\prime}}}{4 \pi} \frac{1}{\left(P S^{\prime}+i \eta_{2}^{\prime} \varepsilon\right)\left(Q S^{\prime}+i \eta_{1}^{\prime} \varepsilon\right)}\nn
&&\hspace{-2.5cm}\times\bigg[\Big\{-\hat{s} \mdot\hat{s}^{\prime}-\hat{p}_{\epsilon}\mdot\hat{q}_{\epsilon} \, \hat{s}\mdot\hat{s}^{\prime}+\hat{p}_{\epsilon}\mdot\hat{s} \, \hat{s}\mdot\hat{s}^{\prime}+\hat{q}_{\epsilon}\mdot\hat{s} \, \hat{s}\mdot\hat{s}^{\prime} + \hat{p}_{\epsilon}\mdot\hat{s}^{\prime}\, \hat{s}\mdot\hat{s}^{\prime}+\hat{q}_{\epsilon}\!\cdot\!\hat{s}^{\prime} \, \hat{s}\mdot \hat{s}^{\prime}-\hat{p}_{\epsilon}\mdot\hat{s} \, \hat{q}_{\epsilon}\!\cdot\!\hat{s}^{\prime} \, \hat{s}\!\cdot\! \hat{s}^{\prime} 
-\hat{q}_{\epsilon}\mdot\hat{s} \, \hat{p}_{\epsilon}\!\cdot\!\hat{s}^{\prime} \, \hat{s}\mdot \hat{s}^{\prime}\nn
&&\hspace{-2.5cm}-\left(\hat{s}\mdot\hat{s}^{\prime}\right)^{2}+\hat{p}_{\epsilon}\mdot\hat{q}_{\epsilon} \, \left(\hat{s} . \hat{s}^{\prime}\right)^ {2} +  \hat{k}\mdot\hat{s}^{\prime}\, \hat{k}\!\cdot\!\hat{s}+\hat{p}_{\epsilon}\mdot\hat{q}_{\epsilon}\, \hat{k}\mdot\hat{s}' \, \hat{k}\!\cdot\!\hat{s}-\hat{p}_{\epsilon}\mdot\hat{s} \, \hat{k}\mdot\hat{s}'\, \hat{k}\mdot\hat{s}
-\hat{q}_{\epsilon}\mdot\hat{s} \, \hat{k}\mdot\hat{s} \, \hat{k}\mdot\hat{s}^{\prime}
-\hat{p}_{\epsilon}\mdot\hat{s}^{\prime}\, \hat{k}\mdot\hat{s} \,\hat{k}.\hat{s}^{\prime}\nn
 &&\hspace{-2.5cm}-\hat{q}_{\epsilon}\mdot\hat{s}^{\prime} \,\hat{k}\mdot\hat{s}\, \hat{k}\mdot\hat{s}^{\prime} +\hat{p}_{\epsilon}\mdot\hat{s} \, \hat{q}_{\epsilon}\mdot\hat{s}^{\prime}\, \hat{k}.\hat{s} \,\hat{k}\mdot\hat{s}^{\prime}+\hat{q}_{\epsilon}\mdot\hat{s} \,\hat{p}_{\epsilon}\mdot\hat{s}^{\prime} \,\hat{k}\mdot\hat{s}\, \hat{k}\mdot\hat{s}^{\prime} 
+ \, \hat{s}\mdot\hat{s}^{\prime} \, \hat{k}.\hat{s} \,\hat{k}\mdot\hat{s}^{\prime}
-\hat{p}_{\epsilon}\mdot\hat{q}_{\epsilon} \, \hat{s}\mdot\hat{s}^{\prime} \,\hat{k}\mdot\hat{s} \,\hat{k}.\hat{s}^{\prime}\Big\} D_{T}^{\mathrm{X}}(K) \Delta_{-\epsilon_{q}}^{\mathrm{Y}}(Q)\nn
&&\hspace{-2.5cm}-\Big\{k^{2}\left(1+\hat{p}_{\epsilon}.\hat{q}_{\epsilon}-\hat{p}_{\epsilon}\mdot\hat{s}-\hat{q}_{\epsilon}\mdot\hat{s}-\hat{p}_{\epsilon}\mdot\hat{s}^{\prime}-\hat{q}_{\epsilon}\mdot\hat{s}^{\prime}+\hat{p}_{\epsilon}\mdot\hat{s} \,\hat{q}_{\epsilon}\mdot\hat{s}^{\prime}+\hat{q}_{\epsilon}\mdot\hat{s} \,\hat{p}_{\epsilon}\mdot\hat{s}^{\prime}+\hat{s}.\hat{s}^{\prime}-\hat{p}_{\epsilon}. \hat{q}_{\epsilon}\,\hat{s}.\hat{s}^{\prime}\right)- k_{0} k\left(\hat{k}\cdot\hat{s} + \hat{k}.\hat{s}^{\prime}\right.\nn
&&\left.\hspace{-2.5cm}+\hat{p}_{\epsilon}.\hat{q}_{\epsilon} \,\hat{k}. \hat{s} + \hat{p}_{\epsilon}.\hat{q}_{\epsilon} \,\hat{k}.\hat{s}^{\prime}-\hat{p}_{\epsilon}.\hat{s} \,\hat{k}\mdot\hat{s}-\hat{p}_{\epsilon}.\hat{s} \,\hat{k}.\hat{s}^{\prime}-\hat{q}_{\epsilon}.\hat{s}\, \hat{k}\mdot\hat{s}-\hat{q}_{\epsilon}.\hat{s} \,\hat{k}.\hat{s}^{\prime}\right.
-\hat{p}_{\epsilon}.\hat{s}^{\prime} \,\hat{k}. \hat{s}-\hat{p}_{\epsilon}.\hat{s}^{\prime} \,\hat{k}\mdot\hat{s}^{\prime}-\hat{q}_{\epsilon}.\hat{s}^{\prime} \,\hat{k}\mdot\hat{s}-\hat{q}_{\epsilon}\!\cdot\!\hat{s}^{\prime} \,\hat{k}.\hat{s}^{\prime}\nn
&&\hspace{-2.5cm}
+\hat{p}_{\epsilon}.\hat{s} \, \hat{q}_{\epsilon} .\hat{s}^{\prime} \,\hat{k}.\hat{s}+\hat{p}_{\epsilon}\!\cdot\!\hat{s} \,\hat{q}_{\epsilon}\!\cdot\!\hat{s}^{\prime} \,\hat{k}\!\cdot\!\hat{s}^{\prime}
+\left.\hat{q}_{\epsilon}\!\cdot\!\hat{s} \,\hat{p}_{\epsilon}.\hat{s}^{\prime} \,\hat{k}\mdot\hat{s}+\hat{q}_{\epsilon}.\hat{s} \,\hat{p}_{\epsilon} .\hat{s}^{\prime} \,\hat{k}\mdot\hat{s}^{\prime}+\hat{k}.\hat{s} \,\hat{s}.\hat{s}^{\prime}+\hat{k}.\hat{s}^{\prime} \,\hat{s}\mdot\hat{s}^{\prime}-\hat{p}_{\epsilon}.\hat{q}_{\epsilon} \,\hat{k}.\hat{s} \,\hat{s}\mdot\hat{s}^{\prime}-\hat{p}_{\epsilon}.\hat{q}_{\epsilon} \,\hat{k}.\hat{s}^{\prime} \,\hat{s}. \hat{s}^{\prime}\right)\nn
&&\hspace{-2.5cm}
+\left.\left.k_{0}^{2}\left(\hat{k}\mdot\hat{s}\,\hat{k}\mdot\hat{s}^{\prime}+\hat{p}_{\epsilon}\mdot\hat{q}_{\epsilon}\,\hat{k}\mdot\hat{s}\,\hat{k}\mdot\hat{s}^{\prime}-\hat{p}_{\epsilon}.\hat{s}\,\hat{k}.\hat{s}\,\hat{k}\mdot\hat{s}^{\prime}-\hat{q}_{\epsilon}\mdot\hat{s}\,\hat{k}\mdot\hat{s}\,\hat{k}\mdot\hat{s}^{\prime}-\hat{p}_{\epsilon}\mdot\hat{s}^{\prime}\, \hat{k}\mdot\hat{s}\,\hat{k}.\hat{s}^{\prime}-\hat{q}_{\epsilon}\!\cdot\!\hat{s}^{\prime}\,\hat{k}\mdot\hat{s}\,\hat{k}\mdot\hat{s}^{\prime}\right.\right.\right.\nn
&&\hspace{-2.5cm}
+\left.\left.\left.\hat{p}_{\epsilon}\mdot\hat{s} \,\hat{q}_{\epsilon}\mdot\hat{s}^{\prime} \,\hat{k}\mdot \hat{s} \,\hat{k}\mdot\hat{s}^{\prime}+\hat{q}_{\epsilon}\mdot\hat{s} \,\hat{p}_{\epsilon}\mdot\hat{s}^{\prime} \,\hat{k}\mdot\hat{s} \,\hat{k}\mdot\hat{s}^{\prime}+\hat{k}\mdot\hat{s} \,\hat{k}\mdot\hat{s}^{\prime} \,\hat{s}\mdot\hat{s}^{\prime}-\hat{p}_{\epsilon}\mdot\hat{q}_{\epsilon} \,\hat{k}\mdot\hat{s} \,\hat{s}\mdot\hat{s}^{\prime}\,\hat{k}\!\cdot\!\hat{s}^{\prime} \right)\right\}\tilde{D}^{X}_{L}(K)\Delta^{Y}_{-\epsilon_{q}}(Q)\right].
\label{htl_terms_2}
\ea
The integrand in $\Sigma^{(1)}_{2}$ contains only one HTL and can be written from eq.~\eqref{Sigma2pm} as
\ba
\Sigma_{(2)\pm}^{(1)}(P)&=& \frac{-i g^{2} C_{F}}{2} \int \frac{d^{4} K}{(2 \pi)^{4}}\ G_{\pm ;--}^{\mathrm{S}}(P, K),
\ea
where\footnote{A typo in~\cite{Abada:2014bma} is corrected}
\ba
G_{\epsilon_{p} ; \eta_{1} \eta_{2}}^{\mathrm{S}}(P, K) &\equiv & \frac{1}{2} \operatorname{tr}\left(\gamma_{\epsilon_{p}} I_{\eta_{1} \eta_{2}}^{\mu \nu}\right) \Delta_{\mu \nu}^{\mathrm{X}}(K)\nn
&=&m_{q}^{2} \int \frac{d \Omega_{s}}{4 \pi} \frac{1}{\left[P S+i \eta_{1} \varepsilon\right]\left[P S+i \eta_{2} \varepsilon\right]}  \left[\frac{1}{(P+K) S+i \eta_{1} \varepsilon}+\frac{1}{(P-K) S+i \eta_{2} \varepsilon}\right] \nn
&&\hspace{-0cm}\times\Bigg[\left(1-\hat{p}_{\epsilon}\mdot\hat{s}\right)\left(1-(\hat{k}\mdot \hat{s})^{2}\right)D_{T}^{\mathrm{X}}(K) 
+\ \Big\{k^{2}-2 k_{0} k\hat{k}.\hat{s}\ +\ k_{0}^{2} \, \,(\hat{k}.\hat{s})^{2}\Big\}\left(1-\hat{p}_{\epsilon}\mdot\hat{s}\right)\tilde{D}_{L}^{\mathrm{X}}(K)\Bigg].\label{sigma2_htl}
\ea
Using the above terms, the NLO one-loop HTL-summed quark self-energy can be expressed in compact form as
\ba
\Sigma_{\pm}^{(1)}(P)&=& -\frac{i g^{2} C_{F}}{2} \int \frac{d^{4} K}{(2 \pi)^{4}}\left[F_{\pm ; 0}^{\mathrm{SR}}(P, K)+F_{\pm ; 0}^{\mathrm{AS}}(P, K)+2 F_{\pm ;--}^{\mathrm{SR}}(P, K)+F_{\pm ;--}^{\mathrm{AS}}(P, K)\right.\nn
&+&\left.F_{\pm ;-+}^{\mathrm{AS}}(P, K)+F_{\pm ;--;--}^{\mathrm{SR}}(P, K)+F_{\pm ;--;+-}^{\mathrm{AS}}(P, K)+G_{\pm ;--}^{\mathrm{S}}(P, K)\right].\label{sigma_final}
\ea
\section{Evaluation of NLO quark self-energy}\label{sec4}
In this section, we will consider all the terms present in eq.~\eqref{sigma_final} more elaborately and show how these terms have been evaluated. We can write eq.~\eqref{sigma_final} as
\ba
\Sigma_{\pm}^{(1)}(P)=\Sigma_{1\pm}^{(1)}(P)+\Sigma_{2\pm}^{(1)}(P)+\Sigma_{3\pm}^{(1)}(P)+\Sigma_{4\pm}^{(1)}(P)+\Sigma_{5\pm}^{(1)}(P)+\Sigma_{6\pm}^{(1)}(P)+\Sigma_{7\pm}^{(1)}(P)+\Sigma_{8\pm}^{(1)}(P).\label{Sigma1-8}
\ea
We consider quark thermal mass to be $1$, i.e., $m_q$ = $1$. Thus, we will get
\begin{equation}
\frac{m_{g}^{2}}{m_{q}^{2}}=\frac{16 N_{c}\left(N_{c}+s_{F}\right)}{6\left(N_{c}^{2}-1\right)}=\left(3+s_{F}\right),\label{ratio}
\end{equation}
Here, we will fix the number of flavor and color charges. In the above equation, we have considered the value of $N_{c}$ to be $3$, and for the value of $N_{f} = 2$, this ratio in eq.~\eqref{ratio} is equal to $4$, which we will consider in further computation. Now, we define a variable $t$ as the ratio of $ p/p_0$. The scaled quark momentum can be written in terms of the variable $t$ as 
\begin{equation}\label{t_vari}
p(t) / m_{q}=\sqrt{\frac{t}{1-t}-\frac{1}{2} \ln \left(\frac{1+t}{1-t}\right)} . 
\end{equation}
Since we have considered quark thermal mass $m_q = 1 $. Equation~\eqref{t_vari} can be derived by using the leading-order quark dispersion relation $ \Delta_{-}^{-1}(p_0,p) = 0 $ 
The variation of the scaled quark momentum $p(t)$ and mass $p_0(t) = \Omega^{(0)}_{-}(p(t))$ is shown in figure~\ref{fig5}. Also, for a slow-moving quark, one would get $p \lesssim m_q$, which gives us the limit on $t$ variable, i.e., $t \lesssim 0.64$. Beyond this value of the t variable, quarks will be considered fast-moving, not an interesting region.
%
\begin{figure}
	\centering
	{\includegraphics[scale=0.8,keepaspectratio]{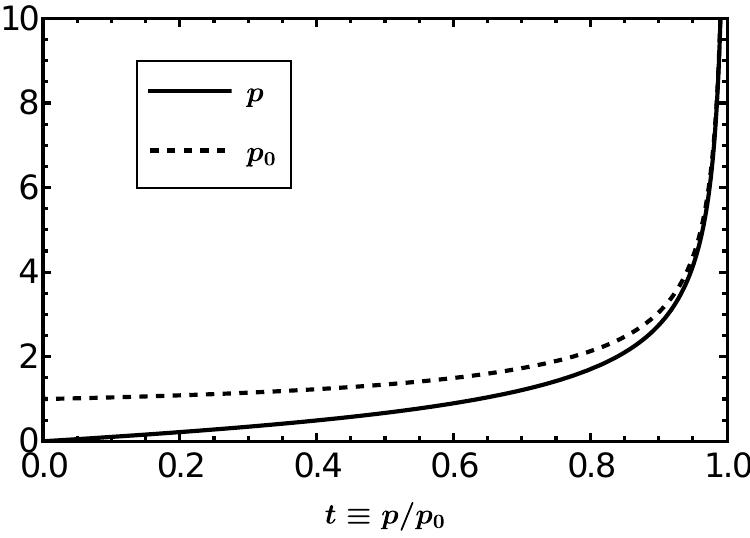}}
	\captionsetup{singlelinecheck=off}
	\caption[.]{Variation of scaled quark momentum $p$(solid) and mass $p_{0}$ (dashed) w.r.t. $``\text{t}\equiv p/p_{0}"$ variable. For slow-moving quarks $t \lesssim 0.64$.}
	\label{fig5}
\end{figure} 

In order to evaluate eq.~\eqref{sigma_final}, we need retarded transverse $D_{T}^{R}(k,k_0,\varepsilon)$, retarded longitudinal $D_{L}^{R}(k,k_0,\varepsilon)$ gluon propagators which are derived by using eqs.~\eqref{glu_prop} and~\eqref{RAS_B} in  appendix~\ref{sec:appendixC}.
The other quantity required is the retarded quark propagators $\Delta_{\pm}^{R}(q,q_{0},\varepsilon)$ and can be obtained by using eqs.~\eqref{q_prop_htl} and~\eqref{RAS_F} (For details see appendix C).
Now, let us consider the first term of eq.~\eqref{sigma_final}, which is 
\begin{equation}
\Sigma_{1\pm}^{(1)}(P)= \frac{-ig^2 C_F}{2} \int \frac{d^4 K}{(2\pi)^4}\Big[F_{\pm ; 0}^{\mathrm{SR}}(P, K)\Big],
\label{Sigma1_1}
\end{equation}
where $F_{\pm ; 0}^{\mathrm{SR}}(P, K) $ is given in eq.~\eqref{bare_terms}. The inclination between $\vec{p}$ and $\vec{k}$ is $x (=\cos{\theta})$. Also, the angle between $\hat{p}$ and $\hat{q}$ is
\begin{equation}
\hat{p}\cdot\!\hat{q} = \frac{\vec{p}\!\cdot\!\vec{q}}{p q} =   \frac{\vec{p}.(\vec{p}-\vec{k})}{p|\vec{p}-\vec{k}|} = \frac{p^2 - p k x}{p q} = \frac{p}{q} - \frac{k}{q}x,\label{p.q}
\end{equation}
and
\begin{equation}
q = |\vec{q}| =  \sqrt{p^2+k^2-2pkx}.
\end{equation}
Similarly, the dot product of the $\hat{k}$ and $\hat{q}$ is
\begin{equation}
\hat{k}\!\cdot\!\hat{q} = \frac{\vec{k}\!\cdot\!\vec{q}}{k q} =   \frac{\vec{k}.(\vec{p}-\vec{k})}{k|\vec{p}-\vec{k}|} = \frac{pkx-k^2}{kq} = \frac{p}{q}x - \frac{k}{q};\label{k.q}
\end{equation}
%
The first term of eq.~\eqref{Sigma1-8} can be written using eq.~\eqref{bare_terms} and after using  eqs.~\eqref{p.q}~\eqref{k.q}, eq.~\eqref{Sigma1_1} becomes
\ba
\Sigma _{1\pm}^{(1)}(P) &=&  \frac{-ig^2 C_F}{2(2\pi)^4}  (2\pi) \int_{-\infty}^\infty dk_0 \int_{0}^\infty k^2 dk  \int_{-1}^1 dx \left[-2\left(1-x\left(\frac{p}{q}x - \frac{k}{q}\right)\right) D^{S}_{T}(K) \Delta^{R}_{\mp}(Q) \right.\nn
&&\hspace{-1cm}-\left.\bigg\{k_{0}^{2}\left(1-\frac{p}{q} + \frac{k}{q}x +2 x\left(\frac{p}{q}x - \frac{k}{q}\right)\right)\mp 2k_{0} k\right.
\left(x+ \frac{px}{q} - \frac{k}{q} \right)+k^{2}\left(1+\frac{p}{q} - \frac{k}{q}x\right)\bigg\}
\tilde{D}_{L}^{\mathrm{S}}(K) \Delta_{\mp}^{\mathrm{R}}(Q)\bigg]
\label{Sigma11_final}
\ea
Similarly, the second term of eq.~\eqref{Sigma1-8} can be written using eq.~\eqref{Sigma11_final} as
\ba
\Sigma _{2\pm}^{(1)}(P) &=&  \frac{-ig^2 C_F}{2(2\pi)^4}  (2\pi) \int_{-\infty}^\infty dk_0 \int_{0}^\infty k^2 dk  \int_{-1}^1 dx \left[-2\left(1-x\left(\frac{p}{q}x - \frac{k}{q}\right)\right) D^{A}_{T}(K) \Delta^{S}_{\mp}(Q) \right.\nn
&-&\left.\left.\bigg\{k_{0}^{2}\left(1-\frac{p}{q} - \frac{k}{q}x+2 x\left(\frac{px}{q} - \frac{k}{q}\right)\right)\mp 2k_{0} k\right.
\left(x+\frac{px}{q} - \frac{k}{q}\right)+k^{2}\left(1+\frac{p}{q} - \frac{kx}{q}\right)\bigg\} \tilde{D}_{L}^{\mathrm{A}}(K) \Delta_{\mp}^{\mathrm{S}}(Q)\right]
\label{Sigma12_final}
\ea
To evaluate eqs.~\eqref{Sigma11_final} and~\eqref{Sigma12_final} numerically, we encounter a few issues. One of those issues is that the integrand has a discontinuity because of the terms $\arctan$ in the propagators and the B.E distribution function at $k_{0} = 0$. Such discontinuity would cause fatal issues in any integration method. So, the numerical outputs are either unreliable or produce unsatisfactory results. This instability in the results is more prolonged if we consider our tuning parameter $\varepsilon$ too small. So, to make further progress, we have partitioned the integration region into the domains bounded by the lines, which causes discontinuity as
\begin{equation}\label{diverg_pts}
\begin{aligned}
&k_{0}=0 ; \quad k_{0}=\pm k ; 
\quad k=k_{t} \equiv \frac{1}{2} \frac{p_{0}^{2}-p^{2}}{p_{0}-x p}=\frac{1}{2 t} \frac{1-t^{2}}{1-x t} \sqrt{\frac{t}{1-t}-\frac{1}{2} \ln \left(\frac{1+t}{1-t}\right)}.
\end{aligned} 
\end{equation}
By doing the swapping of variables $\theta = \tan^{-1}{k}$ and $\phi = \tan^{-1}{k_0}$, these domains are shown in figure~\ref{fig6}. The last (vertical) line shown in figure~\ref{fig6} is discontinuity line $ k = p_{0}-q $.
We evaluate eqs.~\eqref{Sigma11_final} and~\eqref{Sigma12_final} in each of the domains as shown in figure~\ref{fig6} separately or in the domains shown in eq.~\eqref{diverg_pts} numerically and summed up to get the results. 
Now, the third term of eq.~\eqref{Sigma1-8} is given by
\begin{equation}
\Sigma _{3\pm}^{(1)}(P) = \frac{-ig^2 C_F}{2} \int \frac{d^4K}{(2\pi)^4}\left[2F_{\pm ; --}^{\mathrm{SR}}(P, K)\right], \label{Sigma13}
\end{equation}
where $F_{\pm ; --}^{\mathrm{SR}}(P, K)$ can be written using eq.~\eqref{htl_terms_1}. If $F_{1\pm;--}^{SR}(P,K)$ denote the terms in eq.~\eqref{Sigma13} without $ \hat{s}$, then
\begin{figure}[t]
	\centering
	{\includegraphics[scale=0.8,keepaspectratio]{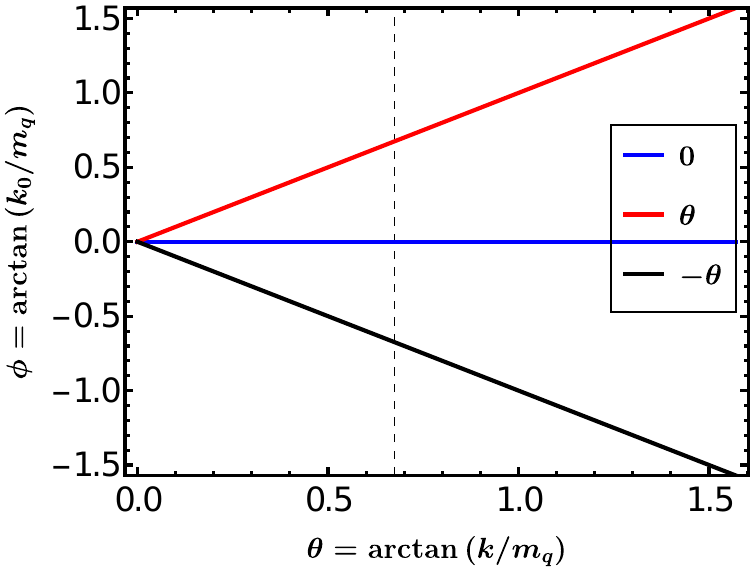}}
	\captionsetup{singlelinecheck=off}
	\caption[.]{Domains in ($k,k_0$) plane at which the integrand in eq.~\eqref{Sigma11_final} has sharp jumps. Here, we have used $ t \equiv p/p_{0} = 0.45 $ and $x\equiv \cos\theta = 0.8 $. }
	\label{fig6} 
\end{figure}
%
\begin{eqnarray}\label{F1_pm_SR_--}
F_{1\pm;--}^{SR}(P,K) &=& m_{q}^{2} \int \frac{d \Omega_{s}}{4 \pi} \frac{1}{\left(P S-i \varepsilon\right)\left(Q S-i \varepsilon\right)}\nn
&\times& \bigg[\left(1-\hat{p}_{\epsilon}\mdot\hat{q}_{\epsilon}\right){D}_{T}^{\mathrm{S}}(K) \Delta_{\mp}^{\mathrm{R}}(Q) -\left\{k^{2}\left(1+\hat{p}_{\epsilon}\mdot\hat{q}_{\epsilon}\right) \right.
\left.- k_{0} k \left(\hat{p}_{\epsilon}\mdot\hat{k}+\hat{q}_{\epsilon}\mdot\hat{k}\right) \right\}\tilde{D}_{L}^{\mathrm{S}}(K) \Delta_{\mp}^{\mathrm{R}}(Q)\bigg].
\end{eqnarray}
The solid angle integration in eq.~\eqref{F1_pm_SR_--} can be computed using eq.~\eqref{J00_2} as 
\ba
\int \frac{d \Omega_{s}}{4 \pi} \frac{1}{\left(P S-i \varepsilon\right)\left(Q S-i \varepsilon\right)} = V\left(t,k,k_0,x,\varepsilon\right),\label{V_k_k0}
\ea
where
\ba
V\left(t, k, k_0, x, \varepsilon\right)&=& \int_{0}^{1} d u \frac{1}{Z u^{2}+2 Y u + X} \nn
&&\hspace{-2.cm}= \frac{1}{2 \sqrt{\Delta}}\left[\frac { 1 } { 2 } \ln \frac{\left(\left(r_{1}-1\right)^{2}+i_{1}^{2}\right)\left(r_{2}^{2}+i_{2}^{2}\right)}{\left(\left(r_{2}-1\right)^{2}+i_{2}^{2}\right)\left(r_{1}^{2}+i_{1}^{2}\right)}\right.
+ \ i\left\{\tan^{-1} \left(\frac{r_{1}}{i_{1}}\right)-\tan^{-1} \left(\frac{r_{1}-1}{i_{1}}\right)
-\left.\left.\tan^{-1} \left(\frac{r_{2}}{i_{2}}\right)+\tan^{-1} \left(\frac{r_{2}-1}{i_{2}}\right)\right\}\right.\right].\hspace{1.5cm} \label{func_V}
\ea
Here, $r_{i}=\operatorname{Re} s_{i}$ and $i_{i}=\operatorname{Im} s_{i}$ with
$
s_{1}=\frac{-Y+\sqrt{\Delta}}{Z} ; \quad s_{2}=\frac{-Y-\sqrt{\Delta}}{Z} ; \quad \Delta=Y^{2}-X Z,
$
and
\begin{equation}
\begin{aligned}
&X=\left(p_{0}-i \varepsilon\right)^{2}-p^{2}; \quad
&Y=\vec{k}\!\cdot\!\vec{p}-\left(k_{0}\right)\left(p_{0}-i \varepsilon\right); \quad Z=k_{0}^{2}-k^{2};
\end{aligned}
\end{equation}
Thus, eq.~\eqref{F1_pm_SR_--} becomes
\begin{equation}
\begin{aligned}
F_{1\pm;--}^{SR}(P,K) = V\left(t,k,k_0,x,\varepsilon\right) \bigg[\left(1-\frac{p}{q} + \frac{k}{q}x\right) D^{S}_{T}(K) \Delta^{R}_{\mp}(Q) -\bigg\{k^{2}\bigg(1+\frac{p}{q} - \frac{k}{q}x \bigg)\\
 \mp k_{0} k \left(x+ \frac{p}{q}x - \frac{k}{q} \right)\bigg\} \tilde{D}_{L}^{\mathrm{S}}(K) \Delta_{\mp}^{\mathrm{R}}(Q) \bigg] .\label{F1pm}
\end{aligned}
\end{equation}
With the expression of $F_{1\pm;--}^{SR}(P,K)$ from eq.~\eqref{F1pm} , the terms of eq.~\eqref{Sigma13} without s becomes
\ba
\Sigma _{3(1)\pm}^{(1)}(P) &=& \frac{-ig^2 C_F}{(2\pi)^4}  (2\pi) \int_{-\infty}^\infty dk_0 \int_{0}^\infty k^2 dk \int_{-1}^1 dx \,\, V\left(t,k,k_0,x,\varepsilon\right) \bigg[ \left(1-\frac{p}{q} + \frac{k}{q}x\right) D^{S}_{T}(K) \Delta^{R}_{\mp}(Q) \nn 
&-&\bigg\{k^{2}\bigg(1+ \frac{p}{q} - \frac{k}{q}x \bigg) \mp k_{0} k \left(x+ \frac{p}{q}x - \frac{k}{q} \right)\bigg\}\tilde{D}_{L}^{\mathrm{S}}(K) \Delta_{\mp}^{\mathrm{R}}(Q) \bigg] .\label{Sigma13_1}
\ea
Now consider the terms with $ \hat{s}$ in eq.~\eqref{Sigma13}
\ba
F_{2\pm; - -}^{SR}(P,K) &=& \int \frac{d \Omega_{s}}{4 \pi} \frac{1}{\left(P S-i\varepsilon\right)\left(Q S-i\varepsilon\right)}\bigg[\Big(-\hat{p}_{\epsilon} \mdot \hat{s}-\hat{q}_{\epsilon}.\hat{s} +\ \hat{p}_{\epsilon}.\hat{k} \,\hat{k}.\hat{s}+\hat{q}_{\epsilon}.\hat{k} \,\hat{k}.\hat{s}\Big){D}_{T}^{\mathrm{S}}(K) \Delta_{\mp}^{\mathrm{R}}(Q) \nn
&&\hspace{-2cm}-\left\{k_{0}^{2}\left(\hat{p}_{\epsilon}\mdot\hat{k} \, \hat{k}\mdot\hat{s}+\hat{q}_{\epsilon}\mdot\hat{k} \,\hat{k}\mdot\hat{s}\right) - k_{0} k \left(2 \hat{p}_{\epsilon}\mdot\hat{q}_{\epsilon} \,\hat{k}\mdot\hat{s} -\hat{q}_{\epsilon}\mdot\hat{k} \,\hat{p}_{\epsilon}\mdot\hat{s}-\hat{p}_{\epsilon}\mdot\hat{k} \, \hat{q}_{\epsilon}\mdot\hat{s}\right)-k^{2}\left(\hat{p}_{\epsilon}\mdot\hat{s}+\hat{q}_{\epsilon}\mdot\hat{s}\right)\right\}
 \tilde{D}_{L}^{\mathrm{S}}(K) \Delta_{\mp}^{\mathrm{R}}(Q)\bigg]\label{F2_pm_SR_--}
\ea
By using eq.~\eqref{J0i}, we get
\ba
\int \frac{d \Omega_{s}}{4 \pi} \frac{\hat{s}^{i}}{\left(P S-i \varepsilon\right)\left(Q S-i \varepsilon\right)} &=& 
\int_{0}^{1} d u\left[\frac{r_{0}}{r_{0}^{2}-r^{2}}-\frac{1}{2 r} \ln \frac{r_{0}+r}{r_{0}-r}\right] \frac{r^{i}}{r^{2}} =
 \int_{0}^{1} d u \, V_1 \, \hat{r}^{i},\label{1s_int}
\ea
where \begin{equation}
V_1(r_0,r) = \left[\frac{r_{0}}{r_{0}^{2}-r^{2}}-\frac{1}{2 r} \ln \frac{r_{0}+r}{r_{0}-r}\right] \frac{1}{r},
\end{equation}
with
\begin{equation}
r_0 = p_0 - i\varepsilon - k_0 u  ; \quad r = \sqrt{p^2+k^2u^2-2pkux}.
\end{equation}
Now, the required angles in order to solve eq.~\eqref{F2_pm_SR_--} are
\ba
\hat{p}\!\cdot\!\hat{r} &=& \frac{p}{r} - \frac{k u}{r}x, \quad \quad
\hat{k}\!\cdot\!\hat{r} = \frac{p}{r}x - \frac{k u}{r}, \quad \quad
\hat{q}\!\cdot\!\hat{r} =  \frac{p^2}{qr} - \frac{pkux}{qr}-\frac{pkx}{qr} + \frac{k^2 u}{qr}.
\ea
%
Thus, eq.~\eqref{F2_pm_SR_--} becomes
\ba
F_{2\pm;--}^{SR}(P,K)&=&\pm \int_{0}^{1} d u \,V_1(r_0,r)\,\bigg[
 \frac{p(p+q)(x^2-1)}{qr}
{D}_{T}^{\mathrm{S}}(K) \Delta_{\mp}^{\mathrm{R}}(Q)
\bigg\{ k^{2}\bigg(\frac{p}{r}-\frac{ku}{r}x +\frac{p^2}{qr} - \frac{pkux}{qr}-\frac{pkx}{qr} 
+ \frac{k^2 u}{qr}\bigg)\nn
&& \mp \frac{k_0k^2p(2u-1)(x^2-1)}{qr}  - k_{0}^{2}\bigg(x+\frac{px}{q} -\frac{k}{q}\bigg)\left(\frac{px}{r}-\frac{ku}{r}\right)  \bigg\}\tilde{D}_{L}^{\mathrm{S}}(K) \Delta_{\mp}^{\mathrm{R}}(Q)\bigg]
\label{F2_PM_SR_--_Sigma13}
\ea
So, the $\hat{s}$ contribution in the third term of eq.~\eqref{Sigma1-8} is 
\begin{equation}\label{Sigma13_2}
\Sigma _{3(2)\pm}^{(1)}(P) = \frac{-ig^2 C_F}{(2\pi)^4}  (2\pi) \int_{-\infty}^\infty dk_0 \int_{0}^\infty k^2 dk  \int_{-1}^1 dx \, F_{2\pm;--}^{SR}(P,K)
\end{equation}
%
Let us consider terms with two $\hat{s}$ in eq.~\eqref{Sigma13}
\ba
F_{3\pm;--}^{SR}(P,K) &=& \int \frac{d \Omega_{s}}{4 \pi} \frac{\hat{s}^{i} \hat{s}^{j}}{\left(P S-i\varepsilon\right)\left(Q S-i\varepsilon\right)}\bigg[\bigg(-\hat{k}_{i}\hat{k}_{j}+\hat{p}.\hat{q}\, \hat{k}_{i}\hat{k}_{j}+2 \hat{p}_{i} \, \hat{q}_{j}-\hat{p}.\hat{k} \, \hat{q}_{i}\hat{k}_{j}-\hat{q}\!\cdot\!\hat{k} \, \hat{p}_{i} \,\hat{k}_{j}\bigg) {D}_{T}^{\mathrm{S}}(K) \Delta_{\mp}^{\mathrm{R}}(Q)\nn
&&\hspace{-2cm}+\bigg\{k_{0}^{2}\bigg((\hat{k}\mdot\hat{s})^{2} - \hat{p}_{\epsilon}\mdot\hat{q}_{\epsilon} \,(\hat{k}\mdot\hat{s})^{2} + \hat{p}_{\epsilon}\mdot\hat{k} \,\hat{q}_{\epsilon}\mdot\hat{s} \,\hat{k}.\hat{s}+\hat{q}_{\epsilon}\mdot\hat{k} \,\hat{p}_{\epsilon}\mdot\hat{s}\, \hat{k}\mdot\hat{s}\bigg) - k_{0} k \left(\hat{q}_{\epsilon}\mdot\hat{s} \, \hat{k}\mdot\hat{s} + \hat{p}_{\epsilon}\mdot\hat{s} \,\hat{k}\mdot\hat{s}\right)\bigg\}
\tilde{D}_{L}^{\mathrm{S}}(K) \Delta_{\mp}^{\mathrm{R}}(Q) \bigg]\label{F3sr2}
\ea
From eq.~\eqref{Jij_int}, we can write
\begin{equation}\label{2s_int}
\begin{aligned}
\int \frac{d \Omega_{s}}{4 \pi} \frac{\hat{s}^{i} \hat{s}^{j}}{\left(P S-i \varepsilon\right)\left(Q S-i \varepsilon\right)} = \int_{0}^{1} d u\left(A \delta^{i j}+B \hat{r}^{i} \hat{r}^{j}\right),
\end{aligned}
\end{equation}
where \begin{equation}\label{A_B_Def}
\begin{aligned}
A &=-\frac{1}{r^{2}}\left(1-\frac{r_{0}}{2 r} \ln \frac{r_{0}+r}{r_{0}-r}\right) ,\quad \quad
B &=\frac{1}{r_{0}^{2}-r^{2}}+\frac{3}{r^{2}}\left(1-\frac{r_{0}}{2 r} \ln \frac{r_{0}+r}{r_{0}-r}\right)
\end{aligned}
\end{equation}
Using eq.~\eqref{2s_int}, eq.~\eqref{F3sr2} becomes
\ba
F_{3\pm;--}^{SR}(P,K) &=& \int_{0}^{1} d u \bigg[\bigg\{ A \left(3 \hat{p}\mdot\hat{q} - 1 -2 x \,\hat{q} \mdot \hat{k}\right) + B \left(\big(\hat{k}\!\cdot\!\hat{r}\big)^2 \,\left(\hat{p}\!\cdot\!\hat{q} -1\right) + 2\hat{p}.\hat{r} \, \hat{q}.\hat{r} - x \, \hat{q}.\hat{r} \,\hat{k}.\hat{r} - \hat{q}\!\cdot\!\hat{k} \, \hat{p}\!\cdot\!\hat{r} \, \hat{k}\!\cdot\!\hat{r} \right)\bigg\}\nn
&\times&{D}_{T}^{\mathrm{S}}(K) \Delta_{\mp}^{\mathrm{R}}(Q)-\bigg\{A\bigg(k_{0}^{2} \big(\hat{p}\!\cdot\!\hat{q} - 1 -2 x \,\hat{q}\!\cdot\!\hat{k}\big)\pm k_{0}k \big(x+\,\hat{q}.\hat{k}\big) \bigg)+ B\bigg(k_{0}^{2}\bigg(\big(\hat{k}\!\cdot\!\hat{r}\big)^2 \,\left(\hat{p}\!\cdot\!\hat{q} -1\right) \nn
&-& \hat{q}.\hat{k} \,\hat{p}\mdot\hat{r} \, \hat{k}\mdot\hat{r}  - x \, \hat{q}\!\cdot\!\hat{r} \,\hat{k}\!\cdot\!\hat{r}\bigg)\pm k_{0}k \, \hat{k}\!\cdot\!\hat{r} \left(\hat{p}\!\cdot\!\hat{r} + \hat{q}\!\cdot\!\hat{r} \right)\bigg)\bigg\}\tilde{D}_{L}^{\mathrm{S}}(K) \Delta_{\mp}^{\mathrm{R}}(Q)\bigg] 
\ea
Thus, the contribution coming from 2s terms of eq.~\eqref{Sigma13} is 
\begin{equation}\label{Sigma13_3}
\Sigma _{3(3)\pm}^{(1)}(P) = \frac{-ig^2 C_F}{(2\pi)^4}  (2\pi) \int_{-\infty}^\infty dk_0 \int_{0}^\infty k^2 dk  \int_{-1}^1 dx \, F_{3\pm;--}^{SR}(P,K)
\end{equation}
As we have seen in eq.~\eqref{Sigma11_final}, there are sudden jumps in the integrand from the gluon propagator, which causes instability. In this term, an additional divergence will come from eq.~\eqref{func_V} i.e., from $V(t,k,k_0,x,\varepsilon)$. The divergence that comes from eqs.~\eqref{1s_int} and eq.\eqref{2s_int} is the same as we get from eq.~\eqref{func_V}. Thus, all the lines of discontinuity are
\begin{equation}
\begin{aligned}
&k_{0}&=&0 ; \quad k_{0}=\pm k ; \quad
k_0 &=& \hspace{1mm} p_0 \pm \sqrt{p^2+k^2-2pkx} ; \quad   
k& = & \hspace{1mm} k_{t} =\frac{1}{2 t} \frac{1-t^{2}}{1-x t} \sqrt{\frac{t}{1-t}-\frac{1}{2} \ln \left(\frac{1+t}{1-t}\right)}
\end{aligned}
\label{diverg_pts_1}
\end{equation}
These domains are shown in figure~\ref{fig7}.
We have evaluated  eqs.~\eqref{Sigma13_1}, \eqref{Sigma13_2}, \eqref{Sigma13_3} in each of the domains of figure~\ref{fig7} numerically and summed up the results. 
Let us consider the fourth term of eq.~\eqref{sigma_final}, which is 
\begin{equation}\label{Sigma1_4}
\Sigma _{4\pm}^{(1)}(P) = \frac{-ig^2 C_F}{2} \int \frac{d^4K}{(2\pi)^4}\Big[F_{\pm ; --}^{\mathrm{AS}}(P, K)\Big]
\end{equation}
This term is analogous to the third term of Eq.~\eqref{sigma_final} except for the change of real-time field indices (i.e., $SR \rightarrow AS$) and a factor of $1/2$. Thus, by using eq.~\eqref{Sigma13_1}, we will get  
\ba
\Sigma _{4(1)\pm}^{(1)}(P) &=& \frac{-ig^2 C_F}{2(2\pi)^4}  (2\pi) \int_{-\infty}^\infty dk_0 \int_{0}^\infty k^2 dk \int_{-1}^1 dx \,\,V\left(t,k,k_0,x,\varepsilon\right) \bigg[\left(1-\frac{p}{q} + \frac{k}{q}x\right) D^{A}_{T}(K) \Delta^{S}_{\mp}(Q)\nn 
&-&\bigg\{k^{2}\bigg(1+\frac{p}{q} - \frac{k}{q}x\bigg) \mp k_{0} k \left(x+\frac{p}{q}x - \frac{k}{q}\right)\bigg\}\tilde{D}_{L}^{\mathrm{A}}(K) \Delta_{\mp}^{\mathrm{S}}(Q) \bigg] .\label{Sigma14_1}
\ea
\begin{figure}[t]
	\centering
	{\includegraphics[height=9cm, width=12cm]{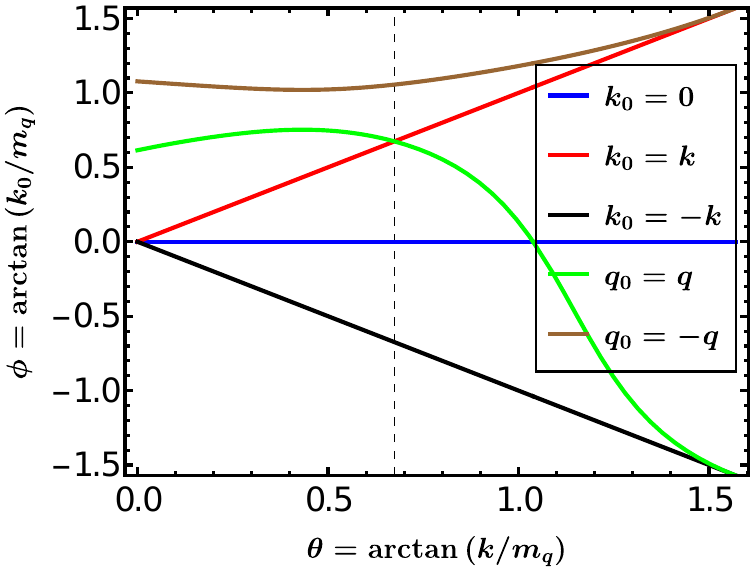}}
	\captionsetup{singlelinecheck=off}
	\caption[.]{Domains in ($k,k_0$) plane at which the integrand in eq.(\ref{Sigma13}) has divergences. Here we have used $ t \equiv p/p_{0} = 0.45 $ and $x \equiv \cos\theta = 0.8 $.} 
	\label{fig7}    
\end{figure}  
By using eq.~\eqref{Sigma13_2}, we get
\begin{equation}\label{sigma14_2}
\Sigma _{4(2)\pm}^{(1)}(P) = \frac{-ig^2 C_F}{2(2\pi)^4}  (2\pi) \int_{-\infty}^\infty dk_0 \int_{0}^\infty k^2 dk  \int_{-1}^1 dx \, F_{2\pm;--}^{AS}(P,K)
\end{equation}
where
\ba
F_{2\pm;--}^{AS}(P,K)&=& \int_{0}^{1} d u \,V_1(r_0,r)\,\bigg[
\pm \frac{p(p+q)(x^2-1)}{qr}
{D}_{T}^{\mathrm{A}}(K) \Delta_{\mp}^{\mathrm{S}}(Q)
-\bigg\{\mp k^{2}\bigg(\frac{p}{r}-\frac{ku}{r}x +\frac{p^2}{qr} - \frac{pkux}{qr}-\frac{pkx}{qr} 
+ \frac{k^2 u}{qr}\bigg)\nn
&& -  \frac{k_0k^2p(2u-1)(x^2-1)}{qr}  \pm k_{0}^{2}\bigg(x+\frac{px}{q} -\frac{k}{q}\bigg)\left(\frac{px}{r}-\frac{ku}{r}\right)  \bigg\}\tilde{D}_{L}^{\mathrm{A}}(K) \Delta_{\mp}^{\mathrm{S}}(Q)\bigg]
\ea
Similarly, by using eq.~\eqref{Sigma13_3}, we will get
\begin{equation}
\Sigma _{4(3)\pm}^{(1)}(P) = \frac{-ig^2 C_F}{2(2\pi)^4}  (2\pi) \int_{-\infty}^\infty dk_0 \int_{0}^\infty k^2 dk  \int_{-1}^1 dx \, F_{3\pm;--}^{AS}(P,K)
\end{equation}
where
\ba
F_{3\pm;--}^{AS}(P,K) &=& \int_{0}^{1} d u \bigg[\bigg\{ A \left(3 \hat{p}\!\cdot\!\hat{q} - 1 -2 x \,\hat{q}\!\cdot\!\hat{k}\right) + B \bigg(\left(\hat{k}\!\cdot\!\hat{r}\right)^2 \left(\hat{p}\!\cdot\!\hat{q} -1\right) + 2\hat{p}\!\cdot\!\hat{r} \, \hat{q}\!\cdot\!\hat{r} - x \, \hat{q}\!\cdot\!\hat{r} \,\hat{k}\!\cdot\!\hat{r} - \hat{q}\!\cdot\!\hat{k} \, \hat{p}.\hat{r} \, \hat{k}.\hat{r} \bigg)\bigg\}\nn
&\times&{D}_{T}^{\mathrm{A}}(K) \Delta_{\mp}^{\mathrm{S}}(Q)-k_0^2\bigg\{A\bigg( \hat{p}\!\cdot\!\hat{q} - 1 -2 x \,\hat{q}\!\cdot\!\hat{k} \pm \frac{k}{k_0} \left(x+\,\hat{q}\!\cdot\!\hat{k}\right) \bigg)+ B\bigg(\left(\hat{k}\!\cdot\!\hat{r}\right)^2 \,\left(\hat{p}\!\cdot\!\hat{q} -1\right) \nn
&-& \hat{q}\!\cdot\!\hat{k} \,\hat{p}\!\cdot\!\hat{r} \, \hat{k}\!\cdot\!\hat{r}  - x \, \hat{q}.\hat{r} \,\hat{k}\!\cdot\!\hat{r} \pm \frac{k}{k_0} \, \hat{k}\!\cdot\!\hat{r} \left(\hat{p}.\hat{r} + \hat{q}\!\cdot\!\hat{r} \right)\bigg)\bigg\}\tilde{D}_{L}^{\mathrm{A}}(K) \Delta_{\mp}^{\mathrm{S}}(Q)\bigg] 
\ea
Let us consider the fifth term of eq.~\eqref{sigma_final}, which is 
\begin{equation}\label{Sigma1_5}
\Sigma _{5\pm}^{(1)}(P)= \frac{-ig^2 C_F}{2} \int \frac{d^4K}{(2\pi)^4}\Big[F_{\pm ; -+}^{\mathrm{AS}}(P, K)\Big]
\end{equation}
This term is analogous to the fourth term of eq.~\eqref{sigma_final}) except for the change of $ k_0$. Here, $k_0 \rightarrow k_0 - 2 i \varepsilon $ in the definition of the function $V(t,k,k_{0},x,\varepsilon) , A$ and $B$. So, $\Sigma 5_{\pm}^{(1)}(P)$ can be calculated numerically analogulsly of fourth term.
Now the sixth term of eq.~\eqref{sigma_final} is 
\begin{equation}\label{Sigma1_6}
\Sigma _{6\pm}^{(1)}(P) = \frac{-ig^2 C_F}{2} \int \frac{d^4K}{(2\pi)^4}\left[F_{\pm ; --;--}^{\mathrm{SR}}(P, K)\right]
\end{equation}
The terms involved in $ F_{\pm ; --;--}^{\mathrm{SR}}(P, K) $ can be deduced from the eq.~\eqref{htl_terms_2}. So let us consider the terms of eq.~\eqref{Sigma1_6} without $s$, which can be written in a compact form after doing some simplification using eq.~\eqref{V_k_k0}, as
\ba 
F_{1\pm ; -- ; --}^{\mathrm{SR}}(P, K) &=&
-k^{2}\left[1+\left(\frac{p}{q}-\frac{kx}{q}\right)\right]\,V(t,k,k_{0},x,\varepsilon)^{2}\,\tilde{D}^{S}_{L}(K)\Delta^{R}_{\mp}(Q)
\ea
Now the terms of eq.~\eqref{Sigma1_6} consisting of $1s$, can be simplified using eq.~\eqref{1s_int}, can be written in a compact form as
\ba
F_{2\pm ; -- ; --}^{\mathrm{SR}}(P, K) &=& -2V_{1}(r_{0},r) \,V(t,k,k_{0},x,\varepsilon)\,\tilde{D}^{S}_{L}(K)\Delta^{R}_{\mp}(Q)\left[\mp k^{2}\left(\hat{p}\mdot \hat{r} + \hat{q}\mdot\hat{r}\right)
- k_{0}k \,\hat{k}\mdot\hat{r}\left(1+\hat{p}\mdot\hat{q}\right)\right]
\ea
Let us consider the terms of eq.~\eqref{Sigma1_6} with $2s$, which can be simplified using the eqs.~\eqref{V_k_k0}~\eqref{1s_int}~\eqref{2s_int}, having the final form as 
\ba
F_{3\pm ; -- ; --}^{\mathrm{SR}}(P, K) &=& \int_{0}^{1} d u \left(V_1 \left(r_0,r \right)\right)^2  D_{T}^{\mathrm{S}}(K) \Delta_{\mp}^{\mathrm{R}}(Q) \left(1 + \hat{p}\mdot\hat{q}\right)\left\{\left(\hat{k}\mdot\hat{r}\right)^2 - 1 \right\}-\int_{0}^{1} d u\bigg(\tilde{D}^{S}_{L}(K)\Delta^{R}_{\mp}(Q)\bigg)\nn
&\times&\bigg[\left(V_1 \left(r_0,r \right)\right)^2 \bigg\{k^{2}\bigg(2\hat{p}\mdot\hat{r}\,\hat{q}\mdot\hat{r}-\hat{p}\mdot\hat{q}+1\bigg) +k_{0}^{2} \bigg(\hat{k}\mdot\hat{r}\bigg)^{2}\bigg(1+\hat{p}\mdot\hat{q}\bigg)\pm 2k_{0}k \bigg(\hat{k}\mdot\hat{r}\big(\hat{p}\mdot\hat{r}+\hat{q}\mdot\hat{r}\big)\bigg)\bigg\}\nn
&\pm& 2V(t,k,k_{0},x,\varepsilon)\, k_{0}k \,\bigg\{A\left(x+\hat{q}\mdot\hat{k}\right)+B\, \hat{k}\mdot\hat{r}\,\big(\hat{p}\mdot\hat{r}+\hat{q}\mdot\hat{r}\big) \bigg\}\bigg]
\label{F1_int}
\ea
Similarly, the terms of eq.~\eqref{Sigma1_6} with $3s$ can be simplified using eqs.~\eqref{1s_int},\eqref{2s_int} and we will get 
\ba
F_{4\pm ; -- ; --}^{\mathrm{SR}}(P, K) &=& \pm 2 \int_{0}^{1} d u \,  V_1(r_0,r) \left[ A \left(\hat{p}\mdot\hat{r} + \hat{q}\mdot\hat{r} -\hat{p}\mdot\hat{k} \, \hat{r}\mdot\hat{k}-\hat{q}\mdot\hat{k} \, \hat{r}\mdot\hat{k}\right)\right. +\left. B \left(\hat{p}\mdot\hat{r} + \hat{q}\mdot\hat{r} -\hat{p}.\hat{r} \, (\hat{k}.\hat{r})^2 - \hat{q}.\hat{r} \, (\hat{k}.\hat{r})^2 \right)\right] \nn
&\times& D_{T}^{\mathrm{S}}(K) \Delta_{\mp}^{\mathrm{R}}(Q) +2 \int_{0}^{1} d u\ k_0^2 \,  V_1(r_0,r)\,\tilde{D}^{S}_{L}(K)\Delta^{R}_{\mp}(Q) \bigg\{\frac{k}{k_0} \, \hat{k}\mdot \hat{r} \, \big(A+B \big) \big(1-\hat{p}\mdot\hat{q}\big)\nn
&&\hspace{-0.5cm}+ A \bigg( \frac{k}{k_0} \, \big(\hat{p}\mdot \hat{k} \, \hat{q}\mdot \hat{r} + \hat{q}\mdot \hat{k} \, \hat{p}\mdot \hat{r}\big) \pm  \big(\hat{p}\mdot \hat{k} \, \hat{k}\mdot \hat{r}+\hat{q}\mdot \hat{k} \, \hat{k}\mdot \hat{r}\big) \bigg)
+ B \bigg(\frac{k}{ k_0} \, \big(2\hat{p}\mdot \hat{r} \, \hat{q}\mdot \hat{r}\, \hat{k}\mdot \hat{r} \big) \pm   \big(\hat{k}\mdot \hat{r}\big)^{2} \left(\hat{p}\mdot \hat{r} +\hat{q}\mdot \hat{r}\right) \bigg)\! \bigg\}
\ea
Finally, the terms of eq.~\eqref{Sigma1_6} with $4s$, can be simplified in terms of $A$ and $B$ as defined in eq.~\eqref{A_B_Def} as
\ba
F_{5\pm ; -- ; --}^{\mathrm{SR}}(P, K)
& =&  \int_{0}^{1} du \, \Bigg[2A^2 \left(   x\, \hat{q}\mdot\hat{k} -1 \right) + B^2 \left(\hat{p}\mdot\hat{q} - 2 \hat{p}\mdot\hat{r} \, \hat{q}\mdot\hat{r}  - 1 + 2 \hat{p}\mdot\hat{r} \, \hat{q}\mdot\hat{r} \, \big(\hat{k}\mdot\hat{r}\big)^2+  \big(\hat{k}\mdot\hat{r}\big)^2 \left( 1 - \hat{p}.\hat{q}\right) \right) \nn
&& \hspace{-1cm}+AB \left\{-4 \hat{p}.\hat{r} \, \hat{q}.\hat{r} + 2 \left(\hat{p}.\hat{q} - 1 \right) + 2 x \, \hat{q}.\hat{r} \, \hat{k}.\hat{r} + 2 \hat{q}.\hat{k} \, \hat{p}.\hat{r} \,\hat{k}\cdot\hat{r} + 2 \left(\hat{k}.\hat{r}\right)^2 \left( 1 - \hat{p}\mdot\hat{q}\right) \right\}\Bigg]  D_{T}^{\mathrm{S}}(K) \Delta_{\mp}^{\mathrm{R}}(Q)\nn
&& - \int_{0}^{1} du \,k_{0}^{2}\,\tilde{D}^{S}_{L}(K)\Delta^{R}_{\mp}(Q) \bigg[\left(1-\hat{p}\mdot\hat{q}\right)\bigg(A^{2} + \left(B^{2}+2AB\right)\big(\hat{k}\mdot\hat{r}\big)^{2}\bigg)\nn
&& \hspace{0.5cm}+\ 2A^{2} \, \hat{p}\mdot\hat{k}\, \hat{q}\mdot\hat{k} + 2B^{2} \, \hat{p}\mdot\hat{r} \, \hat{q}\mdot\hat{r} \, \big(\hat{k}\mdot\hat{r}\big)^{2} + 2AB \, \left(\hat{p}\mdot\hat{k} \, \hat{q}\mdot\hat{r} \, \hat{k}\mdot\hat{r} + \hat{q}\mdot\hat{k} \, \hat{p}\mdot\hat{r} \, \hat{k}\mdot\hat{r} \right)  \bigg]
\ea
Adding all the five individual contributions, eq.~\eqref{Sigma1_6} becomes
\ba
\Sigma _{6\pm}^{(1)}(P) &=& \frac{-ig^2 C_F}{2(2\pi)^4}  (2\pi) \int_{-\infty}^\infty dk_0 \int_{0}^\infty k^2 dk  \int_{-1}^1 dx \, \,\bigg[F_{1\pm;--;--}^{SR}(P,K)\nn
&+& F_{2\pm;--;--}^{SR}(P,K) +F_{3\pm;--;--}^{SR}(P,K)+F_{4\pm;--;--}^{SR}(P,K)+F_{5\pm;--;--}^{SR}(P,K)\bigg].
\ea
Let us consider the seventh term of eq.~\eqref{sigma_final}, which is 
\begin{equation}
\Sigma _{7\pm}^{(1)}(P) = \frac{-ig^2 C_F}{2} \int \frac{d^4K}{(2\pi)^4}\Big[F_{\pm ; --;+-}^{\mathrm{AS}}(P, K)\Big]
\label{Sigma1_7}
\end{equation}
The terms involved in $ F_{\pm ; --;+-}^{\mathrm{AS}}(P, K) $ can be read from using eq.~\eqref{htl_terms_2}. The terms of eq.~\eqref{Sigma1_7} without $s$ can be simplified using eq.~\eqref{V_k_k0} and thus we will get
\ba 
F_{1\pm ; -- ; +-}^{\mathrm{AS}}(P, K) &=&
-k^{2}\left[1+\bigg(\frac{p}{q}-\frac{kx}{q}\bigg)\right] V(t,k,k_{0},x,\varepsilon)\,V^{\prime}(t,k,k_{0},x,\varepsilon)\,\tilde{D}^{A}_{L}(K)\Delta^{S}_{\mp}(Q)
\ea
The expression of $ V^{\prime}\left(t,k,k_{0},x,\varepsilon\right)$ is same as $V\left(t,k,k_{0},x,\varepsilon\right)$ except the change of $k_0 \rightarrow k_0 + 2 i \varepsilon$. Let us take the terms of eq.~\eqref{Sigma1_7} consisting of $1s$, which can be solved using the eqs.~\eqref{V_k_k0}~\eqref{1s_int} and we get the expression as
\ba
F_{2\pm ; -- ; +-}^{\mathrm{AS}}(P, K) &=& -\bigg[\bigg\{\bigg(V_{1}(r_{0},r) \,V^{\prime}(t,k,k_{0},x,\varepsilon)+V^{\prime}_{1}(r_{0},r) \,V(t,k,k_{0},x,\varepsilon)\bigg)\,\tilde{D}^{A}_{L}(K)\Delta^{S}_{\mp}(Q)\bigg\}\nn
&&\times\left\{\mp k^{2} \left(\hat{p}\mdot \hat{r} + \hat{q}\mdot\hat{r}\right) - k_{0}k \,\hat{k}\mdot\hat{r}\left(1+\hat{p}\mdot\hat{q}\right)\right\}\bigg]
\ea
The expression of $ V_{1}^{\prime} \left(r_0,r \right)$ is same as $V_{1} \left(r_0,r \right)$ except the change of $k_0 \rightarrow k_0 + 2 i \varepsilon$. The terms of eq.~\eqref{Sigma1_7} with $2s$, may be simplified using eqs.~\eqref{V_k_k0}~\eqref{1s_int}~\eqref{2s_int}, can be written in the compact form as  
\ba
F_{3\pm ; -- ; +-}^{\mathrm{AS}}(P, K) &=& \int_{0}^{1} d u \left(V_1 \left(r_0,r \right)\right)\left(V^{\prime}_1 \left(r_0,r \right)\right)  D_{T}^{\mathrm{A}}(K) \Delta_{\mp}^{\mathrm{S}}(Q) \left[\left(1 + \hat{p}\mdot\hat{q}\right)\left\{\big(\hat{k}\mdot\hat{r}\big)^2 - 1 \right\}\right]-\int_{0}^{1} d u \tilde{D}^{A}_{L}(K)\Delta^{S}_{\mp}(Q) \nn
&\times&\bigg[ V_1 \left(r_0,r \right) \, V^{\prime}_1 \left(r_0,r \right) \left\{k^{2}\left(2\hat{p}\mdot\hat{r}\,\hat{q}\mdot\hat{r}-\hat{p}\mdot\hat{q}+1\right) +k_{0}^{2} \big(\hat{k}\mdot\hat{r}\big)^{2}\left(1+\hat{p}\mdot\hat{q}\right)\pm 2k_{0}k \,\hat{k}\mdot\hat{r}\big(\hat{p}\mdot\hat{r}+\hat{q}\mdot\hat{r}\big) \right\}\nn
&&\hspace{-2cm}\pm V^{\prime}(t,k,k_{0},x,\varepsilon) k_{0}k \left\{A\big(x+\hat{q}\mdot\hat{k}\big)+B\hat{k}\mdot\hat{r}\left(\hat{p}\mdot\hat{r}+\hat{q}\mdot\hat{r}\right)\right\}
\pm V(t,k,k_{0},x,\varepsilon) k_{0}k \left\{A^{\prime}\big(x+\hat{q}\mdot\hat{k}\big)+B^{\prime}\hat{k}\mdot\hat{r}\left(\hat{p}\mdot\hat{r}+\hat{q}\mdot\hat{r}\right)\right\}\bigg]\qquad\quad
\label{F3_int_AS_--_+-}
\ea
The expression of $ A^{\prime}$ and $B^{\prime}$ are the same as $A$ and $B$ except the change of $k_0 \rightarrow k_0 + 2 i \varepsilon$. Now, let us consider the terms of eq.~\eqref{Sigma1_7} with $3s$, which can be simplified using eqs.~\eqref{1s_int},\eqref{2s_int}, and we will get
\ba
F_{4\pm ; -- ; +-}^{\mathrm{AS}}(P, K) &=& \pm \int_{0}^{1} d u \, D_{T}^{\mathrm{A}}(K) \Delta_{\mp}^{\mathrm{S}}(Q)\, \bigg [\left\{V_{1}^{\prime}(r_0,r) A + V_1(r_0,r) A^{\prime}\right\}
\left(\hat{p}.\hat{r} + \hat{q}.\hat{r} -\hat{p}.\hat{k} \, \hat{r}.\hat{k} -\hat{q}.\hat{k} \, \hat{r}.\hat{k}\right)+\left\{V_{1}^{\prime}(r_0,r) B \right. \nn
&+&\left. V_1(r_0,r) B^{\prime}\right\} \left(\hat{p}\mdot\hat{r} + \hat{q}.\hat{r} -\hat{p}.\hat{r} \, (\hat{k}.\hat{r})^2 - \hat{q}.\hat{r} \, (\hat{k}.\hat{r})^2 \right)\bigg] + \int_{0}^{1} d u \,\bigg[\bigg(V_{1}^{\prime}(r_0,r) A + V_1(r_0,r) A^{\prime}\bigg) \nn
&\times& \left(k_{0} k \left(\hat{p}\mdot \hat{k} \, \hat{q}\mdot \hat{r} + \hat{q}\mdot \hat{k} \, \hat{p}\mdot \hat{r} +\hat{k}\mdot \hat{r} \big(1-\hat{p}\mdot\hat{q}\big)\right)  \pm k_{0}^{2} \big(\hat{p}\mdot \hat{k} \, \hat{k}\mdot \hat{r}+\hat{q}\mdot \hat{k} \, \hat{k}\mdot \hat{r}\big)\right)+ \left(V_{1}^{\prime}(r_0,r) B + V_1(r_0,r) B^{\prime}\right)\nn
&\times&\bigg(k_{0} k \left(2\hat{p}\mdot \hat{r} \, \hat{q}\mdot \hat{r}\, \hat{k}\mdot \hat{r} +\hat{k}\mdot \hat{r} \big(1-\hat{p}\mdot\hat{q}\big)\right) \pm k_{0}^{2} \big(\hat{k}\mdot\hat{r}\big)^{2}\big(\hat{p}\mdot \hat{r}+\hat{q}\mdot \hat{r}\big)\bigg)\bigg]\tilde{D}^{A}_{L}(K)\Delta^{S}_{\mp}(Q)
\ea
Let us consider the terms of eq.~\eqref{Sigma1_7} with $4s$ which can be rewritten in terms of $A,B,A'$ and $B'$ variables as defined in eq.~\eqref{A_B_Def} and gives the final form as
\ba
F_{5\pm ; -- ; +-}^{\mathrm{AS}}(P, K)
& =&  \int_{0}^{1} du \, \Bigg[2AA^{\prime} \left( x\, \hat{q}\mdot\hat{k} -1\right) + BB^{\prime} \left(\hat{p}\mdot\hat{q} - 2 \hat{p}.\hat{r} \, \hat{q}\mdot\hat{r}  - 1 + 2 \hat{p}\mdot\hat{r} \, \hat{q}\mdot\hat{r} \, \big(\hat{k}\mdot\hat{r}\big)^2+  \big(\hat{k}\mdot\hat{r}\big)^2 \left( 1 - \hat{p}\mdot\hat{q}\right) \right) \nn
&&\hspace{-1.5cm}+ \big(A^{\prime}B+AB^{\prime}\big) \left\{-2 \hat{p}.\hat{r} \, \hat{q}\mdot\hat{r} +  \left(\hat{p}\mdot\hat{q} - 1 \right) +  x \, \hat{q}\mdot\hat{r} \, \hat{k}\mdot\hat{r} +  \hat{q}.\hat{k} \, \hat{p}\mdot\hat{r} \,\hat{k}\mdot\hat{r} +  \left(\hat{k}\mdot\hat{r}\right)^2 \left( 1 - \hat{p}\mdot\hat{q}\right) \right\}\Bigg] D_{T}^{\mathrm{A}}(K) \Delta_{\mp}^{\mathrm{S}}(Q) \nn
&&\hspace{-1.5cm}- \int_{0}^{1} du \,k_{0}^{2}\,\tilde{D}^{A}_{L}(K)\Delta^{S}_{\mp}(Q) \bigg[\left(1-\hat{p}\mdot\hat{q}\right)\bigg(AA^{\prime} + \left(BB^{\prime}+A^{\prime}B+AB^{\prime}\right)\big(\hat{k}\mdot\hat{r}\big)^{2}\bigg) +  2AA^{\prime} \, \hat{p}\mdot\hat{k}\, \hat{q}\mdot\hat{k} \nn
&+& 2BB^{\prime} \, \hat{p}\mdot\hat{r} \, \hat{q}\mdot\hat{r} \, \big(\hat{k}\mdot\hat{r}\big)^{2} + \big(A^{\prime}B+AB^{\prime}\big) \, \left(\hat{p}\mdot\hat{k} \, \hat{q}\mdot\hat{r} \, \hat{k}\mdot\hat{r} + \hat{q}\mdot\hat{k} \, \hat{p}\mdot\hat{r} \, \hat{k}\mdot\hat{r} \right)  \bigg]
\ea
Adding all the five individual contributions, eq.~\eqref{Sigma1_7} becomes
\ba
\Sigma _{7\pm}^{(1)}(P) &=& \frac{-ig^2 C_F}{2(2\pi)^4}  (2\pi) \int_{-\infty}^\infty dk_0 \int_{0}^\infty k^2 dk  \int_{-1}^1 dx \, \,\bigg[F_{1\pm;--;+-}^{AS}(P,K)\nn
&+& F_{2\pm;--;+-}^{AS}(P,K) +F_{3\pm;--;+-}^{AS}(P,K)+F_{4\pm;--;+-}^{AS}(P,K)+F_{5\pm;--;+-}^{AS}(P,K)\bigg].
\ea
Equations~\eqref{Sigma1_4},~\eqref{Sigma1_5},~\eqref{Sigma1_6} and~\eqref{Sigma1_7} have the same kind of divergences as we seen in eq.~\eqref{Sigma13}. Thus we get the same sudden jumps as we mentioned in eq.~\eqref{diverg_pts_1}. So, these equations are also evaluated numerically in each of the figure~\ref{fig7} domains and summed up. Now, the eighth term of eq.~\eqref{sigma_final} is 
\begin{equation}\label{Sigma1_8}
\Sigma _{8\pm}^{(1)}(P) = \frac{-ig^2 C_F}{2} \int \frac{d^4K}{(2\pi)^4}\Big[G_{\pm ; --}^{\mathrm{S}}(P, K)\Big]
\end{equation}
The complete terms of $ G_{\pm ; --}^{\mathrm{S}}(P, K) $ are given in eq.~\eqref{sigma2_htl}. 
The term of eq.~\eqref{Sigma1_8} without any s can be simplified using the eq.~\eqref{J_mu_nu_alpha} as
\begin{equation}
\begin{aligned}
G_{1\pm ; - - }^{\mathrm{S}}(P, K) &=& \left[J^{000}_{--} \left(P,K\right) + J^{000}_{--} \left(P,-K\right) \right] \bigg\{ {D}_{T}^{\mathrm{S}}(K) + k^{2} \,\tilde{D}_{L}^{\mathrm{S}}(K)\bigg\} 
\label{G1pm}
\end{aligned}
\end{equation}
Now, using eq.~\eqref{J000}, we get
\begin{equation}
\begin{aligned}
J_{--}^{000}(P, K) &=2 \int_{0}^{1} d u_{1} u_{1} \int_{0}^{1} d u_{2} \frac{t_{0}}{\left(t_{0}^{2}-t^{2}\right)^{2}}
\end{aligned}
\end{equation}
where \begin{equation}
\begin{aligned}
t_0 = u_1 u_2 k_0 + p_0 - i \varepsilon , \hspace{5mm}\vec{t} = u_1 u_2 \vec{k} + \vec{p}\ \text{ and } \
 t = \sqrt{p^2+k^2 u_{1}^2 u_{2}^2 + 2 p k u_1 u_2 x}.
\end{aligned}
\end{equation}
Thus, eq.~\eqref{G1pm} becomes
\begin{equation}
\begin{aligned}
G_{1\pm ; - - }^{\mathrm{S}}(P, K) = 2 \int_{0}^{1} d u_{1} u_{1} \int_{0}^{1} d u_{2} \,\bigg[\frac{t_{0}}{\left(t_{0}^{2}-t^{2}\right)^{2}}+ \frac{t^{\prime}_{0}}{\left(t^{\prime 2}_{0}-t^{\prime 2}\right)^{2}}\bigg]\, \bigg\{ {D}_{T}^{\mathrm{S}}(K) + k^{2} \,\tilde{D}_{L}^{\mathrm{S}}(K)\bigg\} 
\end{aligned}
\end{equation}
The variables $t^{\prime}_{0}$ and $t^{\prime}$ have the same expression as $t_{0}$ and $t$ with negative gluon four-momentum $K$. Now, the required angle in order to solve the other terms of eq.~\eqref{sigma2_htl} are
\begin{equation*}
\hat{p}\cdot\!\hat{t} = \frac{p}{t} + \frac{k u_1 u_2}{t}x \quad \quad
\hat{k}\cdot\!\hat{t} = \frac{p}{t}x + \frac{k u_1 u_2}{t}
\end{equation*}
Let us consider the terms of eq.~\eqref{Sigma1_8} with 1s i.e. $G_{2\pm ; - - }^{\mathrm{S}}(P, K)$. This term can be simplified using eqs.~\eqref{J_mu_nu_alpha} and~\eqref{J00i} and having the final form as 
\ba
G_{2\pm ; - - }^{\mathrm{S}}(P, K) 
&=& \mp \, 2 \int_{0}^{1} d u_{1} u_{1} \int_{0}^{1} d u_{2} \bigg(\frac{1}{\left(t_{0}^{2}-t^{2}\right)^{2}}+\frac{1}{\left(t_{0}^{\prime 2}-t^{\prime 2}\right)^{2}}\bigg) \left[\hat{p}\mdot\hat{t} \, {D}_{T}^{\mathrm{S}}(K)
+ \bigg( k^{2} \,\hat{p}\mdot\hat{t}\pm 2k_{0}k \, \hat{k}\mdot\hat{t}\bigg)\,\tilde{D}_{L}^{\mathrm{S}}(K)
\right]\label{G2pmS}
\end{eqnarray}
In the similar manner, the term with 2s of eq.~\eqref{Sigma1_8} can be expressed using eqs.~\eqref{J_mu_nu_alpha} and ~\eqref{J0ij} as
\ba
G_{3\pm ; - - }^{\mathrm{S}}(P, K) 
&=& 2 \int_{0}^{1} d u_{1} u_{1} \int_{0}^{1} d u_{2} \, \bigg[-{D}_{T}^{\mathrm{S}}(K) \bigg\{C(P,K) + C (P,-K)+\left(D(P,K) + D(P,-K)\right)  \big(\hat{k}.\hat{t}\big)^2 \bigg\}\nn
&&\hspace{-2cm}+\tilde{D}_{L}^{\mathrm{S}}(K) \bigg\{\left(C(P,K) + C (P,-K)\right)\left( k_{0}^{2} \pm 2k_{0}k \,\hat{p}\mdot\hat{k} \right) +\left(D(P,K) + D(P,-K)\right)
\left( k_{0}^{2} \,\big(\hat{k}\mdot\hat{t}\big)^{2} \pm 2k_{0}k \,\hat{p}\mdot\hat{t}\,\hat{k}\mdot\hat{t} \right)\bigg\}              
\bigg]\ \hspace{1cm}
\ea
where
\begin{equation*}
\begin{aligned}
C (P,K) = \frac{t_{0}}{2 t^{2}\left(t_{0}^{2}-t^{2}\right)}-\frac{1}{4 t^{3}} \ln \frac{t_{0}+t}{t_{0}-t}, \quad \quad  \quad \quad
D (P,K )  =\frac{t_{0}\left(5 t^{2}-3 t_{0}^{2}\right)}{2 t^{2}\left(t_{0}^{2}-t^{2}\right)^{2}}+\frac{3}{4 t^{3}} \ln \frac{t_{0}+t}{t_{0}-t}
\end{aligned}
\end{equation*}
Let us take the terms of eq.~\eqref{Sigma1_8} consisting of 3s terms which take the form after doing some simplification 
\ba
G_{4\pm ; - - }^{\mathrm{S}}(P, K) 
&=& \pm 2 \int_{0}^{1} d u_{1} u_{1} \int_{0}^{1} d u_{2} \, \bigg[{D}_{T}^{\mathrm{S}}(K) \Big\{\left(E(P,K) + E (P,-K)\right)\left(\hat{p}\mdot\hat{t} + 2x \hat{k}\mdot\hat{t} \right) \nn
&+& \left(F(P,K) + F(P,-K)\right)  \hat{p}\mdot\hat{t} \big(\hat{k}\mdot\hat{t}\big)^2 \Big\}-\tilde{D}_{L}^{\mathrm{S}}(K) \bigg\{\left(E(P,K) + E (P,-K)\right)\left(\hat{p}\mdot\hat{t} + 2x \hat{k}\mdot\hat{t} \right) \nn
&+& \left(F(P,K) + F(P,-K)\right)\  \hat{p}\mdot\hat{t} \left(\hat{k}\mdot\hat{t}\right)^2 
\bigg\}\bigg]
\ea
with
\begin{equation*}
\begin{aligned}
E (P,K ) &=&\frac{1}{2 t^{3}}\left[2+\frac{t_{0}^{2}}{t_{0}^{2}-t^{2}}-\frac{3 t_{0}}{2 t} \ln \frac{t_{0}+t}{t_{0}-t}\right] ,\quad \quad \text{and}  \quad
F(P,K) &=&\frac{t}{\left(t_{0}^{2}-t^{2}\right)^{2}}-\frac{5}{2 t^{3}}\left[2+\frac{t_{0}^{2}}{t_{0}^{2}-t^{2}}-\frac{3 t_{0}}{2 t} \ln \frac{t_{0}+t}{t_{0}-t}\right].
\end{aligned}
\end{equation*}
\begin{figure}
	\centering
	{\includegraphics[scale=0.8,keepaspectratio]{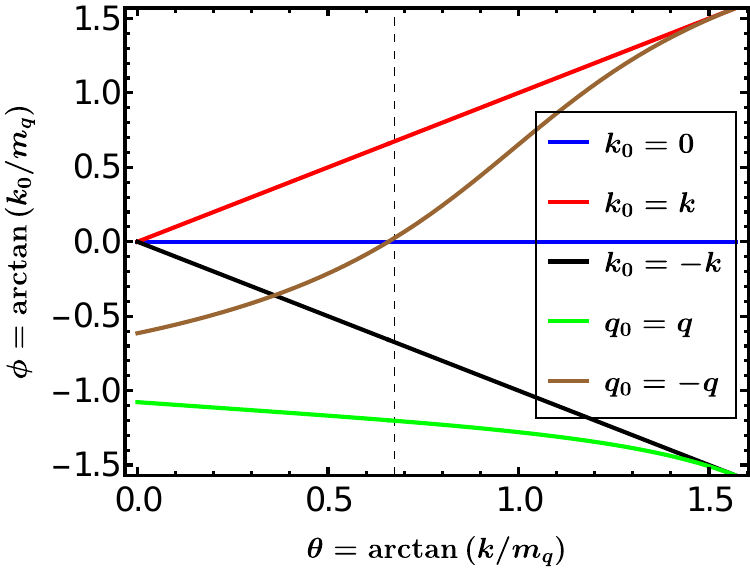}}
	\captionsetup{singlelinecheck=off}
	\caption[.]{Domains in $(k,k_0)$ plane at which the integrand in eq.~\eqref{Sigma8_final} has sudden jumps. Here we have used $t\equiv p/p_{0} = 0.45 $ and $x \equiv \cos\theta= 0.8 $} 
	\label{fig8}    
\end{figure} 
After adding four-contributions of $\Sigma 8_{\pm}^{(1)}(P)$, eq.~\eqref{Sigma1_8} becomes
\ba
\Sigma _{8\pm}^{(1)}(P) &=& \frac{-ig^2 C_F}{(2\pi)^3}  \int\limits_{-\infty}^\infty dk_0 \int\limits_{0}^\infty k^2 dk  \int_{-1}^1 dx \Big[G_{1\pm;--}^{S}(P,K) +G_{2\pm;--}^{S}(P,K)+G_{3\pm;--}^{S}(P,K)+G_{4\pm;--}^{S}(P,K)\Big].\label{Sigma8_final}
\ea
Now, to evaluate eq.~\eqref{Sigma8_final}, we find sudden jumps in the integrand at the following points:
\ba
k_{0}&=&0 ; \quad k_{0}=\pm k ; \quad
k_0 = -p_0 \pm \sqrt{p^2+k^2+2pkx} ;  \quad
k = \frac{1}{2 t} \frac{1-t^{2}}{1-x t} \sqrt{\frac{t}{1-t}-\frac{1}{2} \ln \left(\frac{1+t}{1-t}\right)}\label{domain2}
\ea
Figure~\ref{fig8} depicts the domains of eq.~\eqref{domain2}. We have evaluated eq.~\eqref{Sigma8_final} numerically in each of the individual domains of figure~\ref{fig8} and added the individual contribution to get the final result.
\section{Results and Discussion}\label{sec:result}
All the terms of NLO quark self-energy in eq.~\eqref{sigma_final} have a non-trivial dependence on $\varepsilon $. So in order to be more precise in the integral results, one needs to check the stability for each of the integrals very carefully, which depends non-trivially on the $\varepsilon$ parameter. Here, we have checked the stability for each term by plotting the integrand of that particular integral with $ -\log_{10}\varepsilon$. This is an essential task because different terms have different stability regions, and if one does the integration beyond those regions, then numerical values lose reliability. Even one needs to perform the integrals, which are divided into the different domains as shown in figures~\ref{fig6}, \ref{fig7}, \ref{fig8}, depending on their stability. For demonstration purposes, we have shown the $\varepsilon$ dependence plot in figure~\ref{fig9} for the transverse part of the integral mentioned in eq.~\eqref{F2_PM_SR_--_Sigma13}. For this particular term, on average, we found the stability region around $\sim 10^{-5}$ for the real part and $\sim 10^{-6}$ for the imaginary part, respectively. Then the integral for this particular term has been done around these stability regions. Similarly, other integrals have been handled in calculating NLO quark self-energy. \\
\begin{figure}
	\centering
	\includegraphics[scale=0.8,keepaspectratio]{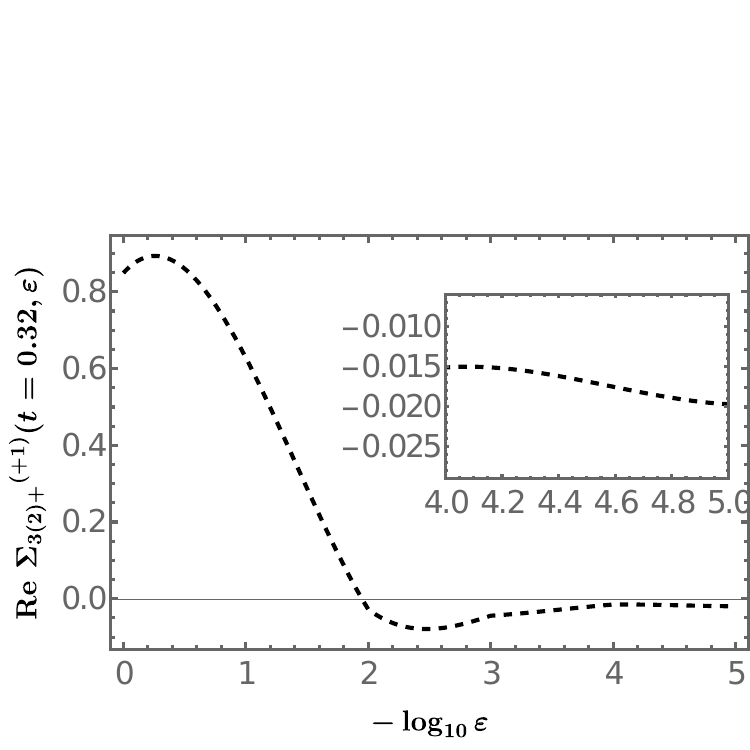}
	\qquad
	\includegraphics[scale=0.8,keepaspectratio]{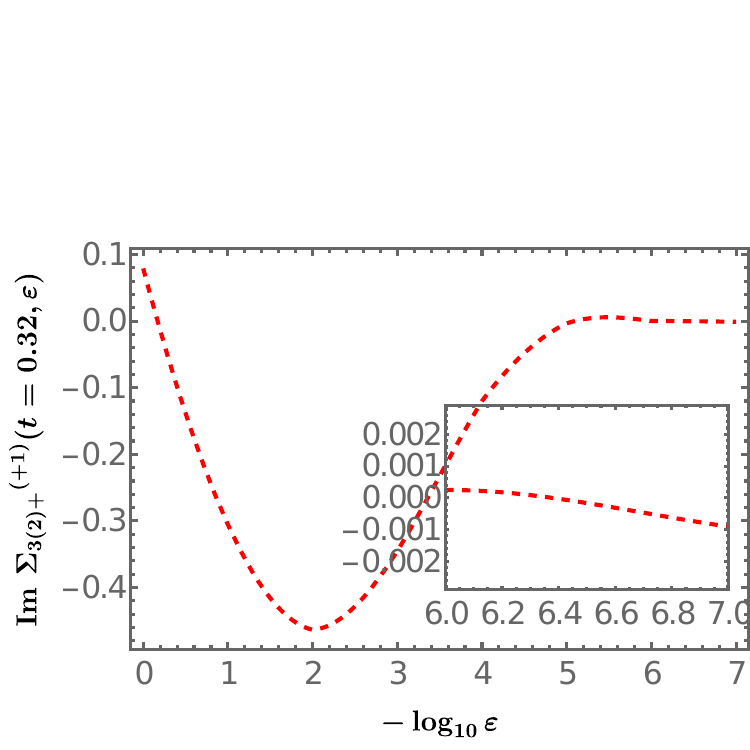}
	\caption{The $\varepsilon$ dependence of the Real and Imaginary part of the integral $\Sigma_{3(2)\pm}^{(+1)}$ at $t=0.32$ containing only transverse contribution of eq.~\eqref{F2_PM_SR_--_Sigma13} in multiple of $gm_{q}$.}
	\label{fig9}
\end{figure}  
{Note that there are a total of 92 terms (46 for $\Sigma_{+}^{(1)}$ and 46 for $\Sigma_{-}^{(1)}$) for which we needed to check the convergence. All these terms have convergence for different values of $\varepsilon$. For example, as we have mentioned, for a particular term in eq.~\eqref{F2_PM_SR_--_Sigma13}, the stability region is around $\sim 10^{-5}$ for the real part and $\sim 10^{-6}$ for the imaginary part. In principle, the $\varepsilon$ value should be zero, but our numerical evaluation can not handle that. Thus, because of the finite value of the $\varepsilon$ parameter, numerical errors are introduced in the evaluation of NLO quark self-energy. We have done the numerical estimation of the percentage of error in the following way: We have taken a few values of $\varepsilon$ in the stability region. Then we have extrapolated the value of that particular $\Sigma_{\pm}^{(1)}$ term to the limit of $\varepsilon$ approaches to $0$. We estimated the error of that particular term from the difference between the considered value and the extrapolated value. The error of 46 terms for $\Sigma_{+}^{(1)}$ is calculated, and we estimated the total error in the evaluation of NLO quark self-energy as $\frac{\delta\Sigma_{+}^{(1)}}{\Sigma_{+}^{(1)}} = \sum_{i} \sqrt{{ \bigg(\frac{\delta \Sigma_{+ i}^{(1)}}{\Sigma_{+ i}^{(1)}}\bigg)}^2}$. We have estimated the maximum error for the measurement of the NLO quark mass due to the finite $\varepsilon$ value being about $11\%$ and for the damping rate, it is about $10\%$ .}
\\
In section~\ref{sec4}, we have given the expressions for each term of eq.~\eqref{sigma_final} more elaborately and evaluated each term numerically. We add the numerical results of all those terms to get the final result for the expression mentioned in eq.~\eqref{sigma_final}. The results shown in figures \ref{fig10}$-$\ref{fig13} are scaled with coefficient $g m_{q}$ on the y-axis and plotted over $p/m_{q}$ on the x-axis. In figure~\ref{fig10}, we have shown how the imaginary part and real part of $\Sigma_{+}^{(1)}$ (transverse contribution + longitudinal contribution) scaled with a coefficient of $g m_{q}$ varies for two and three flavors. Figure~\ref{fig10}(a) shows the variation of imaginary part of $\Sigma_{+}^{(1)}$ with $p/p_0$. From this plot, one can get the NLO damping rate for quarks with $'+'$ mode, i.e., for ordinary quarks for $N_{f}=2$ and $N_{f}=3$, respectively. Figure~\ref{fig10}(b) shows the variation of real part of $\Sigma_{+}^{(1)}$ with $p/p_0$. We will get the  NLO quark energy from this plot with $`+'$ mode. Using the eq.~\eqref{disp_nlo}, damping rate and quark energy for soft momentum, $p$ plotted in figure~\ref{fig10} for '+' quark mode. Figure~\ref{fig11}(a) shows that the damping rate of real quark mode decreases with the increase of soft momentum and then becomes constant, as expected. Similarly, figure~\ref{fig11}(b) shows how the NLO correction to mass for $'+'$ quark mode behaves. 
\begin{figure}
	\centering
	\includegraphics[scale=0.8,keepaspectratio]{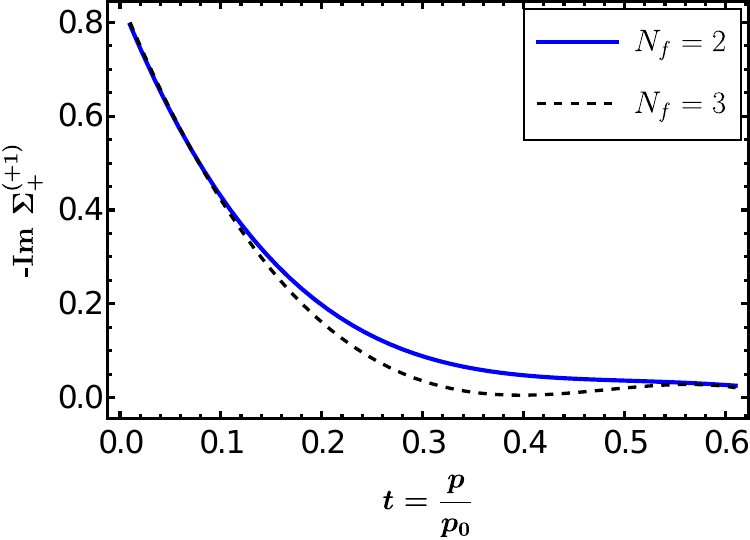}
	\qquad
	\includegraphics[scale=0.8,keepaspectratio]{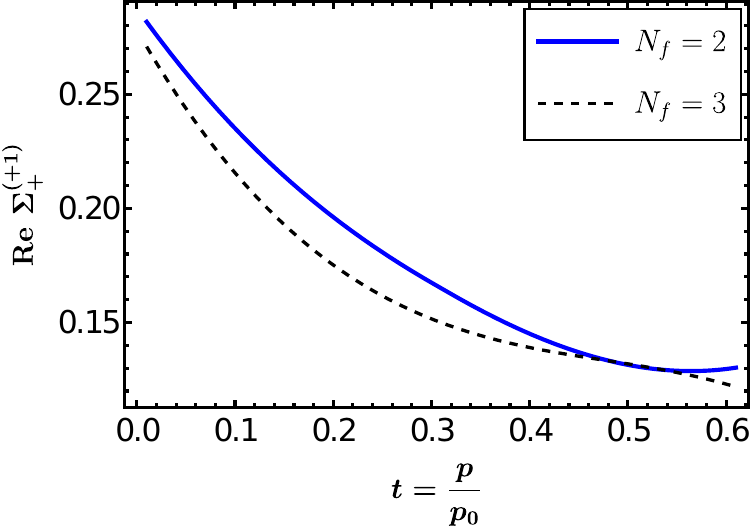}
	\caption{The Imaginary part and Real part of $\Sigma_{+}^{(1)}$, scaled with a coefficient of $g m_{q}$, with respect to parameter $t=p/p_{0}$}
	\label{fig10}
\end{figure}  
\qquad
\begin{figure}
	\centering
	\includegraphics[scale=0.8,keepaspectratio]{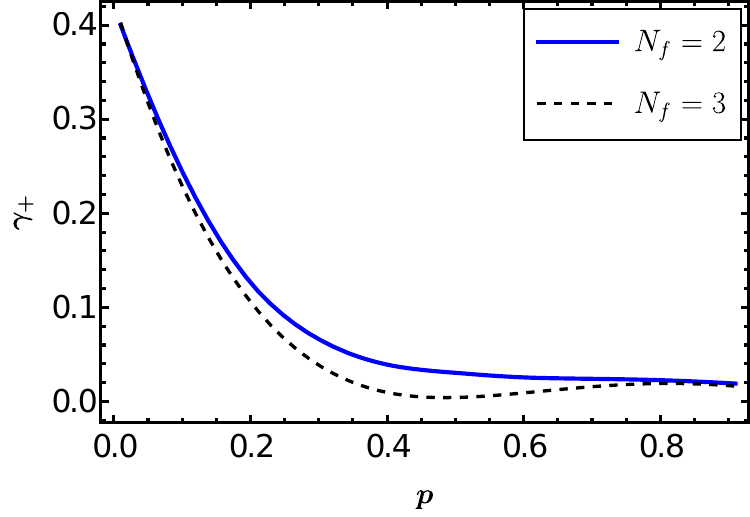} 
	\qquad
	\includegraphics[scale=0.8,keepaspectratio]{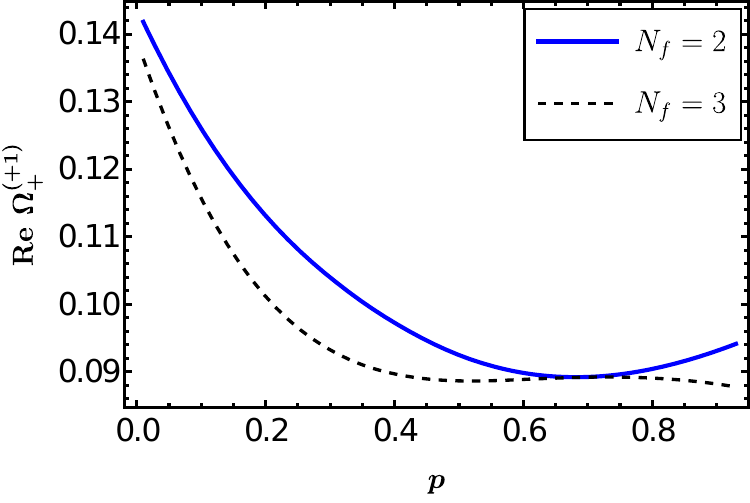}
	\caption{Damping rate and quark energy variation with soft momentum $p/m_{q}$ for `+' mode scaled with a coefficient $g m_{q}$.}
	\label{fig11}
\end{figure}
\hspace{-1.3cm} Now, figure~\ref{fig12} shows the behavior of the imaginary part and real part of $\Sigma_{-}^{(1)}$ (transverse contribution + longitudinal contribution) scaled with a coefficient of $g m_{q}$. Figure~\ref{fig12}(a) shows the variation of imaginary part of $\Sigma_{-}^{(1)}$ with $p/p_0$. One can get the NLO damping rate for plasmino mode from this plot. Figure~\ref{fig12}(b) shows the variation of real part of $\Sigma_{-}^{(1)}$ with $p/p_0$. One can get the NLO quark mass from this plot with $`-'$ mode. Using the expressions from eq.~\eqref{disp_nlo}, we have plotted the damping rate and quark energy w.r.t. soft momentum $p$ in figure~\ref{fig13} for this plasmino mode. Figure~\ref{fig13}(a) shows the behavior of the damping rate for plasmino mode. Figure~\ref{fig13}(b) shows the variation of NLO mass for $`-'$ mode w.r.t soft momentum $p$.
In the limit of zero momentum, one can see that the damping rate and correction to NLO mass approach the same value for both quark modes. The other significant outcome of the above results shows that we can handle the instabilities that arise from the gluon propagator's transverse and longitudinal components, respectively.
\begin{figure}
	\centering
	\includegraphics[scale=0.8,keepaspectratio]{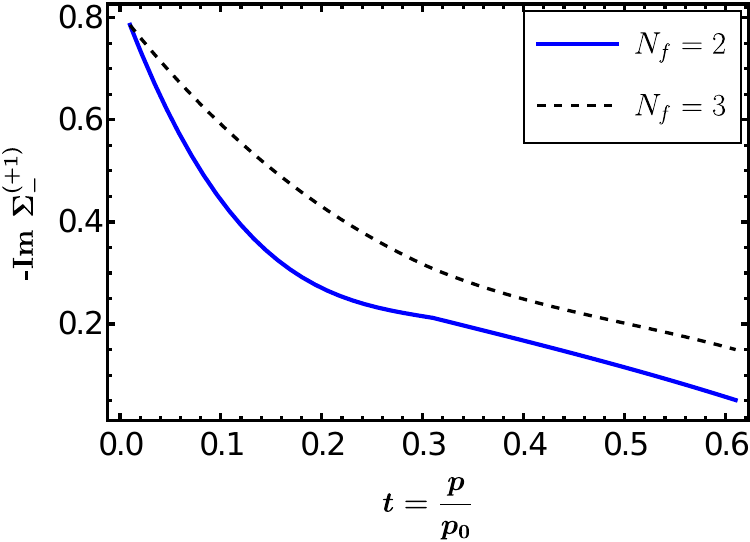} 
	\qquad
	\includegraphics[scale=0.8,keepaspectratio]{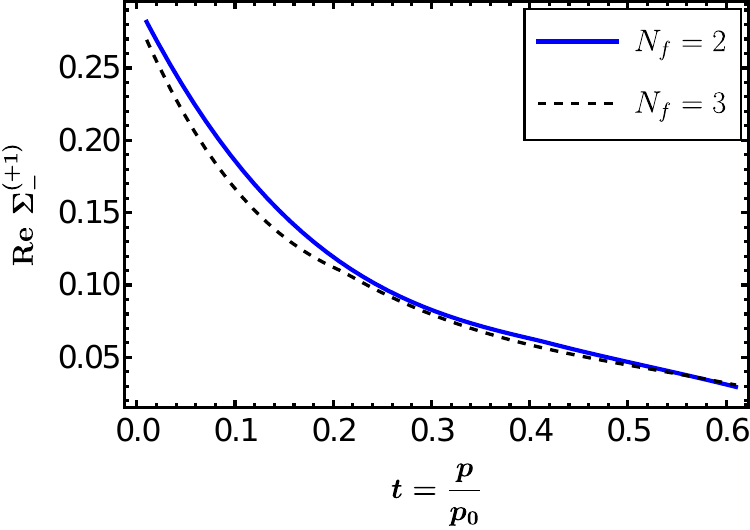}
	\caption{The Imaginary part and Real part of $\Sigma_{-}^{(1)}$, scaled with a coefficient of $g m_{q}$, with respect to parameter $t=p/p_{0}$}
	\label{fig12}
\end{figure}  
\qquad
\begin{figure}
	\centering
	\includegraphics[scale=0.8,keepaspectratio]{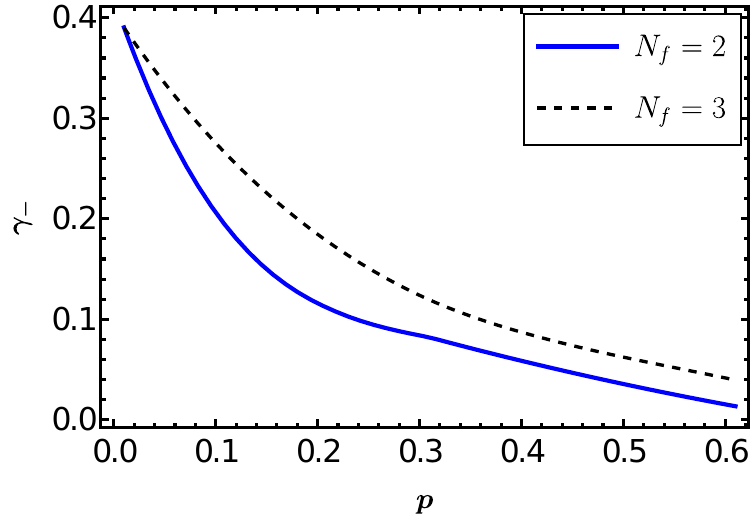} 
	\qquad
	\includegraphics[scale=0.8,keepaspectratio]{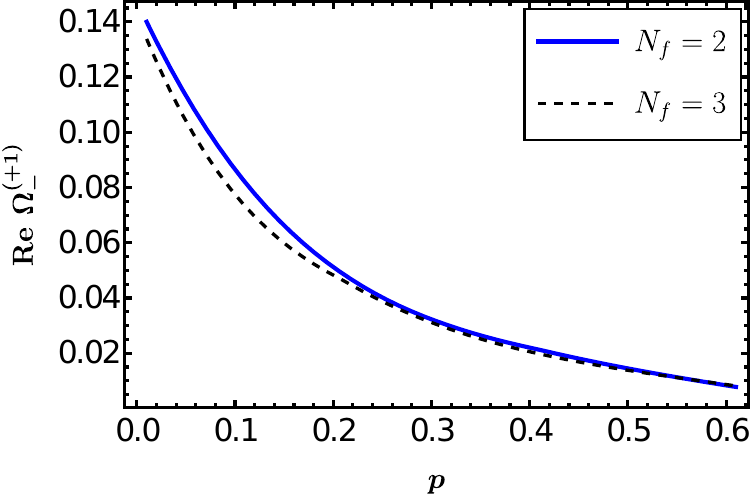}
	\caption{Damping rate and quark energy variation with soft momentum $p/m_{q}$ for plasmino mode scaled with a coefficient $g m_{q}$.}
	\label{fig13}
\end{figure} 
\begin{figure}
	\centering
	\includegraphics[scale=0.8,keepaspectratio]{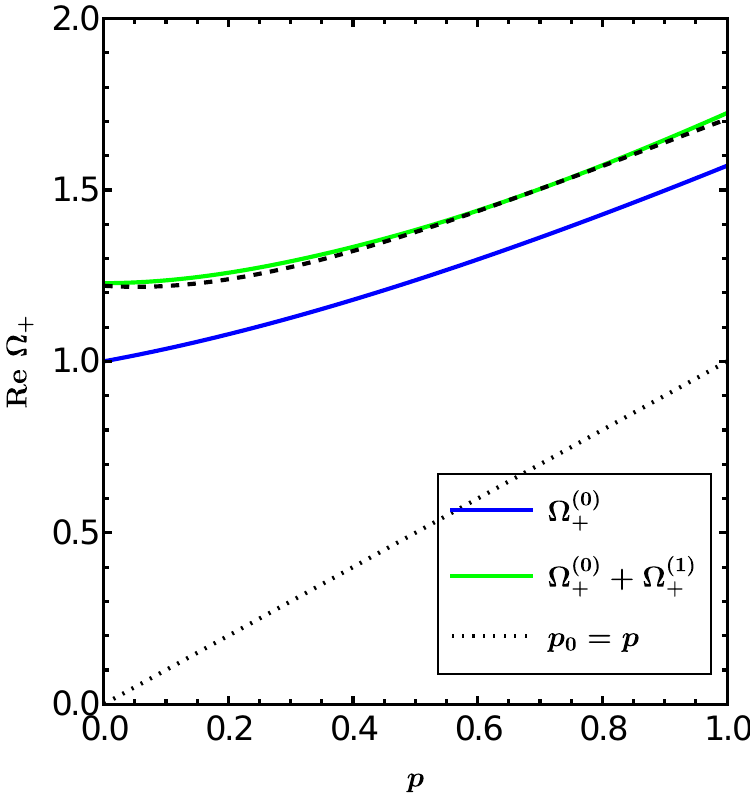} 
	\qquad
	\includegraphics[scale=0.8,keepaspectratio]{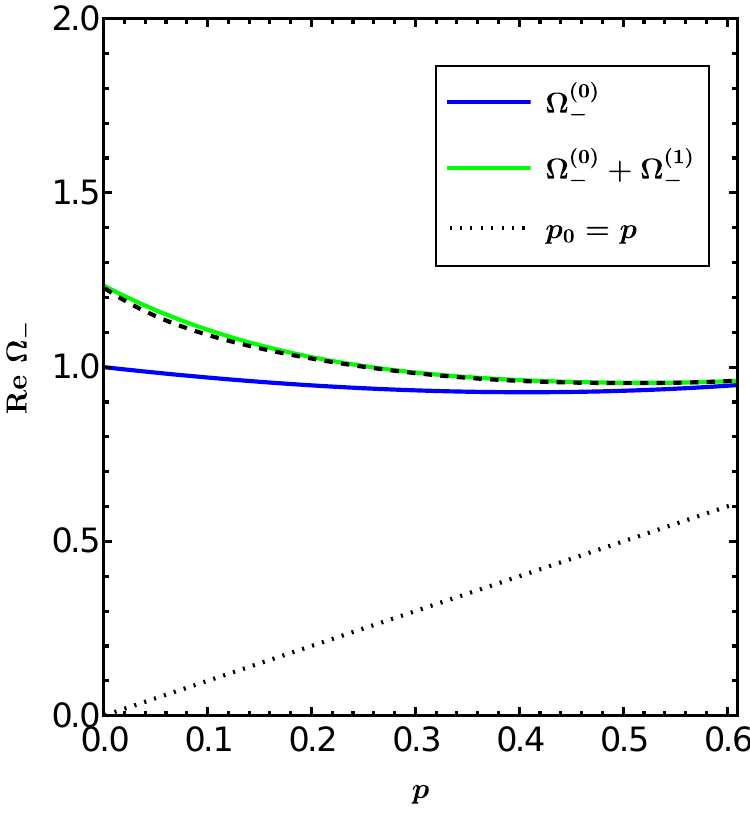}
	\caption{Real part of dispersion relation variation with respect to soft momentum $p/m_{q}$ for both quark modes. The solid line is for $N_{f}=2$ case, while the dotted line shows the results for $N_{f}=3$ flavors.}
	\label{fig_Omega_+1_-1}
\end{figure} 
\begin{figure}
	\centering
	\includegraphics[scale=0.8,keepaspectratio]{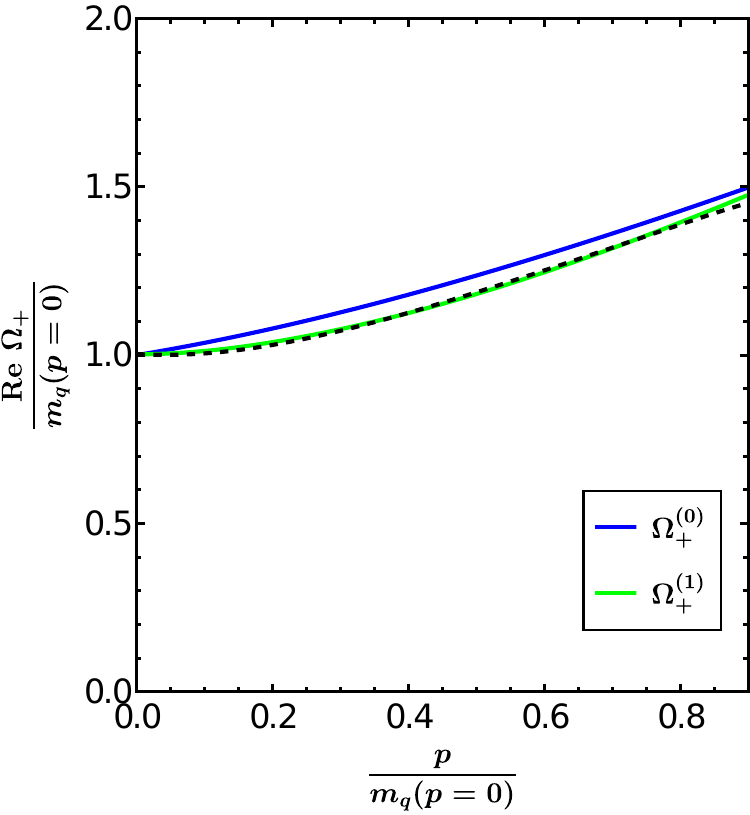} 
	\qquad
	\includegraphics[scale=0.8,keepaspectratio]{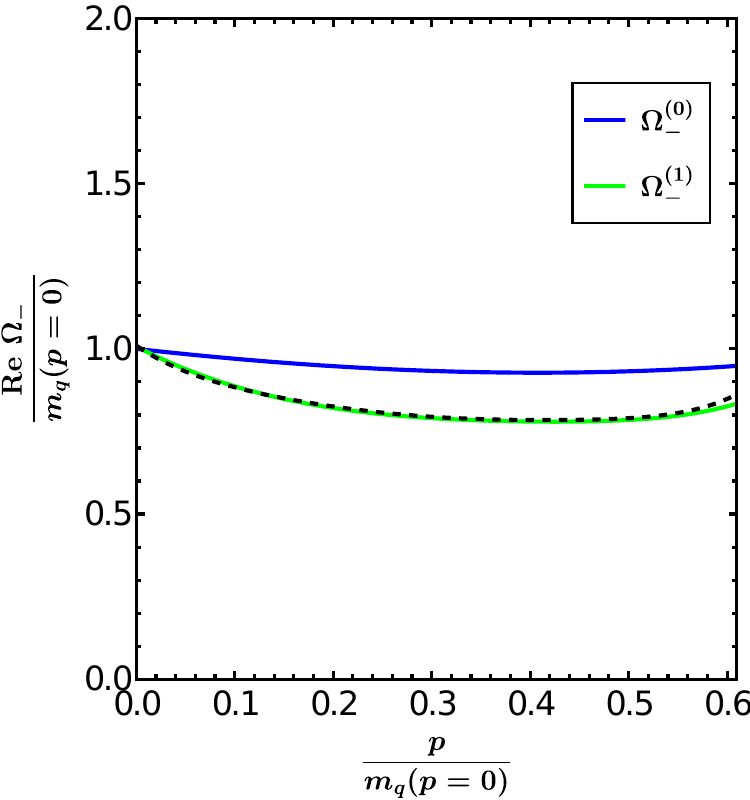}
	\caption{Real part of dispersion relation variation with respect to soft momentum $p/m_{q}$ scaled with their corresponding thermal masses. The solid line shows the results for $N_{f}=2$ flavors, while the dotted line shows the results for $N_{f}=3$ flavors. The value of $\alpha_{s}=0.2$ is used.}
	\label{fig_Omega_+1_-1_scaled}
\end{figure} 
\hspace{-1.5cm} We extract the numerical value of the dispersion relations, i.e., damping rate and mass, by taking the limit of $p \rightarrow 0$ in figures~\ref{fig11} and~\ref{fig13} respectively. In order to compare our results with the existing results in the literature, one can work in the units of $m_{q}$. In the limit of zero momentum, we obtain the values $\gamma_{\pm}(0) \approx 0.159\, g^{2}T$ for $N_{f}=2$ and $\gamma_{\pm}(0) \approx 0.164 g^{2}T$ for $N_{f}=3$ which is, respectively $7\%$ and $9\%$ larger than the existing result obtained in ref.~\cite{Braaten:1992gd}. {Similarly, in the limit of zero momentum for the correction in the mass denoted by $\Delta m$, we obtain the values $\Delta m = 0.142 gm_{q}$ for $N_{f}=2$ which is $\sim 5 \, \%$ smaller as compared with eq.~\eqref{m_q_nlo} and $\Delta m = 0.136 gm_{q}$ for $ N_{f}=3 $. The value of the coupling $\alpha_{s} =0.2$ viz. $g =1.58$ is used in the numerical evaluation of these results.} The real part of the dispersion relation for both modes, i.e., mass, variation w.r.t. soft momentum, is shown in fig.~\ref{fig_Omega_+1_-1}. In fig.~\ref{fig_Omega_+1_-1_scaled}, we have shown the dependence of the real part of dispersion relation viz. mass on the soft momentum scaled with their corresponding thermal masses on both axes for both modes. Also, one can extract the velocity for both quark modes using the graphs in the figure~\ref{fig_Omega_+1_-1}. Figure~\ref{fig_vel_+1_-1} shows the velocity variation w.r.t. soft momentum for real quark and plasmino modes. The result shows that for both quark modes, velocity is less than $c$ as expected. Also, these result shows that velocity for both modes decreases in the limit of zero momentum. The decrease of the velocity in the limit of $ p\rightarrow 0 $ shows that quasi-particles become massive. As the momentum increases, medium effects gradually vanish, as seen in figure~\ref{fig_vel_+1_-1}. Figure~\ref{fig_Mass_+1_-1} compares the real part of the dispersion relation with NLO correction and without NLO correction for soft momentum.    
\qquad
\begin{figure}
	\centering
	\includegraphics[scale=0.9,keepaspectratio]{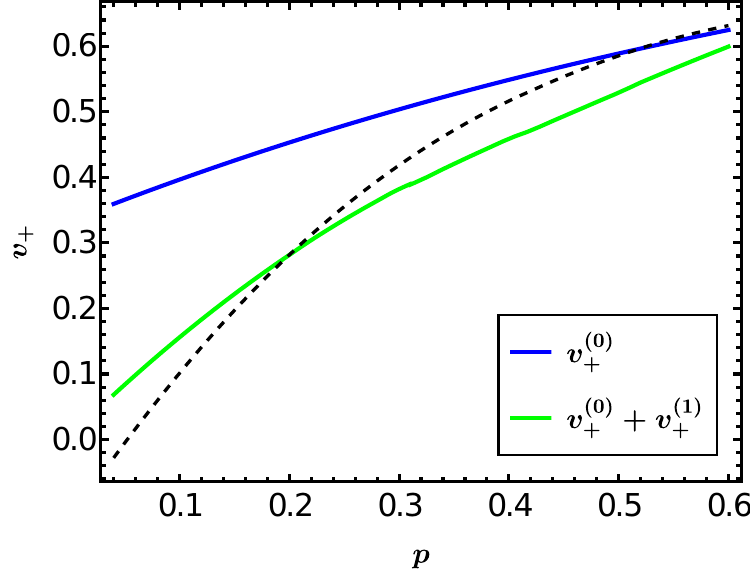} 
	\qquad
	\includegraphics[scale=0.9,keepaspectratio]{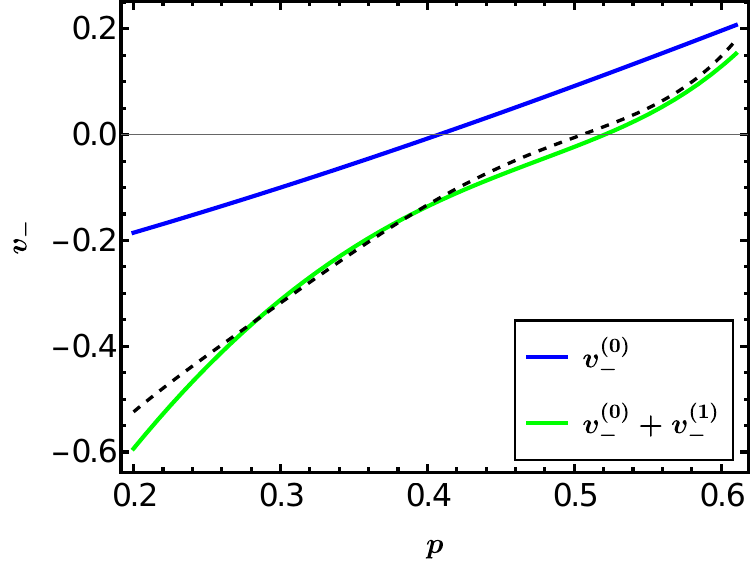}
	\caption{Velocity variation with respect to soft momentum $p/m_{q}$ for quark mode and plasmino mode respectively. The solid and dotted line corresponds to $N_{f} =2,3$ flavors, respectively.}
	\label{fig_vel_+1_-1}
\end{figure} 
\qquad
\begin{figure}
	\centering
	\includegraphics[scale=0.8,keepaspectratio]{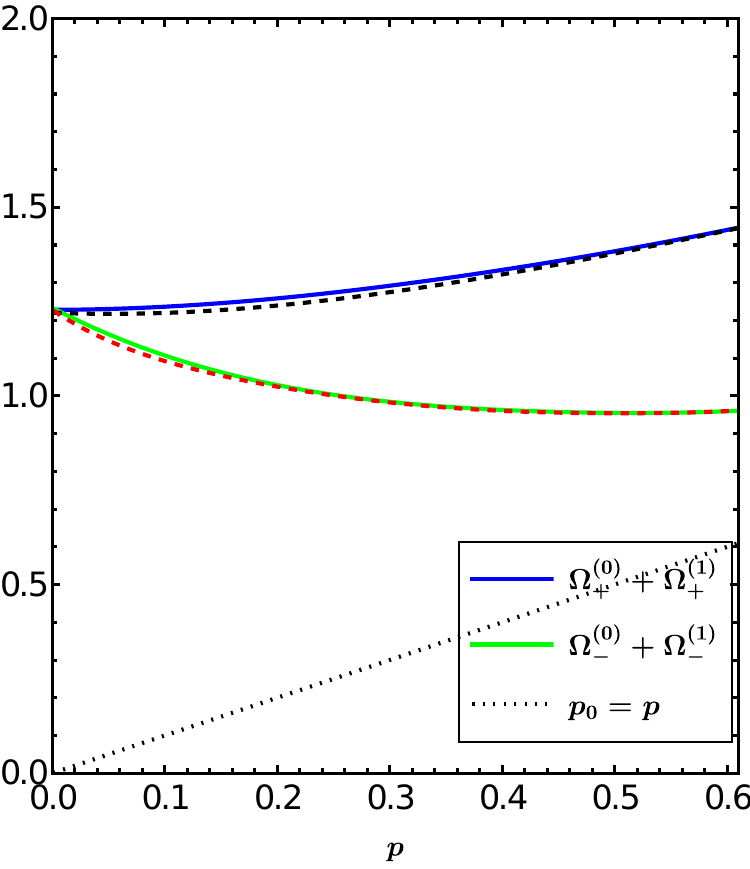} 
	\qquad \qquad
	\includegraphics[scale=0.8,keepaspectratio]{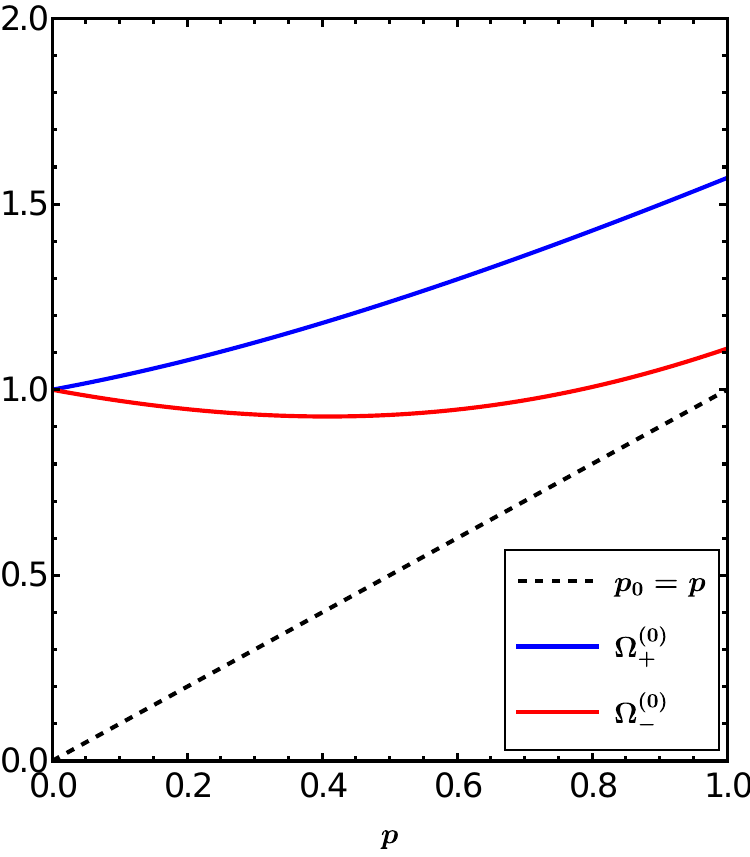}
	\caption{\textit{Left Panel:} Comparison of the real part of dispersion relation variation with respect to soft momentum $p/m_{q}$ for both quark modes. The solid and dotted line corresponds to $N_{f} =2,3$ flavors, respectively.\ \textit{Right Panel:} LO dispersion relations are plotted for the two modes along with the massless free mode.}
	\label{fig_Mass_+1_-1}
\end{figure} 

\section{Summary}\label{sec:summary}
In the present work, we have studied the NLO quark self-energy and their corresponding dispersion relations using the HTL resummation. To study the NLO quark self-energy, we have used the real-time formalism of Keldysh indices, and the considered quarks are the slow-moving ones. The solution of the NLO dispersion laws gives us physical quantities like  NLO damping rate and NLO masses, and these observables come from the zeros of the HTL-dressed quark propagators. In the lowest order, the solution of the quark dispersion relation $\Omega_{\pm}(p)$ is real. To get the NLO contribution of the above-mentioned physical quantities, one needs to evaluate NLO quark self-energy (see eq.~\eqref{disp_nlo}). 
In the current work, the NLO part of the quark self-energies $\Sigma_{\pm}^{(+1)}$ is given using the loop-four momenta integrals, which involves the effective HTL quark, gluon propagators, and three- and four-point vertex functions as done in refs.~\cite{Carrington:2006gb, Abada:2014bma}. The effective three- and four-point vertex functions are derived separately using HTL approximation and expressed in terms of solid-angle integrals using the standard technique. The above ones are rewritten using the standard Feynman parameterization technique to evaluate these integrals. Further, we numerically evaluated the transverse and longitudinal parts of the expression mentioned in eq.~\eqref{sigma_final}. The standard way of doing so is to use the spectral representation of the under-consideration dressed propagators, but we have tackled the integrals directly, which is non-trivial. One main difficulty in computing the integrals directly is the jumps the integrands encounter due to the divergences present in the propagators, more specifically when the fine-tuning parameter of the integrals $\varepsilon$ approaches zero. In evaluating the transverse and longitudinal parts of eq.~\eqref{sigma_final}, we encountered the divergences arising from the integrands. This instability comes mainly from the transverse and longitudinal parts of the gluon propagator. To overcome these divergences, we have broken down the integration into some appropriate domains, and then integration has been done in each of the appropriate domains. The $\varepsilon$  dependence of all terms in NLO quark self-energy has been carried out independently, i.e., we have checked first the stability of all the terms involved in the NLO quark self-energy. After that, the usual integration has been done for that particular value of $\varepsilon$. The domain integration technique comes out to be very useful in order to handle those extensive integrations. In the end, we summed up all the contributions from the transverse and longitudinal terms, and we studied the dependence of the NLO quark self-energy on the variable $t = p/p_0$. Lastly, using the eq.~\eqref{disp_nlo}, we plotted the NLO correction to dispersion relations. Also, the results obtained in the manuscript can be used more precisely to evaluate physical quantities related to transport phenomena.
\section*{Acknowledgments}
We want to thank Abdessamad Abada for clarifying some of the points of ref.~\cite{Abada:2014bma} in the early stage of this work. Sumit would like to acknowledge the hospitality of NISER, where most of the work is done. N. H. is supported in part by the SERB-MATRICS under Grant No. MTR/2021/000939.
\appendix
\section{HTL dressed vertex integrals} \label{sec:appendixA}
This appendix summarises the derivation of three- and four-point vertex functions within HTL approximation in the CTP formalism. The notation in this appendix is the same as in~\cite{Abada:2014bma}.
\subsection{Two quark and one gluon vertex integral}
The one-loop quark-gluon vertex function within HTL approximation is defined as 
\begin{equation}
\Gamma^{\mu} = \gamma^{\mu} + \delta\Gamma^{\mu}.
\end{equation}
where $\gamma^{\mu}$ is the bare vertex contribution and  $\delta\Gamma^{\mu} $ is the one-loop HTL correction.
In the $\{12\}$ basis of the Keldysh indices, the bare vertex $\gamma^{\mu}$ is 
\begin{equation}
\gamma_{i j k}^{\mu}=\left\{\begin{array}{l}
(-1)^{i-1} \gamma^{\mu} \quad \text { when } \quad i=j=k \\
0 \quad \text { otherwise }
\end{array}\right. ,
\end{equation}
with $i,j,k = 1,2$.
The one-loop diagrams which contribute to the quark-gluon three vertex function are shown in figure~\ref{3PT.}. The HTL contributions $\delta\Gamma^{\mu}$ to the three-point vertex can be obtained from the one-loop diagrams shown in figure~\ref{3PT.} as
\begin{equation}
\delta \Gamma_{i j k}^{\mu}(P, Q, R)=4 i g^{2} C_{F} \int \frac{d^{4} K}{(2 \pi)^{4}} K^{\mu} \slashed{K} V_{i j k}^{\prime}-2 i g^{2} N_{c} \int \frac{d^{4} K}{(2 \pi)^{4}} K^{\mu} \slashed{K}\left(V_{i j k}^{\prime}+V_{i j k}\right),\label{Gamma_V}
\end{equation}
We can use HTL approximation to neglect the external momenta compared to the loop momentum $K$. The functions $V_{i j k}$ and $V_{i j k}'$ are
\ba
V_{i j k} &=&(-1)^{i+j+k-3} \bar{D}_{i j}(K) D_{j k}(K-Q) D_{k i}(K+P) ;\nn
V_{i j k}^{\prime} &=&(-1)^{i+j+k-3} D_{i j}(K) \bar{D}_{j k}(K-Q) \bar{D}_{k i}(K+P) .\label{Vijk_def} 
\ea
The functions $D_{ij}(K)$ are the bare bosonic propagator defined in the $\{12\}$ basis as
\ba
D(K) &=&\left(\begin{array}{ll}
	D_{11}(K) & D_{12}(K) \\
	D_{21}(K) & D_{22}(K)
\end{array}\right) \nn
&=&\left(\begin{array}{cc}
	\frac{1+n_{B}\left(k_{0}\right)}{K^{2}+i \varepsilon}-\frac{n_{B}\left(k_{0}\right)}{K^{2}-i \varepsilon} & \frac{\Theta\left(-k_{0}\right)+n_{B}\left(k_{0}\right)}{K^{2}+i \varepsilon}-\frac{\Theta\left(-k_{0}\right)+n_{B}\left(k_{0}\right)}{K^{2}-i \varepsilon} \\
	\frac{\Theta\left(k_{0}\right)+n_{B}\left(k_{0}\right)}{K^{2}+i \varepsilon}-\frac{\Theta\left(k_{0}\right)+n_{B}\left(k_{0}\right)}{K^{2}-i \varepsilon} & \frac{n_{B}\left(k_{0}\right)}{K^{2}+i \varepsilon}-\frac{1+n_{B}\left(k_{0}\right)}{K^{2}-i \varepsilon}
\end{array}\right) .
\ea
The quantity $ \bar{D}_{ij}(K)$ is a fermionic propagator having the same expression as $D_{ij}(K)$ except B.E. distribution function $n_{B}(k_0)$ is replaced by the negative of the F.D. distribution function $-n_{F}(k_0)$. We can express the functions $D_{ij}(K)$ in terms of  symmetric $(F)$ , advanced $(A)$ and retarded $(R)$ propagators as
\ba
&D_{11}=\frac{1}{2}(F+A+R) ; \quad D_{12}=\frac{1}{2}(F+A-R) ;\nn
&D_{21}=\frac{1}{2}(F-A+R) ; \quad D_{22}=\frac{1}{2}(F-A-R) ,\label{gluon_propagator}
\ea
with
\ba
R(K)&=&\frac{\Theta\left(k_{0}\right)}{K^{2}+i \varepsilon}+\frac{\Theta\left(-k_{0}\right)}{K^{2}-i \varepsilon}; \nn
A(K)&=&\frac{\Theta\left(-k_{0}\right)}{K^{2}+i \varepsilon}+\frac{\Theta\left(k_{0}\right)}{K^{2}-i \varepsilon} ;\nn
F(K)&=&\left(1 \pm 2 n_{B, F}\left(k_{0}\right)\right)\left(\frac{1}{K^{2}+i \varepsilon}-\frac{1}{K^{2}-i \varepsilon}\right).
\ea
\begin{figure}[tbh]
	\centering
	\includegraphics[scale=1.5,keepaspectratio]{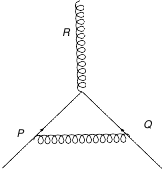} 
	\qquad
	\includegraphics[scale=1.3,keepaspectratio]{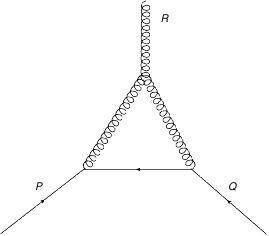}
	\caption{One-loop Feynman diagrams for the two-quarks-one-gluon vertex function.}
	\label{3PT.}
\end{figure}
The vertex functions in the $\{R.A.\}$ basis have linear correspondence with the functions defined in the $\{12\}$ basis. For example, the three-point vertex function $\Gamma_{RAA}$ in $\{R.A.\}$ basis is given by the following relation
\begin{equation}
\Gamma_{RAA} = \Gamma_{111} + \Gamma_{112} +\Gamma_{121} +\Gamma_{122}.\label{RA_12}
\end{equation}
As the vertex functions are related with the functions $V$ and $V^{\prime}$ via the equation~\eqref{Gamma_V},  the relation in eq.~\eqref{RA_12} also applies to the functions $V$ and $V^{\prime}$. Utilizing the expressions in eq.~\eqref{Vijk_def} of the functions $V$ and $ V^{\prime}$, and the relations in eq.~\eqref{gluon_propagator}, we get 
\begin{equation}\label{V_func}
\begin{aligned}
V_{RAA} = \frac{1}{2} \left(A_1 A_2 A_3 + R_1 R_2 R_3 + F_1 A_2 A_3 + R_1 F_2 A_3 + R_1 R_2 F_3  \right) ,
\end{aligned}
\end{equation}
where, for short, $1$, $2$, and $3$ represents the arguments $K$, $K-Q$, and $K+P$, respectively. The integration over the loop momentum $K$ is done in two steps: first, over $k_0$, performed using the residue theorem in the complex $k_0$ plane. In this case, both the terms $A_1 A_2 A_3$ and $R_1 R_2 R_3 $ vanish. The other three terms become
\begin{equation}\label{three_terms}
\begin{aligned}
&F A A=\frac{+i}{8 \pi^{2}} \int_{0}^{+\infty} k d k N_{B, F}(k) \int \frac{d \Omega_{s}}{4 \pi} \frac{S^{\mu} \slashed S}{(P S-i \varepsilon)(Q S+i \varepsilon)} ;\\
&R F A=\frac{-i}{8 \pi^{2}} \int_{0}^{+\infty} k d k N_{B, F}(k) \int \frac{d \Omega_{s}}{4 \pi} \frac{S^{\mu} \slashed S}{(Q S+i \varepsilon)((P+Q) S-i \varepsilon)} ;\\
&R R F=\frac{-i}{8 \pi^{2}} \int_{0}^{+\infty} k d k N_{B, F}(k) \int \frac{d \Omega_{s}}{4 \pi} \frac{S^{\mu} \slashed S}{(PS - i\varepsilon)((P+Q) S-i \varepsilon)} . 
\end{aligned}
\end{equation}
The functions $N_{B,F} (k)$ are defined in eq.~\eqref{dist_func} and $S = (1,\hat{s})$ is a time-like unit four-vector in which $\hat{s}$ is defined as $\vec{k}/k$. Since every external momentum is neglected in front of  the loop momentum $K$, when we sum up all the contributions together as mentioned in eq.~\eqref{three_terms}, we get the final expression as
\ba
\delta \Gamma_{R A A} &=&-m_{q}^{2} \int \frac{d \Omega}{4 \pi} \frac{S^{\mu} \slashed S}{(P S-i \varepsilon)(Q S+i \varepsilon)} , 
\ea
with the thermal quark mass $m_q = \sqrt{C_F/8}gT$. Similarly, the other three-point HTL vertex functions can be obtained in the $\{R.A. \}$ and $\{12\}$ bases analogously to eq.~\eqref{RA_12} as shown in \cite{Abada:2014bma} as well. For each of those other vertex functions, one can perform the same steps as done for $\Gamma_{RAA}$ and eventually, one finds: 
\begin{equation}
\begin{aligned}
&\delta \Gamma_{A R A}(P, Q, R)=-m_{q}^{2} \int \frac{d \Omega_{s}}{4 \pi} \frac{S^{\mu} \slashed S}{(P S+i \varepsilon)(Q S-i \varepsilon)} ;  \\
&\delta \Gamma_{A A R}(P, Q, R)=-m_{q}^{2} \int \frac{d \Omega_{s}}{4 \pi} \frac{S^{\mu} \slashed S}{(P S+i \varepsilon)(Q S+i \varepsilon)} ; \\
&\delta \Gamma_{R R R}(P, Q, R)=-m_{q}^{2} \int \frac{d \Omega_{s}}{4 \pi} \frac{S^{\mu} \slashed S}{(P S-i \varepsilon)(Q S-i \varepsilon)} ; \\
&\delta \Gamma_{R R A}(P, Q, R)=\delta \Gamma_{R A R}(P, Q, R)=\delta \Gamma_{A R R}(P, Q, R)=0 . 
\end{aligned}
\end{equation} 
\subsection{Two quark and two gluon vertex integral}
%
Figure~\ref{two_quark_2_gluons} shows the one-loop diagrams contributing to the quark-gluon four vertex function. The contribution coming from two-gluon-two-quark vertex functions is written in the $\{12\}$ basis as shown in eq.~\eqref{Gamma_mu_nu}.
\ba
\delta \Gamma_{i j k l}^{\mu \nu}(P, Q, R, U)&=&-8 i g^{2}(-1)^{i+j+k+l} \int \frac{d^{4} K}{(2 \pi)^{4}} K^{\mu} K^{\nu} \slashed{K} 
\times \big[ C_{F} D_{i j}(K) \bar{D}_{j k}(K-Q) \bar{D}_{k l}(K+P+U) \bar{D}_{l i}(K+P)\nn
&&\hspace{-3cm}- N_{c} \bar{D}_{i j}(K) D_{j k}(K-Q) D_{k l}(K+P+U) D_{l i}(K+P) 
+\frac{ N_{c}}{2} D_{l i}(K-P) D_{j l}(K-P-U) \bar{D}_{k j}(K+R) \bar{D}_{i k}(K)\big].\hspace{1.5cm}
\label{Gamma_mu_nu}
\ea
\begin{figure}
	\centering
	\includegraphics[width=9.cm,height=8.cm,keepaspectratio]{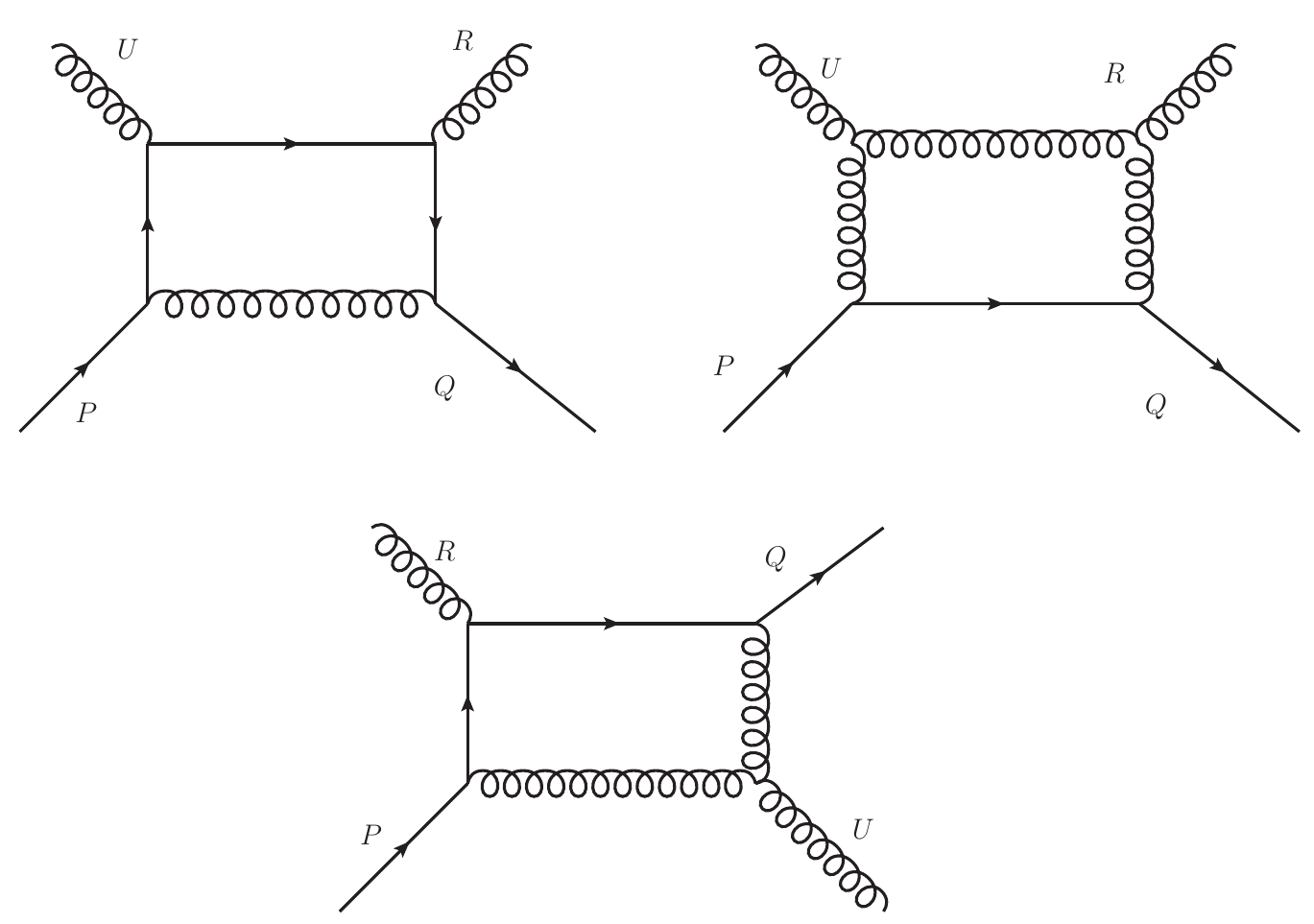} 
	\caption{One-loop Feynman diagrams for the two-quarks-two-gluons vertex function.}
	\label{two_quark_2_gluons}
\end{figure}
Let's consider the $\{RA\}$-component $\delta\Gamma_{RAAA}$, given by 
\begin{equation}
\delta\Gamma_{RAAA} = \delta\Gamma_{1111} + \delta\Gamma_{1112} + \delta\Gamma_{1121} + \delta\Gamma_{1211} + \delta\Gamma_{1122} + \delta\Gamma_{1212} + \delta\Gamma_{1221} + \delta\Gamma_{1222} 
\end{equation}
We denote the terms with the factor $C_F$ in eq.~\eqref{Gamma_mu_nu} as $\delta\Gamma^{1}_{RAAA}$. Using the decomposition as done in eq.~\eqref{gluon_propagator}, we obtain, in a similar symbolic notation as in eq.~\eqref{V_func}, the relation
\begin{equation}\label{Gamma_RAAA}
\delta\Gamma^{1}_{RAAA} = \frac{1}{2}F_1 A_2 A_3 A_4 + \frac{1}{2}R_1 F_2 A_3 A_4 +\frac{1}{2}R_1 R_2 F_3 A_4 +\frac{1}{2}R_1 R_2 R_3 F_4  
\end{equation}
The subscripts $1,2,3,$ and $4$ denote the momenta $K, K-Q, K+P+U$, and $K+P$, respectively. The two contributions $ \frac{1}{2}A_1 A_2 A_3 A_4$ and $\frac{1}{2}R_1 R_2 R_3 R_4$ to $ \delta\Gamma^{1}_{RAAA} $ gives zero each in the $k_0$-complex-plane integration and are not shown explicitly in eq.~\eqref{Gamma_RAAA}. The different terms in eq.~\eqref{Gamma_RAAA} give a contribution as follows
\ba
F A A A &=&\int_{0}^{\infty} \frac{k d k}{16 \pi^{2}} \int \frac{d \Omega_{s}}{4 \pi} \frac{i N_{B}(k) S^{\mu} S^{\nu} \slashed S}{((2 P+U) S+i \varepsilon)((P-R) S+i \varepsilon)(P S+i \varepsilon)}, \nn
R F A A &=&\int_{0}^{\infty} \frac{k d k}{16 \pi^{2}} \int \frac{d \Omega_{s}}{4 \pi} \frac{-i N_{B}(k) K^{\mu} K^{\nu} \slashed S}{((2 P+U) S+i \varepsilon)((R+P+U) S-i \varepsilon)((P+U) S-i \varepsilon)}, \nn
R R F A &=&\int_{0}^{\infty} \frac{k d k}{16 \pi^{2}} \int \frac{d \Omega_{s}}{4 \pi} \frac{i N_{F}(k) S^{\mu} S^{\nu} \slashed S}{((R-P) S-i \varepsilon)((R+P+U) S-i \varepsilon)(R S+i \varepsilon)}, \nn
R R R F &=&\int_{0}^{\infty} \frac{k d k}{16 \pi^{2}} \int \frac{d \Omega_{s}}{4 \pi} \frac{i N_{F}(k) S^{\mu} S^{\nu} \slashed S}{(P S+i \varepsilon)((P+U) S-i \varepsilon)(R S+i \varepsilon)}.\label{four_terms}
\ea
Adding all the terms in eq.~\eqref{four_terms} gives the following results
\begin{equation}
\delta \Gamma_{R A A A}^{1}=\frac{i}{8 \pi^{2}} \int_{0}^{+\infty} k d k \int \frac{d \Omega_{s}}{4 \pi} \frac{\left(n_{B}(k)+n_{F}(k)\right) S^{\mu} S^{\nu} \slashed S}{(P S-i \varepsilon)(Q S+i \varepsilon)((P+U) S-i \varepsilon)}.
\end{equation}
Now, the integrations over the three-momenta $k$ can be computed analytically. We calculate the terms with the factors $N_c$ and $N_{c}/2$ in eq.~\eqref{Gamma_mu_nu} similarly, and they cancel each other. Finally, the expression for the hard-thermal-loop four-vertex function becomes
\ba
\delta \Gamma_{R A A A}(P, Q, R, U)&=&m_{q}^{2} \int \frac{d \Omega_{s}}{4 \pi} \frac{S^{\mu} S^{\nu} \slashed S}{(P S-i \varepsilon)(Q S+i \varepsilon)}
\left[\frac{1}{(P+U) S-i \varepsilon}
+\frac{1}{(P+R) S-i \varepsilon}\right].
\ea
The other $\{RA\}$ four-vertex HTLs can be worked out similarly and 
one finds the following results
\ba
\delta \Gamma_{A R A A}(P, Q, R, U)&=& m_{q}^{2} \int \frac{d \Omega_{s}}{4 \pi} \frac{S^{\mu} S^{\nu} \slashed S}{(P S+i \varepsilon)(Q S-i \varepsilon)} 
\times\left[\frac{1}{(P+U) S+i \varepsilon}+\frac{1}{(P+R) S+i \varepsilon}\right], \nn
\delta \Gamma_{A A R A}(P, Q, R, U)&=& m_{q}^{2} \int \frac{d \Omega_{s}}{4 \pi} \frac{S^{\mu} S^{\nu} \slashed S}{(P S+i \varepsilon)(Q S+i \varepsilon)} 
\times\left[\frac{1}{(P+U) S+i \varepsilon}+\frac{1}{(P+R) S-i \varepsilon}\right], \nn
\delta \Gamma_{\text {AAAR }}(P, Q, R, U)&=& m_{q}^{2} \int \frac{d \Omega_{s}}{4 \pi} \frac{S^{\mu} S^{\nu} \slashed S}{(P S+i \varepsilon)(Q S+i \varepsilon)} 
\times\left[\frac{1}{(P+U) S-i \varepsilon}+\frac{1}{(P+R) S+i \varepsilon}\right] , \nn
\delta \Gamma_{R R A A}(P, Q, R, U)&=& \delta \Gamma_{R A R A}(P, Q, R, U)=\delta \Gamma_{R A A R}(P, Q, R, U)=0 .
\ea
The remaining eight components of four-vertex functions can be computed directly or obtained from the above ones using the KMS conditions.
\subsection{Change of Notations}
The notations used in the main text can be connected with the notation used in this appendix in the following manner
\begin{equation}
\begin{aligned}
\Gamma_{I_{1} I_{2} I_{3}}^{\mu}\left(P_{1}, P_{2}, P_{3}\right) &=\Gamma_{i_{1} i_{3} i_{2}}^{\mu}\left(P_{1},-P_{2}\right); \\
\Gamma_{I_{1} I_{2} I_{3} I_{4}}^{\mu \nu}\left(P_{1}, P_{2}, P_{3}, P_{4}\right) &=\Gamma_{i_{1} i_{4} i_{3} i_{2}}^{\mu \nu}\left(P_{1}, P_{4}, P_{3},-P_{2}\right),
\end{aligned}
\end{equation}
Here, \begin{equation*}
I_{j}=R(A) \leftrightarrow i_{j}=\mathrm{a}(\mathrm{r})
\end{equation*}
So, the three-point vertex functions are
\ba
\Gamma_{\text {arr }}^{\mu}(P, Q)&=&\Gamma_{R A A}^{\mu}(P,-Q, R)=\gamma^{\mu}+I_{--}^{\mu}(P, Q); \nn
\Gamma_{\text {rar }}^{\mu}(P, Q)&=&\Gamma_{A A R}^{\mu}(P,-Q, R)=\gamma^{\mu}+I_{+-}^{\mu}(P, Q);\nn
\Gamma_{\mathrm{aar}}^{\mu}(P, Q) &=&\Gamma_{R A R}^{\mu}(P,-Q, R)=0;\nn
\Gamma_{\mathrm{rra}}^{\mu}(P, Q) &=&\Gamma_{A R A}^{\mu}(P,-Q, R)=\gamma^{\mu}+I_{++}^{\mu}(P, Q); \nn
\Gamma_{\mathrm{ara}}^{\mu}(P, Q) &=&\Gamma_{R R A}^{\mu}(P,-Q, R)=0,
\ea
with the following notation
\begin{equation}\label{I_eta1_eta2}
I_{\eta_{1} \eta_{2}}^{\mu}(P, Q)=m_{q}^{2} \int \frac{d \Omega_{s}}{4 \pi} \frac{S^{\mu}\slashed S}{\left(P S+i \eta_{1} \varepsilon\right)\left(Q S+i \eta_{2} \varepsilon\right)}.
\end{equation}
For the four-point vertex functions, we have
\ba
\Gamma_{\text {arr }}^{\mu \nu}(P, K) &\equiv& \Gamma_{\text {arrr }}^{\mu \nu}(P, K,-K, P)=\Gamma_{R A A A}^{\mu \nu}(P,-P,-Q, Q)=I_{--}^{\mu \nu}(P, K); \nn
\Gamma_{\text {aarr }}^{\mu \nu}(P, K) &\equiv& \Gamma_{\text {aarr }}^{\mu \nu}(P, K,-K, P)=\Gamma_{R A A R}^{\mu \nu}(P,-P,-Q, Q)=0; \nn
\Gamma_{\text {arar }}^{\mu \nu}(P, K) &\equiv& \Gamma_{\text {arar }}^{\mu \nu}(P, K,-K, P)=\Gamma_{R A R A}^{\mu \nu}(P,-P,-Q, Q)=0,
\ea
with the notation
\begin{equation}\label{I_mu_nu}
\begin{aligned}
I_{\eta_{1} \eta_{2}}^{\mu \nu}(P, K)=& m_{q}^{2} \int \frac{d \Omega_{s}}{4 \pi} \frac{-S^{\mu} S^{\nu} \slashed S}{\left[P S+i \eta_{1} \varepsilon\right]\left[P S+i \eta_{2} \varepsilon\right]} 
 \left[\frac{1}{(P+K) S+i \eta_{1} \varepsilon}+\frac{1}{(P-K) S+i \eta_{2} \varepsilon}\right].
\end{aligned}
\end{equation}
Equation~\eqref{I_eta1_eta2} and eq.~\eqref{I_mu_nu} have been utilised in the text, see eqs.~\eqref{solid_ang_int1},~\eqref{solid_ang_int2}.
\section{HTL vertex integrals using Feynman technique}\label{appendixB}
In this appendix, we will calculate the solid-angle integrals present in eqs.~\eqref{solid_ang_int1} and~\eqref{solid_ang_int2}. This will be done by utilizing the Feynman parametrization technique. By using eqs.~\eqref{htl_terms_1},~\eqref{htl_terms_2}, and~\eqref{sigma2_htl}, we have only two types of solid-angle integrals which we need to evaluate, namely,
\ba
J_{\eta_{1} \eta_{2}}^{\mu \alpha}(P, Q)&=& \int \frac{d \Omega_{s}}{4 \pi} \frac{S^{\mu} S^{\alpha}}{\left[P S+i \eta_{1} \varepsilon\right]\left[Q S+i \eta_{2} \varepsilon\right]} ;\label{Jmualpha}\\
I_{\eta_{1} \eta_{2}}^{\mu \nu \alpha}(P, K)&=& \int \frac{d \Omega_{s}}{4 \pi} \frac{S^{\mu} S^{\nu} S^{\alpha}}{\left[P S+i \eta_{1} \varepsilon\right]\left[P S+i \eta_{2} \varepsilon\right]} \left[\frac{1}{(P+K) S+i \eta_{1} \varepsilon}+\frac{1}{(P-K) S+i \eta_{2} \varepsilon}\right].
\ea
The `$00$' component of $J_{\eta_{1} \eta_{2}}^{\mu \alpha}$ is the simplest of all these integrals and becomes
\begin{equation}
J_{\eta_{1} \eta_{2}}^{00}(P, Q)=\int \frac{d \Omega_{s}}{4 \pi} \frac{1}{\left(P S+i \eta_{1} \varepsilon\right)\left(Q S+i \eta_{2} \varepsilon\right)}.\label{J00}
\end{equation}
Here, $S = (1,\hat{s})$ and the integration has to be done over the solid angle of the unit vector $\hat{s}$. We will drop the i$\varepsilon$ prescription for some time. Utilizing the Feynman technique, equation~\eqref{J00} becomes
\begin{equation}
J_{\eta_{1} \eta_{2}}^{00}(P, Q)=\int_{0}^{1} d u \int \frac{d \Omega_{s}}{4 \pi} \frac{1}{[(P-K u) S]^{2}}=\int_{0}^{1} \frac{d u}{(P-K u)^{2}},\label{J00_1}
\end{equation}
where $K = P-Q$. The integration over $u$ in eq.~\eqref{J00_1} can be done analytically to get
\begin{equation}\label{J00_2}
J_{\eta_{1} \eta_{2}}^{00}(P, Q)=\frac{1}{2 \sqrt{\Delta}} \ln \frac{\left(1-u_{1}\right) u_{2}}{\left(1-u_{2}\right) u_{1}},
\end{equation}
where we have used the notation $u_{1,2} = ((PK) \pm \sqrt{\Delta})/K^{2}$
and $\Delta = (PK)^{2} - P^{2}K^{2} $. Note that $ PS+ i\eta _{1} \varepsilon = p_{0} + i\eta_{1}\varepsilon - \vec{p}\cdot \hat{s}$, so i$\varepsilon$'s appearance in the final result~\eqref{J00_2} can be achieved by changing the variables $p_0 \rightarrow p_0 + i\eta_{1}\varepsilon$ and $q_{0} \rightarrow q_{0} + i\eta_{2}\varepsilon$. This will also apply to the next two terms. Now, the `$0i$' component of $J_{\eta_{1} \eta_{2}}^{\mu \alpha}$ is given by
\begin{equation}
J_{\eta_{1} \eta_{2}}^{0 i}(P, Q)=\int \frac{d \Omega_{s}}{4 \pi} \frac{\hat{s}^{i}}{\left(P S+i \eta_{1} \varepsilon\right)\left(Q S+i \eta_{2} \varepsilon\right)}.\label{J0i_def}
\end{equation}
Again, by using the same approach as done above and denoting $R = P -(P - Q)u$, eq.~\eqref{J0i_def} becomes
\begin{equation}\label{J0i}
\begin{aligned}
J_{\eta_{1} \eta_{2}}^{0 i}(P, Q) &=\int_{0}^{1} d u\left[\frac{r_{0}}{r_{0}^{2}-r^{2}}-\frac{1}{2 r} \ln \frac{r_{0}+r}{r_{0}-r}\right] \frac{r^{i}}{r^{2}}.
\end{aligned}
\end{equation}
Similarly the `$ij$' component of $J_{\eta_{1} \eta_{2}}^{\mu \alpha}$ is given by
\begin{equation}
J_{\eta_{1} \eta_{2}}^{i j}(P, Q)=\int \frac{d \Omega_{s}}{4 \pi} \frac{\hat{s}^{i} \hat{s}^{j}}{\left(P S+i \eta_{1} \varepsilon\right)\left(Q S+i \eta_{2} \varepsilon\right)}.\label{Jij_def}
\end{equation}
The spatial component `$ij$' can be computed in the similar way  $J_{\eta_{1} \eta_{2}}^{0 i}(P, Q)$ ws calculated in eq.~\eqref{J0i} and eq.~\eqref{Jij_def} becomes
\ba
J_{\eta_{1} \eta_{2}}^{i j}(P, Q) &=&\int_{0}^{1} d u\left(A_{\eta_{1} \eta_{2}} \delta^{i j}+B_{\eta_{1} \eta_{2}} \hat{r}^{i} \hat{r}^{j}\right), \label{Jij_int}
\ea
where
\ba
A_{\eta_{1} \eta_{2}} &=&-\frac{1}{r^{2}}\left(1-\frac{r_{0}}{2 r} \ln \frac{r_{0}+r}{r_{0}-r}\right) ; \quad \quad
B_{\eta_{1} \eta_{2}} =\frac{1}{r_{0}^{2}-r^{2}}+\frac{3}{r^{2}}\left(1-\frac{r_{0}}{2 r} \ln \frac{r_{0}+r}{r_{0}-r}\right) .
\ea
The integration over $u$ in eq.~\eqref{Jij_int} will be done numerically.
Considering, $P_{1,2} \equiv (p_{0} + i\eta_{1,2}\varepsilon,\vec{p})$, we rewrite
\begin{equation}\label{J_mu_nu_alpha}
\begin{aligned}
I_{\eta_{1} \eta_{2}}^{\mu \nu \alpha}(P, K) &=J_{\eta_{1} \eta_{2}}^{\mu \nu \alpha}(P, K)+J_{\eta_{2} \eta_{1}}^{\mu \nu \alpha}(P,-K) ;\\
J_{\eta_{1} \eta_{2}}^{\mu \nu \alpha}(P, K) &=\int \frac{d \Omega_{s}}{4 \pi} \frac{S^{\mu} S^{\nu} S^{\alpha}}{P_{1} S P_{2} S\left(P_{1}+K\right) S} .
\end{aligned}
\end{equation}
We can denote the scalar products as
$ A=P_{2}S $, $B= P_{1}S$,  $C=(P_{1}+K)S$ , and the four-vector as 
\ba
T \equiv (t_0,\vec{t})= P_{2} + u_{1}(P_{1}-P_{2}) +  u_{1}u_{2}K , \label{T_def}
\ea
with
\ba
t_{0} &=& u_{1}u_{2}k_{0} + iu_{1}(\eta_{1}-\eta_{2})\varepsilon + p_{0} + i\eta_{2} \varepsilon \quad \quad
\vec{t} = u_{1}u_{2}\vec{k} + \vec{p}.
\ea
With the previous definitions, $J_{\eta_{1} \eta_{2}}^{\mu \nu \alpha}$ in eq.~\eqref{J_mu_nu_alpha} becomes
\begin{equation}
J_{\eta_{1} \eta_{2}}^{\mu \nu \alpha}(P, K)=2 \int_{0}^{1} d u_{1} u_{1} \int_{0}^{1} d u_{2} \int \frac{d \Omega_{s}}{4 \pi} \frac{S^{\mu} S^{\nu} S^{\alpha}}{\left(t_{0}-\vec{t} . \hat{s}\right)^{3}}.
\end{equation}
The dependence of $ \eta_{1} \eta_{2}$ in r.h.s. comes through the zeroth component of $T$, defined in eq.~\eqref{T_def}. As one can see that the quantity $J_{\eta_{1} \eta_{2}}^{\mu \nu \alpha}(P, K)$ has symmetry in its Lorentz indices, so there will be only four independent components to work out. Thus, we will get
\begin{equation}
\begin{aligned}
J_{\eta_{1} \eta_{2}}^{000}(P, K) &=2 \int_{0}^{1} d u_{1} u_{1} \int_{0}^{1} d u_{2} \frac{t_{0}}{\left(t_{0}^{2}-t^{2}\right)^{2}}.
\end{aligned}\label{J000}
\end{equation}
\begin{equation}
\begin{aligned}
J_{\eta_{1} \eta_{2}}^{00 i}(P, K) &=2 \int_{0}^{1} d u_{1} u_{1} \int_{0}^{1} d u_{2} \frac{t^{i}}{\left(t_{0}^{2}-t^{2}\right)^{2}}.
\end{aligned}\label{J00i}
\end{equation}
The third solid-angle integral consists of a symmetric tensor of rank two as
\ba
J_{\eta_{1} \eta_{2}}^{0 i j}(P, K) &=&2 \int_{0}^{1} d u_{1} u_{1} \int_{0}^{1} d u_{2}\left(C_{\eta_{1} \eta_{2}} \delta^{i j}+D_{\eta_{1} \eta_{2}} \hat{t}^{i} \hat{t}^{j}\right) ,\label{J0ij} \ea
where
\ba
C_{\eta_{1} \eta_{2}} &=& \frac{t_{0}}{2 t^{2}\left(t_{0}^{2}-t^{2}\right)}-\frac{1}{4 t^{3}} \ln \frac{t_{0}+t}{t_{0}-t} ; \quad \quad
D_{\eta_{1} \eta_{2}} = \frac{t_{0}\left(5 t^{2}-3 t_{0}^{2}\right)}{2 t^{2}\left(t_{0}^{2}-t^{2}\right)^{2}}+\frac{3}{4 t^{3}} \ln \frac{t_{0}+t}{t_{0}-t} .\label{C_D_def}
\ea
Similarly, the fourth solid-angle integral consists of a completely symmetric tensor of rank three.
\ba
J_{\eta_{1} \eta_{2}}^{i j k}(P, K) &=&2 \int_{0}^{1} d u_{1} u_{1} \int_{0}^{1} d u_{2}\left[E_{\eta_{1} \eta_{2}}\left(\hat{t}^{i} \delta^{j k}+\hat{t}^{j} \delta^{k i}+\hat{t}^{k} \delta^{i j}\right)+F_{\eta_{1} \eta_{2}} \hat{t}^{i} \hat{t}^{j} \hat{t}^{k}\right],\label{Jijk}
\ea
with
\ba
E_{\eta_{1} \eta_{2}} &=&\frac{1}{2 t^3}\left(2+\frac{t_{0}^{2}}{t_{0}^{2}-t^{2}}-\frac{3 t_{0}}{2 t} \ln \frac{t_{0}+t}{t_{0}-t}\right) ;\quad
F_{\eta_{1} \eta_{2}} =\frac{t}{\left(t_{0}^{2}-t^{2}\right)^{2}}-\frac{5}{2 t^{3}}\left(2+\frac{t_{0}^{2}}{t_{0}^{2}-t^{2}}-\frac{3 t_{0}}{2 t} \ln \frac{t_{0}+t}{t_{0}-t}\right).\label{E_F_def}
\ea 
\section{HTL dressed propagators}\label{sec:appendixC}
This appendix summarizes the derivation of HTL-dressed transverse and longitudinal gluon and quark propagators.
\subsection{Transverse gluon propagator}
The transverse gluon HTL-dressed propagator ${D}_T (K)$ is  
\begin{equation}
D_{T}^{-1}(K)=K^{2}-m_{g}^{2}\left[1+\frac{K^{2}}{k^{2}}\left(1-\frac{k_{0}}{2 k} \ln \frac{k_{0}+k}{k_{0}-k}\right)\right]
\end{equation}
For retarded gluon propagator 
\begin{equation}
D_{\mu \nu}^{\mathrm{R}}(K) \equiv {D}_{\mu \nu}^{\mathrm{ra}}(K)= {D}_{\mu \nu}\left(k_{0}+i \varepsilon, \vec{k}\right)
\end{equation}
Thus, 
\ba
{D}_{T}^{R{(-1)}}(k,k_0,\varepsilon) &=& k_{0}^2 -\varepsilon^2 + 2k_{0} i \varepsilon -k^2 -m_{g}^{2} X
\label{glu_ret}
\ea
where 
\ba
X & =& \left(\frac{k_{0}^{2}}{k^{2}}-\frac{\varepsilon^2}{k^2}+\frac{2k_0i\varepsilon}{k^2} \right)+ \left(\frac{k_{0}+i\varepsilon}{2 k}\right)\ln A + \left(\frac{-3 i \varepsilon k_0^2}{2 k^3} \ln A -\frac{3k_{0} \varepsilon^2}{2k^3} \ln A \right)- \left(\frac{k_{0}^3 - i \varepsilon^3}{2k^3} \right)\label{X_def}
\ea
and
\begin{equation}
\ln A = \ln \frac{k_0+k+i\varepsilon}{k_0-k+i\varepsilon}
\end{equation}
As for $N_c=3$ and $N_f=2$, $ m_{g}^2 = 4 $, so eq.~\eqref{glu_ret} becomes
\ba
{D}_{T}^{R{(-1)}}(k,k_0,\varepsilon)
&=&  -\left[\left(k^{2} -k_{0}^{2} + \frac{4k_{0}^2}{k^{2}} - 2k_{0} i\varepsilon + \frac{8k_0 i \varepsilon}{k^2} \right) + \left(\frac{2k_{0}}{k} - \frac{2k_{0}^{3}}{k^3} \right) \ln A+ \left(\frac{2 i \varepsilon}{k} - \frac{2 i \varepsilon k_0^2}{k^3} \right) \ln A \right]\label{DelTRinv}
\ea
Now, the $\ln A $ term can explicitly be written as
\ba
- \ln A 
&=& \frac{1}{2} \ln \left[\frac{\left(k_0 - k\right)^2 + \varepsilon^2}{\left(k_0+k\right)^2 + \varepsilon^2} \right] + i \left[\tan^{-1} \frac{\varepsilon}{k_0 - k} -  \tan^{-1} \frac{\varepsilon}{k_0 + k} \right]\label{lnA}
\ea
Now, using the expression of $\ln A$ from eq.~\eqref{lnA}, eq~\eqref{DelTRinv} becomes
\ba
{D}_{T}^{R{(-1)}}(k,k_0,\varepsilon) &=& -\frac{4k_0^{2}}{k^{2}} - \left(k^2 - k_0^{2} \right) \left[1 - \frac{k_0}{k^3}  \ln \frac{\left(k_0 - k\right)^2 + \varepsilon^2}{\left(k_0+k\right)^2 + \varepsilon^2} \right]\nn
&-&  i \left[\frac{2k_{0}}{k^3} \left( k_0^{2} - k^2 \right) \left(\tan^{-1} \frac{\varepsilon}{k_0 - k} -  \tan^{-1} \frac{\varepsilon}{k_0 + k} \right) - \varepsilon \Theta(k_0) \right]
\ea
\subsection{Longitudinal gluon propagator}
The longitudinal part of the HTL gluon propagator ${D}_L (K)$ is given by 
\begin{equation}
{D}_{L}^{-1}(K)=\left[K^{2}+2 m_{g}^{2} \frac{K^{2}}{k^{2}}\left(1-\frac{k_{0}}{2 k} \ln \frac{k_{0}+k}{k_{0}-k}\right)\right].
\end{equation}
As the scaled longitudinal part of the HTL gluon propagator $\tilde{D}_L(K)$ is defined as
\begin{equation*}
\tilde{D}_L(K) = \frac{D_L(K)}{K^2},
\end{equation*}
So, the retarded part of the inverse of the longitudinal gluon propagator becomes
\ba
\tilde{D}_{L}^{R (-1)}(K) 
&=& \left(k_{0}^{2} -k^{2} \right)^2 \left[1+\frac{8}{k^2} - \frac{4k_0}{k^3} \ln A \right] +\mathcal{O}(\varepsilon)^3.\label{DelLR}
\ea
Using the expression of $\ln A$ from eq.~\eqref{lnA}, eq.~\eqref{DelLR} becomes
\ba
\tilde{D}_{L}^{R (-1)}(k,k_0,\varepsilon) =\left(k_{0}^{2} -k^{2} \right)^2 \left[1+ \frac{8}{k^2} + \frac{2k_0}{k^3} \ln \frac{\left(k_0 - k\right)^2 + \varepsilon^2}{\left(k_0+k\right)^2 + \varepsilon^2}  
+ i \left\{\frac{4k_0}{k^3} \left(\tan^{-1} \frac{\varepsilon}{k_0 - k} -  \tan^{-1} \frac{\varepsilon}{k_0 + k} \right) + \varepsilon \Theta(k_0) \right\} \right].\ \qquad
\ea
\subsection{Quark propagator}
The HTL-dressed quark propagator $\Delta_\pm (Q)$ is given by 
\begin{equation}
\Delta_{\pm}^{-1}(Q)=q_{0} \pm q-\frac{m_{q}^{2}}{q}\left[\mp 1+\frac{1}{2 q} m_{q}^{2}\left(q \pm q_{0}\right) \ln \frac{q_{0}+q}{q_{0}-q}\right] .
\end{equation}
The inverse of the retarded quark propagator for plasmino mode can be written as
\ba
\Delta_{-}^{R(-1)}(q,q_0,\varepsilon) 
&=& q_{0} + i \varepsilon - q - \frac{1}{q} + \frac{q_{0} - q}{2q^2}  \ln \frac{q_{0}+q+i \varepsilon}{q_{0}-q+ i \varepsilon}  + \frac{i \varepsilon}{2 q^2}  \ln \frac{q_{0}+q+i \varepsilon}{q_{0}-q+ i \varepsilon}.\label{DelQ_inv}
\ea
Using eq.~\eqref{lnA}, eq.~\eqref{DelQ_inv} becomes
\ba
\Delta_{-}^{R(-1)}(q,q_0,\varepsilon) &=& -\left[\frac{1}{q} + q - q_0 - \frac{q_0-q}{4q^2} \ln \frac{(q_{0}+q)^2+ \varepsilon^2}{(q_{0}-q)^2+ \varepsilon^2}+\frac{\varepsilon}{2 q^2}\right. 
\left.\left\{\tan^{-1} \left(\frac{\varepsilon}{q_0 + q }\right) - \tan^{-1} \left(\frac{\varepsilon}{q_0-q}\right)\right\} \right.\nn
&&\left.- i \left[\varepsilon+  \frac{\varepsilon}{4q^2} \ln  \frac{(q_{0}+q)^2+ \varepsilon^2}{(q_{0}-q)^2+ \varepsilon^2}
+ \frac{q_0-q}{2 q^2} \left\{\tan^{-1} \left( \frac{\varepsilon}{q_0 + q }\right) - \tan^{-1} \left(\frac{\varepsilon}{q_0-q}\right)\right\}\right]\right].
\ea

Similarly, the retarded quark propagator for the real quark mode comes out to be 
\ba
\Delta_{+}^{R(-1)}(q,q_0,\varepsilon) &=& q_0 + q + \frac{1}{q} - \frac{q_0+q}{4q^2} \ln \frac{(q_{0}+q)^2+ \varepsilon^2}{(q_{0}-q)^2+ \varepsilon^2}+\frac{\varepsilon}{2 q^2}\
\left\{\tan^{-1} \left(\frac{\varepsilon}{q_0 + q }\right) - \tan^{-1} \left(\frac{\varepsilon}{q_0-q}\right)\right\}\nn
&& - i \left[-\varepsilon+  \frac{\varepsilon}{4q^2} \ln  \frac{(q_{0}+q)^2+ \varepsilon^2}{(q_{0}-q)^2+ \varepsilon^2}\right.
+\left. \frac{q_0+q}{2 q^2} \left\{\tan^{-1} \left( \frac{\varepsilon}{q_0 + q }\right) - \tan^{-1} \left(\frac{\varepsilon}{q_0-q}\right)\right\}\right].
\ea


\begin{thebibliography}{99}
\bibitem{Kalashnikov:1979cy}
O.~K.~Kalashnikov and V.~V.~Klimov,
``Polarization Tensor in QCD for Finite Temperature and Density,''
Sov. J. Nucl. Phys. \textbf{31}, 699 (1980)

\bibitem{Linde:1978px}
A.~D.~Linde,
``Phase Transitions in Gauge Theories and Cosmology,''
Rept. Prog. Phys. \textbf{42}, 389 (1979)

\bibitem{Linde:1980ts}
A.~D.~Linde,
``Infrared Problem in Thermodynamics of the Yang-Mills Gas,''
Phys. Lett. B \textbf{96}, 289-292 (1980)

\bibitem{Gross:1980br}
D.~J.~Gross, R.~D.~Pisarski and L.~G.~Yaffe,
``QCD and Instantons at Finite Temperature,''
Rev. Mod. Phys. \textbf{53}, 43 (1981)

\bibitem{Klimov:1981ka}
V.~V.~Klimov,
``Spectrum of Elementary Fermi Excitations in Quark Gluon Plasma. (In Russian),''
Sov. J. Nucl. Phys. \textbf{33}, 934-935 (1981)

\bibitem{Klimov:1982bv}
V.~V.~Klimov,
Sov. Phys. JETP \textbf{55}, 199-204 (1982)

\bibitem{Weldon:1982aq}
H.~A.~Weldon,
Phys. Rev. D \textbf{26}, 1394 (1982)

\bibitem{Weldon:1982bn}
H.~A.~Weldon,
Phys. Rev. D \textbf{26}, 2789 (1982)

\bibitem{Kobes:1987bi}
R.~Kobes and G.~Kunstatter,
Phys. Rev. Lett. \textbf{61}, 392 (1988)

\bibitem{Pisarski:1988vd}
R.~D.~Pisarski,
Phys. Rev. Lett. \textbf{63}, 1129 (1989)

\bibitem{Braaten:1989mz}
E.~Braaten and R.~D.~Pisarski,
Nucl. Phys. B \textbf{337}, 569-634 (1990)

\bibitem{Braaten:1989kk}
E.~Braaten and R.~D.~Pisarski,
Phys. Rev. Lett. \textbf{64}, 1338 (1990)

\bibitem{Frenkel:1989br}
J.~Frenkel and J.~C.~Taylor,
``High Temperature Limit of Thermal QCD,''
Nucl. Phys. B \textbf{334}, 199-216 (1990)
doi:10.1016/0550-3213(90)90661-V

\bibitem{Bellac:2011kqa}
M.~L.~Bellac,
``Thermal Field Theory,''
Cambridge University Press, 1996.

\bibitem{Braaten:1990it}
E.~Braaten and R.~D.~Pisarski,
Phys. Rev. D \textbf{42}, 2156-2160 (1990)
doi:10.1103/PhysRevD.42.2156

\bibitem{Burgess:1991wc}
C.~P.~Burgess and A.~L.~Marini,
Phys. Rev. D \textbf{45}, 17-20 (1992)

\bibitem{Rebhan:1992ca}
A.~Rebhan,
Phys. Rev. D \textbf{46}, 482-483 (1992)

\bibitem{Rebhan:1993az}
A.~K.~Rebhan,
Phys. Rev. D \textbf{48} (1993), R3967-R3970
doi:10.1103/PhysRevD.48.R3967
[arXiv:hep-ph/9308232 [hep-ph]].

\bibitem{Haque:2014rua}
N.~Haque, A.~Bandyopadhyay, J.~O.~Andersen, M.~G.~Mustafa, M.~Strickland and N.~Su,
JHEP \textbf{05}, 027 (2014)

\bibitem{Haque:2013sja}
N.~Haque, J.~O.~Andersen, M.~G.~Mustafa, M.~Strickland and N.~Su,
Phys. Rev. D \textbf{89}, 061701 (2014)

\bibitem{Haque:2013qta}
N.~Haque, M.~G.~Mustafa and M.~Strickland,
JHEP \textbf{07}, 184 (2013)

\bibitem{Haque:2012my}
N.~Haque, M.~G.~Mustafa and M.~Strickland,
``Two-loop hard thermal loop pressure at finite temperature and chemical potential,''
Phys. Rev. D \textbf{87}, no.10, 105007 (2013)
doi:10.1103/PhysRevD.87.105007
[arXiv:1212.1797 [hep-ph]].

\bibitem{Andersen:2011sf}
J.~O.~Andersen, L.~E.~Leganger, M.~Strickland and N.~Su,
``Three-loop HTL QCD thermodynamics,''
JHEP \textbf{08}, 053 (2011)
doi:10.1007/JHEP08(2011)053
[arXiv:1103.2528 [hep-ph]].

\bibitem{Andersen:2010wu}
J.~O.~Andersen, L.~E.~Leganger, M.~Strickland and N.~Su,
``NNLO hard-thermal-loop thermodynamics for QCD,''
Phys. Lett. B \textbf{696}, 468-472 (2011)
doi:10.1016/j.physletb.2010.12.070
[arXiv:1009.4644 [hep-ph]].

\bibitem{Jiang:2010jz}
Y.~Jiang, H.~x.~Zhu, W.~m.~Sun and H.~s.~Zong,
J. Phys. G \textbf{37}, 055001 (2010)
doi:10.1088/0954-3899/37/5/055001
[arXiv:1003.5031 [hep-ph]].

\bibitem{Liu:2011if}
J.~Liu, M.~j.~Luo, Q.~Wang and H.~j.~Xu,
``Refractive Index of Light in the Quark-Gluon Plasma with the Hard-Thermal-Loop Perturbation Theory,''
Phys. Rev. D \textbf{84}, 125027 (2011)
doi:10.1103/PhysRevD.84.125027
[arXiv:1109.4083 [hep-ph]].

\bibitem{Muller:2021wri}
B.~M\"uller,
Phys. Rev. D \textbf{104} (2021) no.7, L071501
doi:10.1103/PhysRevD.104.L071501
[arXiv:2107.14775 [hep-ph]].

\bibitem{Carignano:2019ofj}
S.~Carignano, M.~E.~Carrington and J.~Soto,
Phys. Lett. B \textbf{801} (2020), 135193
doi:10.1016/j.physletb.2019.135193
[arXiv:1909.10545 [hep-ph]].

\bibitem{Ekstedt:2023anj}
A.~Ekstedt,
[arXiv:2302.04894 [hep-ph]].

\bibitem{Wang:2022dcw}
Y.~Wang, Q.~Du and Y.~Guo,
[arXiv:2207.06039 [hep-ph]].

\bibitem{Gorda:2022fci}
T.~Gorda, A.~Kurkela, J.~\"Osterman, R.~Paatelainen, S.~S\"appi, P.~Schicho, K.~Sepp\"anen and A.~Vuorinen,
[arXiv:2204.11279 [hep-ph]].

\bibitem{Gorda:2022zyc}
T.~Gorda, A.~Kurkela, J.~\"Osterman, R.~Paatelainen, S.~S\"appi, P.~Schicho, K.~Sepp\"anen and A.~Vuorinen,
[arXiv:2204.11893 [hep-ph]].

\bibitem{Kobes:1992ys}
R.~Kobes, G.~Kunstatter and K.~Mak,
Phys. Rev. D \textbf{45}, 4632-4639 (1992)
doi:10.1103/PhysRevD.45.4632

\bibitem{Braaten:1992gd}
E.~Braaten and R.~D.~Pisarski,
Phys. Rev. D \textbf{46}, 1829-1834 (1992)
doi:10.1103/PhysRevD.46.1829

\bibitem{Mustafa:2022got}
M.~G.~Mustafa,
[arXiv:2207.00534 [hep-ph]].

\bibitem{Kapusta:2006pm} J.I. Kapusta and C. Gale, Finite-temperature field theory: principle and applications, 2nd ed.,Cambridge Monographs on Mathematical Physics, Cambridge University Press, Cambridge U.K. (2011).

\bibitem{Carrington:2006gb}
M.~E.~Carrington,
Phys. Rev. D \textbf{75}, 045019 (2007)
doi:10.1103/PhysRevD.75.045019
[arXiv:hep-ph/0610372 [hep-ph]].

\bibitem{Abada:1998ue}
A.~Abada and O.~Azi,
Phys. Lett. B \textbf{463}, 117-125 (1999)
doi:10.1016/S0370-2693(99)00959-4
[arXiv:hep-ph/9807439 [hep-ph]].

\bibitem{Abada:1997vm}
A.~Abada, O.~Azi and K.~Benchallal,
Phys. Lett. B \textbf{425}, 158-165 (1998)
doi:10.1016/S0370-2693(98)00221-4
[arXiv:hep-ph/9712210 [hep-ph]].

\bibitem{Abada:2004dr}
A.~Abada, K.~Bouakaz and O.~Azi,
Phys. Scripta \textbf{74}, 77-103 (2006)
doi:10.1088/0031-8949/74/1/011
[arXiv:hep-ph/0402041 [hep-ph]].

\bibitem{Abada:2007opj}
A.~Abada, N.~Daira-Aifa and K.~Bouakaz,
Int. J. Mod. Phys. A \textbf{22}, 6033-6042 (2007)
doi:10.1142/S0217751X07039213

\bibitem{Abada:2005na}
A.~Abada, N.~Daira-Aifa and K.~Bouakaz,
Int. J. Mod. Phys. A \textbf{21}, 5317-5332 (2006)
doi:10.1142/S0217751X06033052
[arXiv:hep-ph/0511258 [hep-ph]].

\bibitem{Abada:2000hh}
A.~Abada, K.~Bouakaz and N.~Daira-Aifa,
Eur. Phys. J. C \textbf{18}, 765-777 (2001)
doi:10.1007/s100520100554
[arXiv:hep-ph/0008335 [hep-ph]].

\bibitem{Abada:2007zz}
A.~Abada, K.~Bouakaz and D.~Deghiche,
Mod. Phys. Lett. A \textbf{22}, 903-914 (2007)
doi:10.1142/S021773230702124X

\bibitem{Abada:2011cc}
A.~Abada and N.~Daira-Aifa,
JHEP \textbf{04}, 071 (2012)
doi:10.1007/JHEP04(2012)071
[arXiv:1112.6065 [hep-ph]].

\bibitem{Abada:2005jq}
A.~Abada and K.~Bouakaz,
JHEP \textbf{01}, 161 (2006)
doi:10.1088/1126-6708/2006/01/161
[arXiv:hep-ph/0510330 [hep-ph]].

\bibitem{Schulz:1993gf}
H.~Schulz,
Nucl. Phys. B \textbf{413}, 353-395 (1994)
doi:10.1016/0550-3213(94)90624-6
[arXiv:hep-ph/9306298 [hep-ph]].

\bibitem{Carrington:2006xj}
M.~E.~Carrington, T.~Fugleberg, D.~S.~Irvine and D.~Pickering,
Eur. Phys. J. C \textbf{50}, 711-727 (2007)
doi:10.1140/epjc/s10052-007-0276-9
[arXiv:hep-ph/0608298 [hep-ph]].

\bibitem{Carrington:2008dw}
M.~E.~Carrington, A.~Gynther and D.~Pickering,
Phys. Rev. D \textbf{78}, 045018 (2008)
doi:10.1103/PhysRevD.78.045018
[arXiv:0805.0170 [hep-ph]].

\bibitem{Landsman:1986uw}
N.~P.~Landsman and C.~G.~van Weert,
Phys. Rept. \textbf{145}, 141 (1987)
doi:10.1016/0370-1573(87)90121-9

\bibitem{Chou:1984es}
K.~c.~Chou, Z.~b.~Su, B.~l.~Hao and L.~Yu,
Phys. Rept. \textbf{118}, 1-131 (1985)
doi:10.1016/0370-1573(85)90136-X

\bibitem{Mirza:2013ula}
A.~Mirza and M.~E.~Carrington,
Phys. Rev. D \textbf{87}, 065008 (2013)
doi:10.1103/PhysRevD.87.065008
[arXiv:1302.3796 [hep-ph]].

\bibitem{Abada:2014bma}
A.~Abada, K.~Benchallal and K.~Bouakaz,
JHEP \textbf{03}, 058 (2015)
doi:10.1007/JHEP03(2015)058
[arXiv:1501.00140 [hep-ph]].

\bibitem{Martin:1959jp}
P.~C.~Martin and J.~S.~Schwinger,
Phys. Rev. \textbf{115}, 1342-1373 (1959)
doi:10.1103/PhysRev.115.1342

\bibitem{Keldysh:1964ud}
L.~V.~Keldysh,
Zh. Eksp. Teor. Fiz. \textbf{47}, 1515-1527 (1964)

\bibitem{Mustafa:2002pb}
M.~G.~Mustafa and M.~H.~Thoma,
Pramana \textbf{60}, 711-724 (2003)
doi:10.1007/BF02705170
[arXiv:hep-ph/0201060 [hep-ph]].

\bibitem{Binosi:2003yf}
D.~Binosi and L.~Theussl,
Comput. Phys. Commun. \textbf{161}, 76-86 (2004)
doi:10.1016/j.cpc.2004.05.001
[arXiv:hep-ph/0309015 [hep-ph]].

\end{thebibliography}
\end{document}